\documentclass[fleqn,usenatbib]{mnras}

\usepackage{newtxtext,newtxmath}

\usepackage[T1]{fontenc}
\usepackage{ae,aecompl}
\usepackage{array,longtable}

\usepackage{graphicx}	
\usepackage{amsmath}	
\usepackage{amssymb}	
\usepackage{xcolor}
\usepackage{float} 
\usepackage{placeins}
\usepackage{pdflscape}
\usepackage[normalem]{ulem}

\title[Parameters for 45 open clusters with Gaia DR2]{Fundamental parameters for 45 open clusters with Gaia DR2, an improved extinction correction and a metallicity gradient prior}

\author[H. Monteiro et al.]{
H. Monteiro,$^{1}$\thanks{E-mail:hektor.monteiro@gmail.com},  W. S. Dias$^{1}$, 
A. Moitinho$^{2}$, 
T. Cantat-Gaudin,$^{3}$,
J. R. D. L\'epine$^{4}$,
\newauthor G., Carraro,$^{5}$
and E. Paunzen$^{6}$ 
\\
$^{1}$Instituto de F\'isica e Qu\'imica, Universidade Federal de Itajub\'a, Av. BPS 1303 Pinheirinho, 37500-903 Itajub\'a, MG, Brazil\\
$^{2}$CENTRA, Faculdade de Ci\^encias, Universidade de Lisboa, Ed. C8, Campo Grande, 1749-016 Lisboa, Portugal\\
$^{3}$Institut de Ciencies del Cosmos, Universitat de Barcelona (IEEC-UB), Marti i Franques 1, E-08028 Barcelona, Spain\\
$^{4}$Universidade de S\~ao Paulo, Instituto de Astronomia, Geof\'isica e Ci\^encias Atmosf\'ericas, S\~ao Paulo,  SP, Brazil\\
$^{5}$Department of Physics and Astronomy, University of Padova, Vicolo dell'Osservatorio 3, I-35122 Padova, Italy 0000-0002-0155-9434\\
$^{6}$Departament of Theoretical Physics and Astrophysics. Marsaryk University. Brno, Czech Republic
}

\date{Accepted XXX. Received YYY; in original form ZZZ}

\pubyear{2020}

\begin{document}
\label{firstpage}
\pagerange{\pageref{firstpage}--\pageref{lastpage}}
\maketitle

\begin{abstract}

Reliable fundamental parameters of open clusters such as distance, age and extinction are key to our understanding of Galactic structure and stellar evolution. In this work we use {\it Gaia} DR2 to investigate 45 open clusters listed in the \emph{New catalogue of optically visible open clusters and candidates} (DAML) but with no previous astrometric membership estimation based on {\it Gaia} DR2. In the process of selecting targets for this study we found that some clusters reported as new discoveries in recent papers based on {\it Gaia} DR2 were already known clusters listed in DAML. Cluster memberships were determined using a maximum likelihood method applied to {\it Gaia} DR2 astrometry. This has allowed us to estimate mean proper motions and mean parallaxes for all investigated clusters. Mean radial velocities were also determined for 12 clusters, 7 of which had no previous published values.  We have improved our isochrone fitting code to account for interstellar extinction using an updated extinction polynomial for the {\it Gaia} DR2 photometric band-passes and the Galactic abundance gradient as a prior for metallicity. The updated procedure was validated with a sample of clusters with high quality $[Fe/H]$ determinations. We then did a critical review of the literature and verified that our cluster parameter determinations represent a substantial improvement over previous values.

\end{abstract}




\section{Introduction}

The fundamental parameters of open clusters (OCs) - distance, age, metallicity, interstellar extintion along the line of sight, proper motions and radial velocities - have long been considered key for revealing the structure and evolution of the Milky Way \citep{1970IAUS...38..205B, 1982ApJS...49..425J}.  However, because each cluster contributes with a single point in parameter space, the accumulation of OC data has traditionally been a lengthy process, with leaps in our knowledge of the Galaxy based on OCs taking many years \citep{2010IAUS..266..106M}.  Such a jump has been recently brought by the ESA Gaia mission \citep{2016A&A...595A...1G}. The {\it Gaia} Data Release 2 catalogue \citep[DR2, ][]{GAIA-DR22018} provides precise astrometric and photometric data for more than one billion stars with magnitude $G$ brighter than 21, which are bringing a new era of Galactic research with OCs. A summary of various past, pre-{\it Gaia}, efforts to compile homogeneous OC parameters is given in \citep{martin15} and a review of pre-{\it Gaia} results of Galactic structure with OCs can be found in \citet{2010IAUS..266..106M}.

The richness of {\it Gaia} DR2 has triggered numerous large scale OC studies. Without being exhaustive, we indicate some significant examples:
\citet{Cantat-Gaudin2018A&A...618A..93C} and \citet{Cantat-Gaudin2020A&A...633A..99C} determined proper motions and distances for 1481 open clusters based on membership obtained using the UPMASK membership determination method \citep{Krone-Martins2014A&A...561A..57K}.
\citet{2018A&A...619A.155S} determined proper motions and radial velocities for a kinematic study of 406 OCs.
\citet{ChinesCat} used the Friend of Friend method to flag over two thousand cluster candidates.
\citet{Kounkel2019AJ....158..122K} performed a clustering analysis to study 1900 possible aggregates within 1 kpc. Also in the solar neighborhood, \citet{Sim2019JKAS...52..145S} reported on 655 clusters (proposing 207 new candidates) by visual inspection of the stellar distributions in proper motion space and spatial distributions in the Galactic coordinates $(l,b)$ space. Members were determined using Gaussian mixture model and mean-shift algorithms.
\citet{Monteiro2019MNRAS.487.2385M} determined the parameters of 150 OCs adopting  a maximum likelihood method to estimate cluster memberships. Using the same procedure \citet{Dias2019MNRAS.486.5726D} determined the parameters of several hundreds of OCs, from which they selected 80 younger that 50 Myr for determining the spiral pattern rotation speed of the Galaxy and the corotation radius.
\cite{Bossini19} employed a Bayesian methodology for determining the ages, distances and  interstellar absorption for 269 OCs with membership determinations from \citet{Cantat-Gaudin2018A&A...618A..93C}.
\cite{Castro-Ginard2020A&A...635A..45C}, using a deep learning artificial neural network (ANN), reported the discovery of 588 new OCs for which they estimated distances and proper motions.  Likewise using an ANN to characterise  1867 OCs, \citet{2020A&A...640A...1C} analysed the spiral structure, scale height of the thin disk and warp of the Milky Way. 
It is also worthwhile mentioning that {\it Gaia} DR2 has also been used in combination with ground based observations for smaller scale, but more detailed studies 
of individual objects \citep[e.g.][]{Dias2018MNRAS.481.3887D, 2020A&A...637A..95P}. 

Despite the intense activity enabled by the high quality {\it Gaia} DR2 data, many previously known objects remain with no membership and parameter determinations based on Gaia DR2. The goal of this paper is to present our determinations of the fundamental parameters of these difficult left-over clusters and the methodological improvements that allowed to reach those results.

The remainder of the manuscript is organized as follows. In the next section, we describe the data selection and the sample of the studied objects. 
Section~3 is dedicated to describe the method of astrometric membership determination and to briefly introduce the isochrone fitting procedure. In section~4 we present improvements to our isochrone fitting procedure using a revised treatment of interstellar extinction with updated {\it Gaia} photometric band-passes and constraining metallicity. These improvements are validated with a control sample of clusters from the literature.
In section~5 we discuss the results and in section~6 we compare the values here obtained with those from the literature. Finally, in section~7 we give some concluding remarks.

\section{Cluster sample and data} \label{sec:sample}
We started by cross-matching all 2167 clusters published in the \emph{New catalog of optically visible open clusters and candidates}  \citep[][ hereafter DAML]{Dias2002} with the literature for which membership determinations using {\it Gaia} DR2 data were available \citep{Cantat-Gaudin2018A&A...618A..93C,Castro-Ginard2020A&A...635A..45C, ChinesCat, Sim2019JKAS...52..145S}. This led to a list of 75 clusters for which no previous {\it Gaia} DR2 based memberships were available. For each cluster we selected the stars in {\it Gaia} DR2, using the central coordinates and the radius taken from the DAML catalogue. 
To allow for some uncertainty in the radius and include possible cluster members further away from center, we took a region in the sky with radius 2 arcmin larger than the radius listed in DAML. We note that stars originated in the cluster might be further away due to processes such as dynamical evolution or an underestimated radius. However, for the purposes of this work, complete samples of members are not required, but only enough stars for determining the reddening, distance and age of the clusters. 
Before determining the astrometric membership as detailed in the next section, we filtered the data to assure that only reliable astrometric solutions were used. The filtering was done following the recipe published by \citet{GaiaHRD18}, which takes into account systematic effects of {\it Gaia} data, consistency between $G$ and $G_{BP}+G_{RP}$ filter fluxes, as well as the number of passes in the given field, among other factors. As described in section~\ref{sec:results}, a subsequent quality control step left us with a final sample of 45 clusters for which results are presented.

\section{Method}

\subsection{Membership determination}

The membership analysis follows the method described in \citet{Dias2014}.
We assume that errors in proper motion components and parallaxes are normally distributed and use a maximum likelihood method to obtain the memberships adopting a model which assumes Gaussian distributions for proper motions in both cluster and field stars. The model is described by equation \ref{eq:multinormal} where the uncertainties of the data and their correlations follow the recommendation given by \citet{Gaia-DR2-Luri2018} such that the probability $f(\mathbf {X})$ is given by:

\begin{equation} 
\label{eq:multinormal}
f(\mathbf {X}) = {\frac {\exp \left(-{\frac {1}{2}}({\mathbf {X} }-{\boldsymbol {\mu }})^{\mathrm {T} }{\boldsymbol {\Sigma }}^{-1}({\mathbf {X} }-{\boldsymbol {\mu }})\right)}{\sqrt {(2\pi )^{k}|{\boldsymbol {\Sigma }}|}}}
\end{equation}

where $\mathbf {X}$ is the column vector ($\mu_{\alpha}cos{\delta}$, $\mu_{\delta}$, $\varpi$) composed of the proper motion components and the parallax, $\boldsymbol {\mu }$ the mean column vector and $|{\boldsymbol {\Sigma }}|$ is the co-variance matrix, which incorporates the uncertainties ($\sigma$) and their correlations ($\rho$), given in the {\it Gaia} DR2 catalogue. 

The maximum likelihood solution  provides the distribution of cluster membership probabilities. This allows the determination of the cluster membership probability of each star in the selected field as well the mean proper motions and parallaxes of the clusters, considering as members those stars with cluster membership probability greater than 0.51. The adopted membership cut-off of 0.51 is merely based on the availability of statistical evidence for the pertinence to a given cluster and used as a compromise between completeness and contamination. As discussed in the next section, the isochrone fitting procedure will use the membership probabilities for decreasing the weight of the possible contaminants in the determination of the cluster fundamental parameters. Still, For the open clusters studied here we also ran the fits with a cut-off of 0.8 as a sanity check on the results. The differences with respect the results obtained with the 0.51 cut-off were ($0.04 \pm 0.18$) dex, ($-28.98 \pm 189.86$) pc, ($-0.02 \pm 0.05$) mag, for age, distance and $A_V$ respectively, which are are comparable to the uncertainties obtained in either case, showing that adopting one or the other cut-off is equivalent within the errors.

We also estimate radial velocity as the mean of the radial velocity data with a $3\sigma$ outlier rejection of the members. We note that Gaia DR2 radial velocities are only available for small numbers of cluster members. The estimated uncertainty is given by the standard deviation of the radial velocities. 

\subsection{Isochrone fit}\label{sec:isofits}

It is well known that the stars in an open cluster align along a distinctive sequence in a color-magnitude diagram (CMD). This sequence is most evident when only stars with a sufficiently high stellar membership probability (e.g. as determined by the method described above) are included. In other words, the sequence is most evident when field star contamination is minimum. Likewise, the member stars that form this feature should exhibit a clump in a 3D plot with proper motion and parallax data, since they occupy a limited volume in space and have similar velocities.  
In this context, a net evidence of a cluster sequence in a CMD of member stars is a strong indicator of the presence of a real open cluster and allows the determination of its age, extinction, and an estimate of the cluster distance independent of the parallax measurements.  
Consequently, the next step in our analysis was to use {\it Gaia} DR2 $G_{BP}$ and $G_{RP}$ magnitudes and to perform the isochrone fits to the cluster member stars identified with the method outlined above. 

As discussed in previous works  \citep[e.g.][]{Dias2018MNRAS.481.3887D}, membership knowledge and an objective method for isochrone fitting are determinant to the final results. We note that many isochrone fits performed in the literature, objective or not, were based on limited membership determinations, mainly due to large errors or even absence of stellar proper motions and/or parallax data.

Here we applied the cross-entropy (CE) method to fit theoretical isochrones to the CMDs of cluster member stars as detailed in \citet{Monteiro2017}. This approach has already been successfully applied to {\it Gaia} DR2 data in \citet{Dias2018MNRAS.481.3887D}, \citet{Monteiro2019MNRAS.487.2385M} and \citet{Dias2019MNRAS.486.5726D}. In short, the CE method involves an iterative statistical procedure where in each iteration  the initial sample of the fit parameters is randomly generated using predefined criteria. Then, the code selects the 10$\%$ best fits based on calculated weighted likelihood values taking into account the astrometric membership probabilities. Based on the parameter space defined by the best fits, a new random fit parameter sample is generated and applied in the following run of the code. This procedure continues until a convergence criterion is reached. In other words, the isochrone fit in this technique consists in choosing the best set of points of a model with respect  to the set of points of the observed data. The errors of the fit are estimated by bootstrapping the process. This also reduces the influence of possible field stars contaminating the lists of members.

In our code we adopt a likelihood function given in the usual manner for the maximum likelihood problem as:

\begin{equation}
\mathcal L (D_N|{\bf X}) = \prod_{i=1}^{N}\Phi(I({\bf X}),D_N), 
\label{likelihood}
\end{equation}

where $\Phi(I({\bf X}))$ is a multivariate normal, ${\bf X}$ is the vector of parameters ($A_V$, distance $d$, age $log(t)$ and $[Fe/H]$), $I({\bf X})$ is the synthetic cluster obtained for the isochrone defined by ${\bf X}$ and $D_N$ the data for the $N$ observed stars in the cluster. 

The likelihood function above is used to define the objective function as:

\begin{equation}
S({\bf X}|D_N) = -log( P({\bf X}) \times \mathcal L (D_N|{\bf X}) )
\end{equation}

where the function $P({\bf X})$ is the prior probability for the parameters given by $P({\bf X}) = \prod_{n=0}^{n}P(X_n)$. 
For age we adopt $P(X_n) = 1$, for distance we use $\mathcal{N}(\mu,\,\sigma^{2})$ obtained with Bayesian inference from the parallax ($\varpi$) and its uncertainty ($\sigma_{\varpi}$) and the variance ($\sigma^2$) is obtained from the distance interval calculated from the inference using the uncertainty as $1\sigma_{\varpi}$. The prior in $A_V$ is also adopted as a normal distribution with $\mu$ and variance ($\sigma^{2}$) for each cluster taken from the 3D extinction map produced by \citet{3Debv}\footnote{The 3D extinction map is available online at \url{https://stilism.obspm.fr/}}. The prior for $[Fe/H]$ used the the Galactic gradient from \cite{OCCAMgradient20} as detailed in the Section~\ref{sec:validation}. 
The optimization algorithm then minimizes with respect to ${\bf X}$.

In the present study, our algorithm uses the Padova PARSEC version 1.2S database of stellar evolutionary tracks and isochrones \citep{Bressan2012}, which uses the {\it Gaia} filter passbands of \cite{Maiz18}, is scaled to solar metal content with $Z_{\odot} = 0.0152$  and scans the following parameter space limits: 
\begin{itemize}
\item age: from log(age) =6.60 to log(age) =10.15;
\item distance: from 1 to 25000 pc;
\item $A_V$: from 0.0 to 5.0 mag; 
\item $[Fe/H]$: from -0.90 to +0.70 dex
\end{itemize}

 Since our method uses a synthetic cluster obtained from model isochrones, we include the extinction for each star generated based on a $A_{\lambda}/A_V$ relation of choice.  For each generated star of the synthetic cluster we obtain, in each filter, what would be the reddened observed photometry for the particular model $I({\bf X})$. The synthetic clusters have been generated with a binary fraction of 0.5 and the masses of components drawn from the same IMF. The synthetic cluster is then compared to the observed data through the likelihood defined in Eq. \ref{likelihood}.  


\section{Improvements to the inference of cluster parameters}

When analyzing the clusters with the software of \citet{Monteiro2017} described in the previous section, we noticed that  about 20\% (8 clusters) of the fits would only converge to consistent solutions when only $G_{BP}$ and $G_{RP}$ magnitudes were used, without using $G$. For most of these clusters the extinction was considerable, reaching as high as $A_V=2.9$. We had originally adopted the same polynomial as \cite{Bossini19} to correct for extinction, although they only investigated clusters with low $A_V$ and used the now outdated band passes. Therefore, we decided to redo the extinction polynomial based on the updated {\it Gaia} filter band-passes. In the process, we  analyzed different approaches for constraining another key parameter: metallicity.

\subsection{Revised Gaia extinction polynomial}\label{sec:extinction_poly}
To account for the extinction coefficients dependency on colour and extinction due to the large passbands of {\it Gaia} filters, we followed the procedure described by \citet{Danielski2018} and used in \cite{GaiaHRD18}. We used the same model atmospheres and same value grid: Kurucz model spectra \citep{2003IAUS..210P.A20C} (for 3500~K$<T_{\rm eff}<$10000~K in steps of 250~K and two surfaces gravities: log$g$=2.5 and 4. For the extinction law we adopted the more recent one from \cite{FM19extinction} and a grid of 0.01$<A_V$<5~mag in steps of 0.01~mag for the calculations. We also use the more up-to-date \cite{Maiz18} revised {\it Gaia} photometric passbands, given that these bands provide better agreement between synthetic {\it Gaia} photometry and {\it Gaia} observations.

The model spectra were convolved with the filter passbands and extinction scaled law to construct a grid of reddened photometry. The extinction coefficients k$_m$ were calculated with the equations below:

\begin{equation}
A_m = m - m_0 = -2.5 \log_{10}\left(\frac{\int F_\lambda T_\lambda E_\lambda^{A_V} d\lambda}{\int F_\lambda T_\lambda d\lambda}\right) 
\end{equation} 

and 

\begin{equation}
k_m = \frac{A_m}{A_V }
\end{equation} 

where $E_\lambda^{A_V}$ is the extinction function, which in this case was the \cite{FM19extinction} law.

A polynomial defined as in Eq. \ref{eq:polynomial} was then fit to the $k_m$ versus $A_V$ grid of values using the package \citep{astropy}. In this expression $x$ and $y$ are $A_V$ and $G_{BP} - G_{RP}$, respectively, $k_m$ is the extinction coefficient and the $m$ subscript refers to each of the bands $G$, $G_{BP}$ and $G_{RP}$. Unlike the work in \cite{GaiaHRD18}, here we fit a full 4th degree polynomial to the grid. The results of the fit are given in Table~\ref{tab:polynomial}.

\begin{equation}
\begin{split}
k(x,y)=c_{00}+c_{10}x+...+c_{n0}x^n+c_{01}y+... \\
+c_{0n}y_n+c_{11}xy+c_{12}xy^2+... \\
+c_{1(n-1)}xy^{n-1}+...+c_{(n-1)1}x^{n-1}y
\end{split}
\label{eq:polynomial}
\end{equation} 


\begin{table*}
\caption{Coefficients of the polynomial fit to the $k_m$ versus $A_V$ grid of values.}
\resizebox{\textwidth}{!}{
\begin{tabular}{lrrrrrrrrrrrrrrr}
\hline
Band & 
\multicolumn{1}{c}{$c_{00}$} & 
\multicolumn{1}{c}{$c_{10}$} & 
\multicolumn{1}{c}{$c_{20}$} & 
\multicolumn{1}{c}{$c_{30}$} & 
\multicolumn{1}{c}{$c_{40}$} & 
\multicolumn{1}{c}{$c_{01}$} & 
\multicolumn{1}{c}{$c_{02}$} & 
\multicolumn{1}{c}{$c_{03}$} & 
\multicolumn{1}{c}{$c_{04}$} & 
\multicolumn{1}{c}{$c_{11}$} & 
\multicolumn{1}{c}{$c_{12}$} & 
\multicolumn{1}{c}{$c_{13}$} & 
\multicolumn{1}{c}{$c_{21}$} & 
\multicolumn{1}{c}{$c_{22}$} & 
\multicolumn{1}{c}{$c_{31}$} \\
\hline
\hline
$G_{BP}$ &1.2002  & 0.0599   & 0.0139   & 0.0017   & 0.0001  & -0.1602   & 0.0625  & -0.0317   & 0.0074  & -0.0665   & 0.0433  & -0.0119  & -0.0163   & 0.0066  & -0.0016 \\
$G_{RP}$ & 0.6692   & 0.0172   & 0.0098   & 0.0018   & 0.0001  & -0.0451   & 0.0439  & -0.0259   & 0.0043  & -0.0433   & 0.0336  & -0.0070  & -0.0138   & 0.0040  & -0.0010 \\
$G$ & 0.9937   & 0.0342  & -0.0003  & -0.0008  & -0.0001  & -0.1292  & -0.0217   & 0.0164  & -0.0024  & 0.0051  & -0.0134   & 0.0033   & 0.0050  & -0.0021   & 0.0007 \\
\hline
\end{tabular}}
\label{tab:polynomial}
\end{table*}

Our results agree with the ones obtained by \cite{Wang19}, using a different method. Specifically, they derive their own extinction law and do not fit a polynomial to the $A_V$-color dependence, but do apply corrections for the large filter passbands. They obtain  $1.002 \pm 0.007$ for k$_{BP}$ and $0.589 \pm 0.004$ for k$_{RP}$. Our average results from the polynomial fit  are $1.072 \pm 0.065$ and $0.634 \pm 0.021$ for k$_{BP}$ and k$_{RP}$ respectively. For the G filter we get $0.832 \pm 0.077$ while \cite{Wang19} obtained $0.789 \pm 0.005$ for k$_{G}$.

\subsection{Metallicity}\label{sec:validation}
To validate and to determine possible limitations of the new extinction polynomial we have applied our code to a sample of well studied clusters. The sample was defined with clusters that had $[Fe/H]$ determined from high resolution spectroscopy in  \cite{Netopil16} as well as from APOGEE as published in \cite{apogee_sample}. Both samples have a good coverage of the fundamental parameters age, distance and $A_V$.

We performed four test runs of our fitting procedure: 1) using a prior on distance and $A_V$ only; 2) using a prior in distance, $A_V$ and $[Fe/H]$ based on the Galactic abundance gradient from \cite{OCCAMgradient20}; 3) using a prior in distance, $A_V$ and $[Fe/H]$ fixed at values from \cite{apogee_sample} and 4) using a prior in distance, $A_V$ and $[Fe/H]$ fixed at the solar value.

\begin{figure*}
\centering
\includegraphics[width=\textwidth]{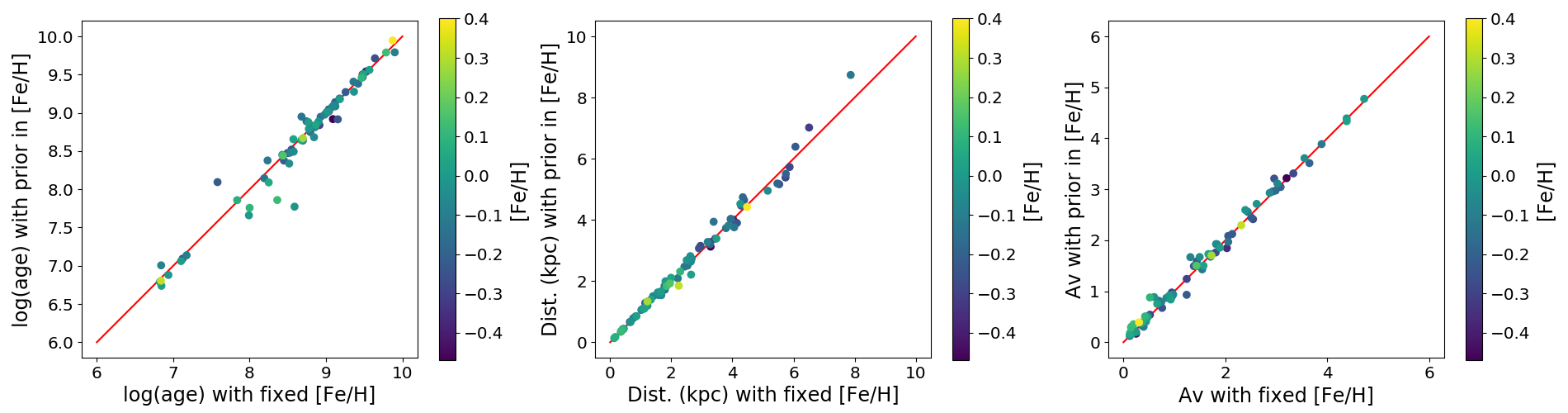}
\caption{Comparison of results obtained for the fundamental parameters with $[Fe/H]$ held fixed at the value from \citet{apogee_sample} and allowed to vary with a prior based on the Galactic metallicity gradient as described in the text. }
\label{fig:FeHsample-parcomp}
\end{figure*}

A first consistency check is to see how the fundamental parameters age, distance and $A_V$ are affected by fixing or not the parameter $[Fe/H]$. This is important for assessing the degree to which the fit results are sensitive to the assumptions made. In Fig. \ref{fig:FeHsample-parcomp} we show the comparison of results obtained for the fundamental parameters with $[Fe/H]$ held fixed at the value from \cite{apogee_sample} and allowed to vary subjected to the Galactic metallicity gradient prior. The results show that the agreement is good between parameters determined using both strategies. There are no detectable systematic effects. Considering the fact that $A_V$ and $[Fe/H]$ are generally hard to untangle based on photometry, this is an indication that the high quality of {\it Gaia} photometry allows for a good definition of CMD shape and this removes some of the degeneracy in these parameters.

\begin{figure*}
\centering
\includegraphics[width=\textwidth]{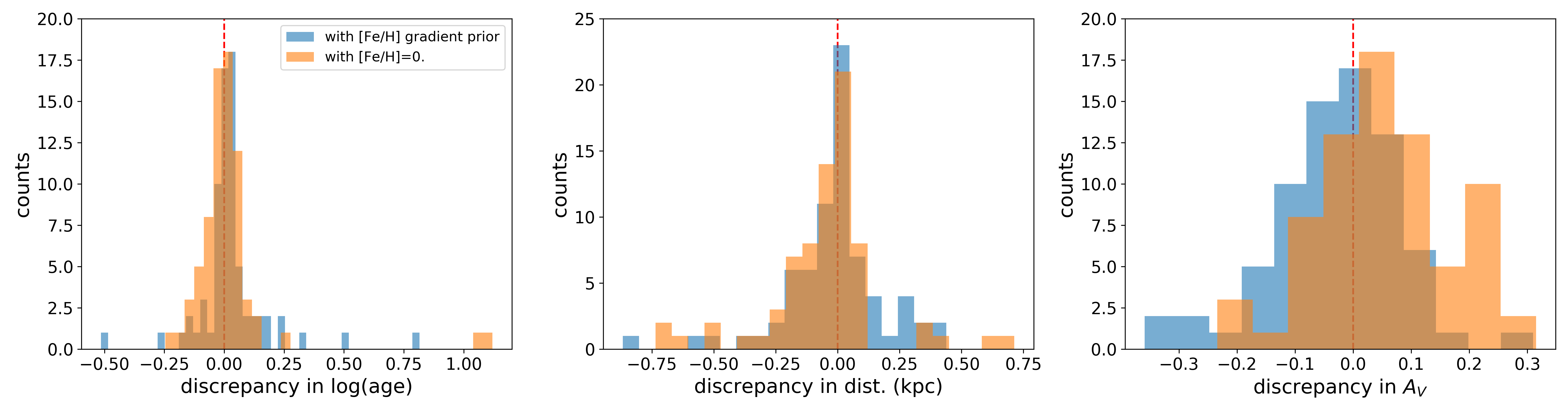}
\caption{Discrepancies in parameter estimates, obtained from fits using a prior for $[Fe/H]$ based on the Galactic metallicity gradient and fits using $[Fe/H]=0.0$ 
when compared to estimates obtained from fits where $[Fe/H]$ is fixed to the values from \citet{apogee_sample}. }
\label{fig:FeHsample-pardiscp}
\end{figure*}

Then we look at how the discrepancies in parameter estimates obtained from fits using a prior for $[Fe/H]$ based on the Galactic abundance gradient and fits using $[Fe/H]=0.0$ (which is the usual procedure adopted when this parameter is unknown), when compared to estimates obtained from fits where $[Fe/H]$ is fixed to the values from \cite{apogee_sample} which we take to be the most accurate. In Fig.~\ref{fig:FeHsample-pardiscp} we show histograms of the discrepancies for log(age), distance and $A_V$ in both situations. The histograms show that assuming $[Fe/H]=0.0$ leads to slightly larger discrepancies in log(age) and similar in distance although some outliers are clearly seen. These outliers are all from clusters with CMDs or turn-offs that are not clearly defined. There is a small systematic overestimation of 0.05 mag in $A_V$ as well.

\begin{figure*}
\centering
\includegraphics[width=\columnwidth]{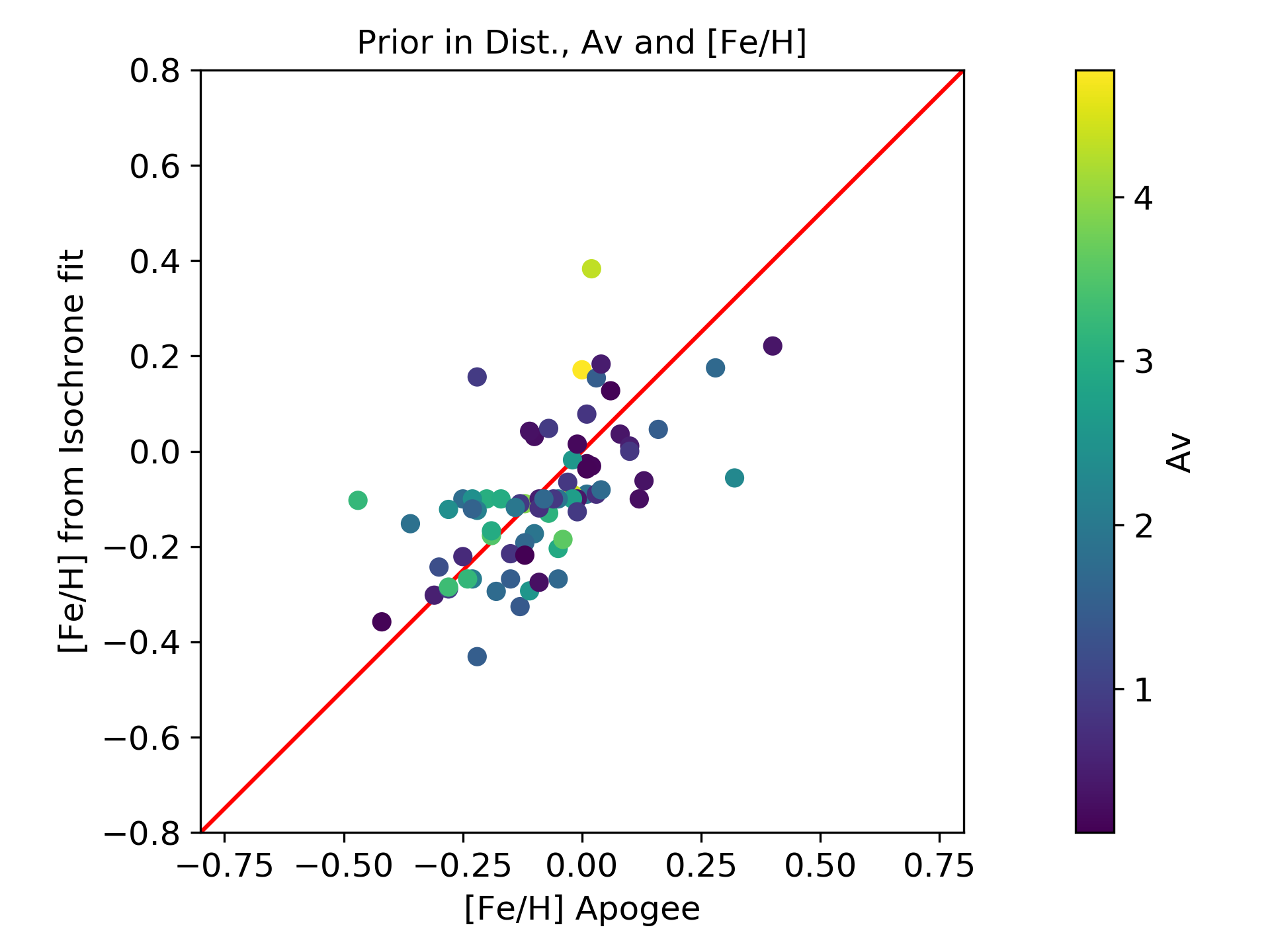}
\includegraphics[width=\columnwidth]{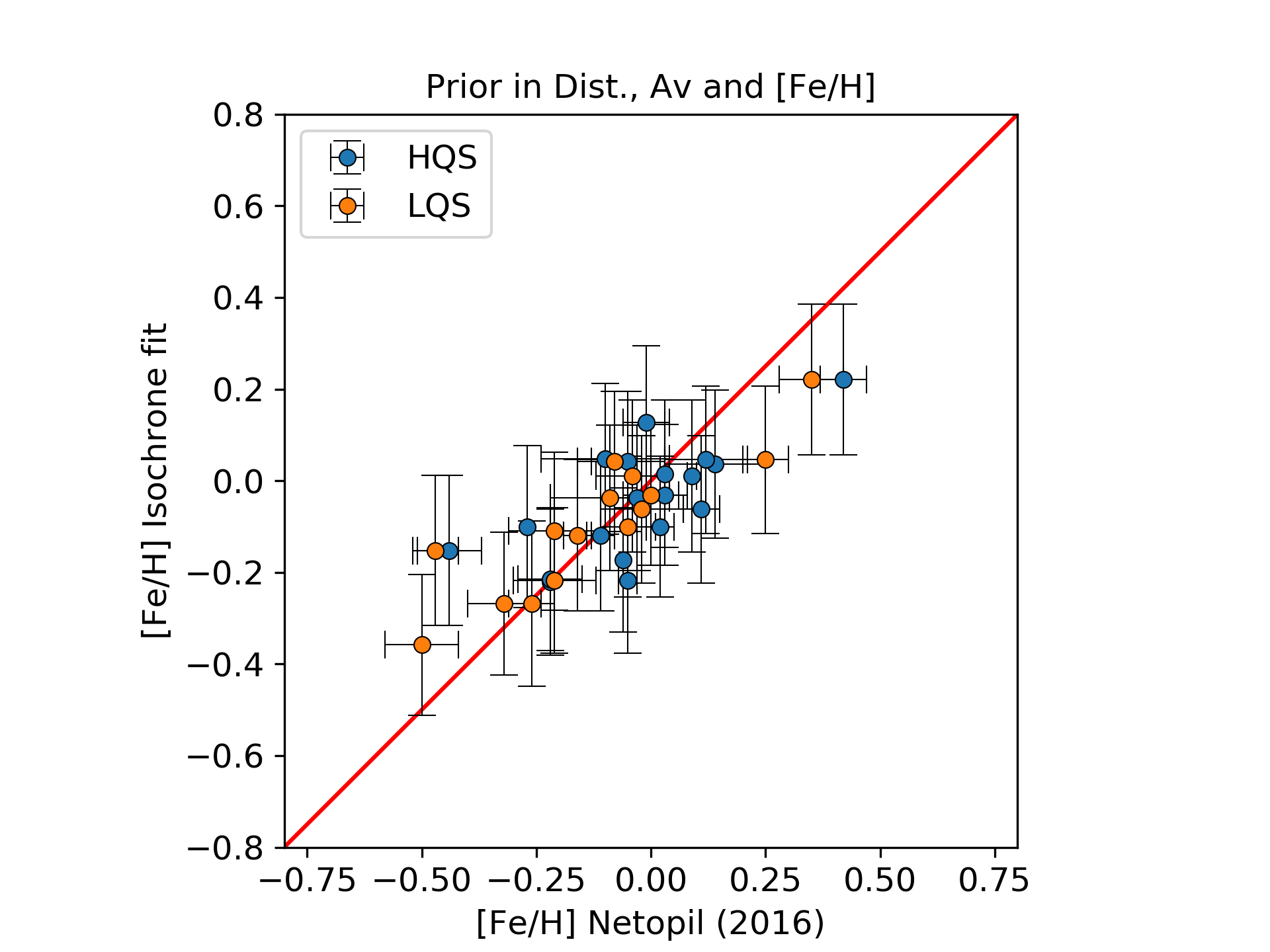}
\caption{Comparison of $[Fe/H]$ estimates obtained from isochrone fits using the Galactic metallicity gradient as prior to values obtained from the literature. Left panel shows comparison to values from APOGEE from \citet{apogee_sample}, where symbols are colored according to $A_V$. No systematic deviations due to $A_V$ are apparent. Right panel shows comparison to values from \citet{Netopil16} indicating errors as described in the text and discriminating between the high quality sample (HQS) and the lower quality sample (LQS) as defined by the authors.}
\label{fig:FeHsample-FeHcomp}
\end{figure*}

The sensitivity of {\it Gaia} data to $[Fe/H]$ can be verified in the results shown in Fig.~\ref{fig:FeHsample-FeHcomp} where the metallicity values obtained from fits using the Galactic abundance gradient prior are compared to values from \cite{apogee_sample} and \cite{Netopil16}. The same behaviour was found for fits where $[Fe/H]$ had no prior albeit with a larger spread, as expected. The average differences from the literature values are $0.014 \pm 0.137$ and $0.015 \pm 0.127$ with respect to \cite{apogee_sample} and \cite{Netopil16} respectively. Based on this result we incorporate a baseline error of 0.15 which is combined quadratically with the fit error to give the final uncertainty in $[Fe/H]$.

It is important to point out that this $[Fe/H]$ estimate should not be used as a proper metallicity determination for the open clusters studied. While the derived values of $[Fe/H]$ are indicative of the metallicity of individual clusters, statistically they are based on the metallicity gradient prior and thus cannot be used as a set for determining the Galactic abundance gradient. We have chosen to use $[Fe/H]$ as a free parameter because, as discussed above, it gives less biased results for $A_V$ when compared to the widespread practice of adopting $[Fe/H]=0.0$. By letting $[Fe/H]$ vary as a free parameter we also get more reliable estimates and uncertainties in the other parameters. Another positive point in adopting this strategy is that it may indicate clusters where interesting or deviant properties may be present allowing for a sample selection for more detailed observational followup campaigns.

As shown above, compared to the fixed $[Fe/H]$ prior from high resolution spectroscopy, adopting the $[Fe/H]$ values determined with the abundance gradient prior does not introduce systematic effects in the other parameters. Based on these results we have adopted the following procedure for the fits in this work: 1) if there is a reliable determination of $[Fe/H]$ in the literature, such as in \cite{apogee_sample} and \cite{Netopil16} we adopt that value and its uncertainty for the metallicity prior; 2) if there are no reliable value to be used as prior we use a prior based on the Galactic metallicity gradient from \cite{OCCAMgradient20}. The results of the isochrone fits, using the  Galactic metallicity gradient prior, to the clusters with high resolution spectroscopy analyzed in this section are given in Table~\ref{tab:fitsfeh}.

\section{Results} \label{sec:results}

\begin{figure}
\centering
\includegraphics[width=0.95\columnwidth]{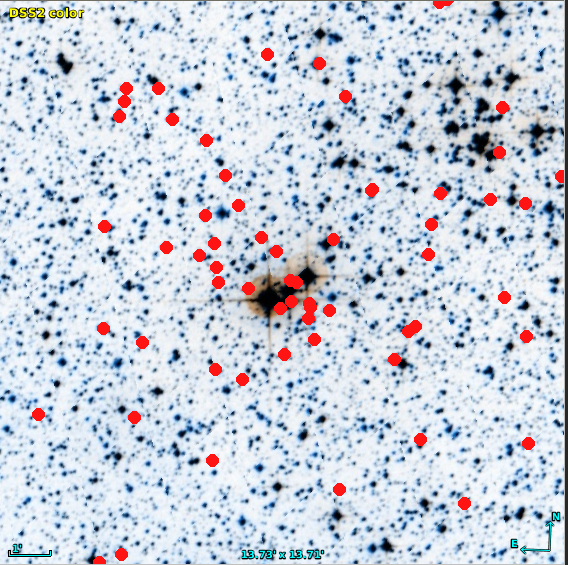}
\caption{Example of a rediscovered star cluster. Field of $13'\times 13'$ centered on Hogg~22, with coordinates from \citet{Dias2002}. The members of the cluster labelled UBC~547 by \citet{Castro-Ginard2020A&A...635A..45C} are marked in red.
In the upper right (N-W) we start seeing a concentration of bright stars which are on the edge of the nearby open cluster NGC 6204.}
\label{fig:castro_guinard}
\end{figure}

Of the 75 clusters selected as described in section~\ref{sec:sample}, the membership results for 30 objects either did not reveal identifiable cluster sequences or the isochrone fits were not a good match to the data and were thus discarded from further consideration. These clusters are identified in Table~\ref{tab:removed}. Typically, the fits failed when the sequences were faint and therefore had a small magnitude range, with higher errors, for fitting. While it may seem that our method is not being able to produce results for a high fraction of the clusters, we note that the selected sample of 75 clusters is composed of the leftovers from previous works. Thus, in fact our pipeline has been able to successfully handle 45 remaining considerably difficult cases. The classifications from  DAML illustrate the type of objects in our sample: 
10 were discovered in infra-red but are visible in the DSS images; Dolidze~1 was classified as possible cluster; Dolidze~35 and ESO~332-13 as a dubious objects; ESO~392-13 was an not found in the DSS images inspection; Sigma Orionis, NGC~1977, and Trapezium were classified as possible OB associations, and ESO~429-02 was classified as a possible open cluster remnant. 

With respect to the Trapezium cluster, the situation is further complicated by the presence of other young stellar populations along the same line of sight \citep{2012A&A...547A..97A, 2019arXiv190511429C}. The cluster here studied is composed of optically revealed, low extinction, elements in the foreground of the embedded Trapezium cluster. It is part of foreground "group 5" in \citet{2019arXiv190511429C}, which includes NGC~1980 and NGC~1981. In this work we identify this stellar aggregate as Trapezium-FG.

During the analysis we found that some clusters reported as new discoveries in recent papers were known clusters listed in DAML. 
The clusters FoF 2316 and FoF 868 found by \citet{ChinesCat} using using {\it Gaia} DR2 have similar positions, parallaxes and proper motions, and coincide with NGC~6530. We note, however, that they are not in the  high quality "Class 1" group defined by those authors.
\citet{Castro-Ginard2020A&A...635A..45C}, also using {\it Gaia} DR2, reported the discovery of 582 clusters which they identify under the designation "UBC". Some are known clusters listed in DAML: Czernik~43 $=$ UBC~399; Dolidze~1 $=$ UBC~367; ESO~429~02 $=$ UBC~464; FSR~0761 $=$ UBC~197; Hogg~22 $=$ UBC~547; IC~1442 $=$ UBC~164; NGC~133 $=$ UBC~185; NGC~1977 $=$ UBC~621; NGC~1980 $=$ UBC~208; NGC~6444 $=$ UBC~329;  Ruprecht~118 $=$ UBC~313. Fig.~\ref{fig:castro_guinard} illustrates the case of Hogg~22 (UBC~547). Curiously, although \citet{Castro-Ginard2020A&A...635A..45C} mention that NGC1980 is listed in DAML, it is included in the list of newly discovered clusters as UBC~208. We note however that these cases are in small number and do not raise concerns over the much broader scope of findings in \citet{Castro-Ginard2020A&A...635A..45C}. They do however highlight how delicate it is to cross-identify open clusters, which are extended objects, not continuous like galaxies, but often sparse discrete groups with irregular shapes, different apparent sizes and without clear boundaries.

For Berkeley 64 we estimated better central coordinates at  $\alpha=02^{\rm h}21^{\rm m}45^{\rm s}$; $\delta=+65^{\rm o}53^{\prime}30^{\prime\prime}$ in J2000.
For IC 1442 improved central coordinates are $\alpha=22^{\rm h}16^{\rm m}04^{\rm s}$; $\delta=+53^{\rm o}59^{\prime}29^{\prime\prime}$ in J2000, similar to the value estimated by \citep{Maurya2020arXiv200507375M}.

In the final analysis we also visually checked the color-magnitude diagrams with the isochrone fitted to the $G_{BP}$ and $G_{RP}$ photometric data from {\it Gaia} DR2 catalogue. The vector proper motion diagram constructed with individual symbol sizes and colours scaled to the kernel density estimated density in proper motion and parallax space as discussed in \citet{Monteiro2019MNRAS.487.2385M}, was also checked since in ($\mu_{\alpha}cos{\delta}$, $\mu_{\delta}$, $\varpi$) space a real clusters must show a concentration of stars.

In Table \ref{tab:astrometric} we present the mean astrometric parameters ($\mu_{\alpha}cos{\delta}$, $\mu_{\delta}$, $\varpi$) provided by the method described in section 3.1. In Table \ref{tab:photometric} the parameters obtained by the isochrone fit are given. In Fig.~\ref{fig:CMD} the final results of the isochrone fit with the stars with membership probability greater than 0.51 are shown.

   We point out that the fitting procedure has limitations in the treatment of very young clusters: On the one hand, variable extinction and age spread within the cluster are not specifically included in the fitting model. On the other hand, the grid of PARSEC isochrones does not include ages younger than $\sim$ 4 Myr ($\log$(age) = 6.60) and may not be particularly suited for pre-main sequence (PMS) evolutionary phases. To assess the adequacy of the fits for the youngest objects, we consider the 12 clusters determined to be younger than 10 Myr. For 10 clusters (ESO~332~08, ESO~332~13, FSR~0224, NGC~1980, NGC~6530, NGC~6604, Sigma~Orionis, Teutsch~132, Trapezium-FG, vdBergh~130), Fig.~\ref{fig:CMD} shows that while the PMS displays some dispersion, this is not evident on the main sequence which has a large fraction of members. The good definition of the main sequence is indicative that there is no significant variable reddening or age spread ($\lesssim$ 1-2 Myr) for those clusters. The PMS portions of the isochrones display a turn-on to the main sequence that match the observations. Then, at lower masses the isochrones tend to define lower envelopes of the cluster sequence. In terms of the quality of the fits, the PMS and main sequence produce consistent results for these clusters, indicating that the PARSEC isochrones are suitable for analyses  of young clusters at least in the Gaia photometric bands \citep[see][ for a discussion on the consistency between PMS and nuclear ages depending on the choice of photometric bands]{2006A&A...453..101L}. In any case, with the current pipeline, given the 4 Myr lower age limit of the isochrone grid and the possibility of age spreads, results for clusters found to be younger than $\sim$ 10 Myr should be visually checked and confirmed. The two remaining clusters in the group (Bica~2, FSR~0236) present a clear dispersion on both the main sequence and PMS branches.  While it is not clear if the dispersion is due to variable reddening, age spread or a contamination of field stars, it is clear that the results for these two clusters may not be very reliable.

\begin{figure}
\centering
\includegraphics[width=\columnwidth]{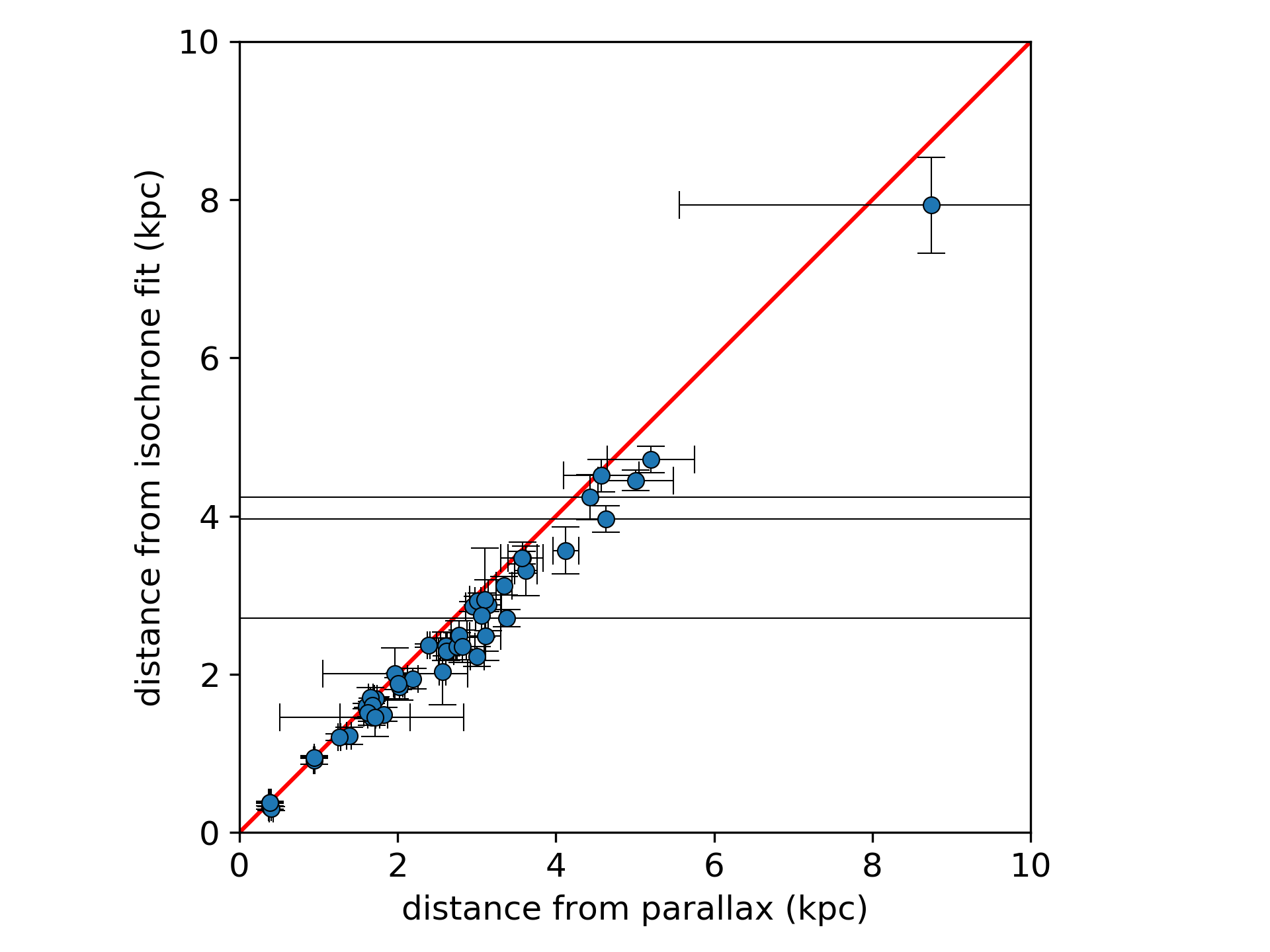}
\caption{Comparison of distances obtained from parallaxes and  isochrone fitting, both based on {\it Gaia} DR2. }
\label{fig:dist_isoXdist_plx}
\end{figure}

In Fig.~\ref{fig:dist_isoXdist_plx} we present the comparison of the distances obtained with the isochrone fits with those obtained by using the parallax of the member stars. 
The distances obtained from parallaxes were determined with a maximum likelihood estimation assuming a normal distribution for individual stars and taking into account individual parallax uncertainties. The standard errors provided in the distance from parallaxes were estimated by considering a symmetric distribution so that $\sigma = r_{95} - r_{5}/(2\times1.645)$, which is equivalent to 1$\sigma$ Gaussian uncertainty, where $r_{5}$ and $r_{95}$ are 5th and 95th percentile confidence intervals. 

The comparison shows a good agreement between the parallax distance and the one obtained via isochrone fitting. After 2.5 kpc a slight tendency for larger distances from parallaxes can be seen, but still within the errors. The result is a clear improvement with respect to the one presented in  \citet{Monteiro2019MNRAS.487.2385M}. While overall the methods are similar in both works, the main difference is that here we use a revised {\it Gaia} extinction correction and constrain  metallicity with the Galactic abundance gradient prior.

The mean difference in the values is of about 218 pc in the sense of distance from parallaxes minus distance from isochrone fit with a standard deviation of 212 pc. In general, the most discrepant cases are clusters more distant than 5 kpc  and whose main sequence is defined below  $G =$ 16. In this region the error in parallax increases considerably going from 0.02 mas ($G \leq$14) to typically 0.15 mas at $G =$ 18, leading to relative uncertainties as high as 75\%.

\begin{figure*}
\centering
\includegraphics[scale = 0.5]{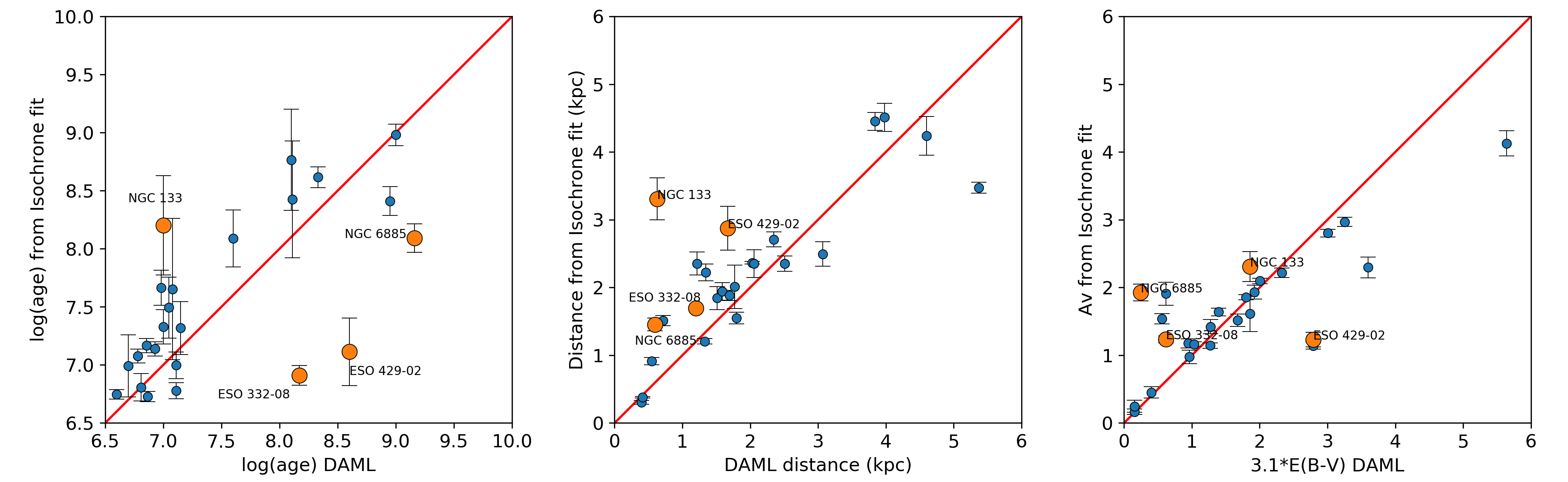}\\
\includegraphics[scale = 0.5]{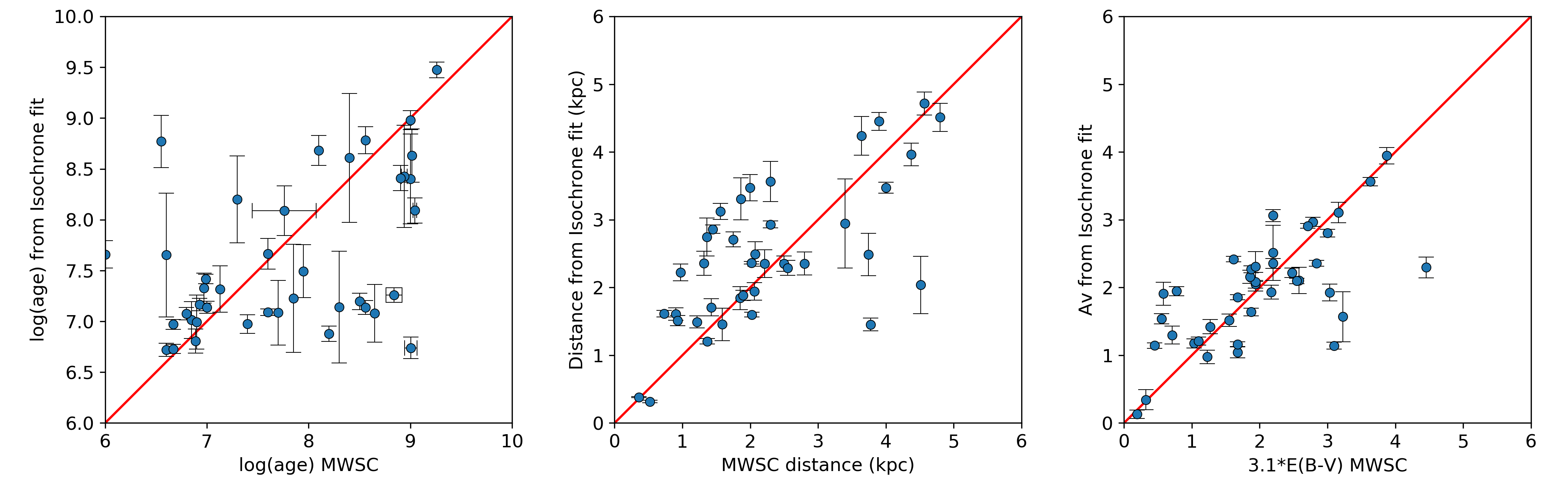}
\caption{Comparison of the values of distance (left panel), age (middle panel) and $A_{V}$ (right panel) obtained by the isochrone fit with those published in DAML (top) and MWSC (bottom).}
\label{fig:comparison}
\end{figure*}

\begin{table*}
\caption[]{Results of mean astrometric parameters obtained using the {\it Gaia} DR2 stellar proper motion and parallaxes.  
The meaning of the
symbols are as follows:
$RA_{ICRS}$ and $DE_{ICRS}$ are the central coordinates of the clusters; $r_{50}$ is the radius in which half of
the identified members are located; 
N is the number of cluster stars;
$\mu_{\alpha}cos{\delta}$ and $\mu_{\delta}$ are the  proper motion components in mas yr$^{-1}$;
$\sigma$ is the dispersion of cluster stars' proper motions;
$\varpi$ is the mean parallax of the cluster and $\sigma \varpi$ is the dispersion of the mean parallax.
RV and $\sigma RV$ are the mean and 1$\sigma$ dispersion radial velocity obtained for the cluster using {\it Gaia} DR2 data and $NRV$ is the number of stars used in the determination of RV after outlier rejection.}
\label{tab:astrometric}
\begin{center}
\begin{tabular}{lrrccrcrcccrrc}
\hline
   Name             &  
   \multicolumn{1}{c}{$RA_{ICRS}$} & 
   \multicolumn{1}{c}{$DE_{ICRS}$}  &  
   $r_{50}$    &   
   N  &  
   \multicolumn{1}{c}{$\mu_{\alpha}cos{\delta}$}   &   
   $\sigma_{\mu_{\alpha}cos{\delta}}$  &  
   \multicolumn{1}{c}{$\mu_{\delta}$}  &   
   $\sigma_{\mu_{\delta}}$  &   
   $\varpi$   &   
   $\sigma_{\varpi}$  &       
   \multicolumn{1}{c}{RV} &                
   \multicolumn{1}{c}{$\sigma_{RV}$} & 
   \multicolumn{1}{c}{$NRV$}\\      
   \multicolumn{1}{c}{$ $} &      
   \multicolumn{1}{c}{$(deg)$} & 
   \multicolumn{1}{c}{$(deg)$}  &  
   \multicolumn{1}{c}{$(deg)$}  &  
   \multicolumn{1}{c}{$ $} &      
   \multicolumn{1}{c}{$(mas)$}   &   
   \multicolumn{1}{c}{$(mas)$}   &   
   \multicolumn{1}{c}{$(mas)$}   &   
   \multicolumn{1}{c}{$(mas)$}   &   
   \multicolumn{1}{c}{$(mas)$}   &   
   \multicolumn{1}{c}{$(mas)$}   &   
   \multicolumn{1}{c}{($km s^{-1}$)} &                
   \multicolumn{1}{c}{($km s^{-1}$)} &                
   \multicolumn{1}{c}{$ $}\\       

\hline
BH~88               &  151.6211 &   -51.5557 &   0.056 &    89 &  -6.086 &   0.377  &   3.602 &    0.345 &   0.386  &  0.165  &      22.751 &           1.100  &   2\\       
Berkeley~64         &   35.3246 &    65.8934 &   0.041 &   138 &  -0.551 &   0.323  &   0.814 &    0.408 &   0.201  &  0.170  &             &                  &    \\       
Bica~2              &  308.3153 &    41.3068 &   0.060 &   140 &  -2.660 &   0.244  &  -4.378 &    0.230 &   0.555  &  0.102  &             &                  &    \\       
Bochum~10           &  160.5040 &   -59.1324 &   0.148 &   264 &  -7.291 &   0.317  &   2.992 &    0.223 &   0.378  &  0.092  &       0.943 &           0.290  &   2\\       
Collinder~104       &   99.1571 &     4.8155 &   0.143 &   179 &  -1.230 &   0.411  &   0.507 &    0.418 &   0.514  &  0.210  &             &                  &    \\       
Czernik~43          &  351.4483 &    61.3294 &   0.057 &   173 &  -3.862 &   0.317  &  -2.078 &    0.244 &   0.350  &  0.134  &             &                  &    \\       
DC~3                &  111.7507 &   -37.5195 &   0.025 &   105 &  -1.214 &   0.331  &   2.645 &    0.472 &   0.081  &  0.189  &             &                  &    \\       
Dolidze~1           &  302.4057 &    36.5052 &   0.075 &   226 &  -2.721 &   0.312  &  -4.961 &    0.326 &   0.288  &  0.118  &             &                  &    \\       
Dolidze~35          &  291.3465 &    11.6414 &   0.064 &    91 &  -1.967 &   0.245  &  -4.322 &    0.392 &   0.288  &  0.181  &      22.577 &           0.174  &   2\\       
ESO~123~26          &  118.1254 &   -60.3348 &   0.105 &    22 &  -3.572 &   0.318  &  10.909 &    0.180 &   1.026  &  0.044  &             &                  &    \\       
ESO~332~08          &  253.6906 &   -40.7299 &   0.080 &   201 &  -0.272 &   0.316  &  -1.348 &    0.313 &   0.529  &  0.200  &             &                  &    \\       
ESO~332~13          &  254.1701 &   -40.5887 &   0.058 &    52 &  -0.080 &   0.225  &  -1.117 &    0.237 &   0.558  &  0.127  &             &                  &    \\       
ESO~392~13          &  261.7178 &   -34.7020 &   0.092 &    21 &   1.690 &   0.262  &  -2.882 &    0.176 &   0.900  &  0.139  &             &                  &    \\       
ESO~429~02          &  113.3481 &   -28.1816 &   0.050 &    54 &  -2.806 &   0.199  &   3.673 &    0.292 &   0.281  &  0.155  &             &                  &    \\       
FSR~0224            &  306.3509 &    40.2243 &   0.021 &    19 &  -3.242 &   0.343  &  -4.373 &    0.268 &   0.566  &  0.068  &             &                  &    \\       
FSR~0236            &  308.1682 &    41.4418 &   0.048 &    89 &  -2.491 &   0.394  &  -4.076 &    0.243 &   0.522  &  0.163  &             &                  &    \\       
FSR~0377            &  338.7186 &    58.3041 &   0.044 &   116 &  -3.219 &   0.366  &  -2.155 &    0.301 &   0.207  &  0.159  &             &                  &    \\       
FSR~0441            &  355.5402 &    58.5480 &   0.040 &    95 &  -2.049 &   0.313  &  -1.160 &    0.220 &   0.239  &  0.169  &             &                  &    \\       
FSR~0591            &   36.9315 &    58.7637 &   0.057 &   222 &  -0.231 &   0.571  &  -0.566 &    0.484 &   0.293  &  0.214  &     -72.562 &           0.864  &   2 \\       
FSR~0674            &   63.0983 &    48.7296 &   0.033 &    49 &  -0.894 &   0.419  &  -0.871 &    0.243 &   0.275  &  0.285  &             &                  &    \\
FSR~0761            &   83.3381 &    39.8388 &   0.034 &    85 &   0.323 &   0.350  &  -1.361 &    0.334 &   0.253  &  0.145  &     -27.299 &           1.923  &   2\\       
FSR~1443            &  129.8570 &   -47.3566 &   0.054 &   154 &  -3.563 &   0.382  &   4.147 &    0.439 &   0.220  &  0.171  &      37.746 &           1.384  &   2\\       
FSR~1698            &  230.2346 &   -59.6270 &   0.044 &   161 &  -4.033 &   0.341  &  -3.524 &    0.307 &   0.247  &  0.224  &             &                  &    \\       
Hogg~16             &  202.2997 &   -61.2087 &   0.047 &    46 &  -3.479 &   0.095  &  -1.645 &    0.146 &   0.431  &  0.065  &             &                  &    \\       
Hogg~22             &  251.6599 &   -47.0782 &   0.044 &   117 &  -0.750 &   0.254  &  -2.013 &    0.339 &   0.343  &  0.138  &             &                  &    \\       
IC~1442             &  334.0070 &    53.9900 &   0.058 &   333 &  -3.083 &   0.484  &  -2.884 &    0.476 &   0.240  &  0.198  &             &                  &    \\       
Majaess~65          &   87.4284 &    27.0746 &   0.120 &    51 &  -0.258 &   0.229  &  -1.063 &    0.320 &   0.974  &  0.160  &             &                  &    \\       
NGC~133             &    7.8324 &    63.3583 &   0.068 &   284 &  -2.324 &   0.431  &  -0.410 &    0.250 &   0.223  &  0.158  &     -86.916 &           0.405  &   5\\       
NGC~1977            &   83.7945 &    -4.8018 &   0.145 &    93 &   1.260 &   0.453  &  -0.569 &    0.520 &   2.590  &  0.185  &      27.392 &           2.361  &   6\\       
NGC~1980            &   83.8212 &    -5.9207 &   0.125 &   120 &   1.192 &   0.388  &   0.511 &    0.385 &   2.583  &  0.128  &      25.264 &           7.055  &   8\\       
NGC~2384            &  111.2913 &   -21.0211 &   0.063 &    80 &  -2.303 &   0.185  &   3.118 &    0.220 &   0.330  &  0.132  &             &                  &    \\       
NGC~6200            &  251.0322 &   -47.4582 &   0.109 &   433 &  -0.950 &   0.333  &  -2.244 &    0.351 &   0.307  &  0.265  &             &                  &    \\       
NGC~6444            &  267.3950 &   -34.8221 &   0.059 &    47 &  -0.934 &   0.114  &  -0.929 &    0.096 &   0.521  &  0.073  &             &                  &    \\       
NGC~6530            &  271.1088 &   -24.3572 &   0.087 &    80 &   1.375 &   0.352  &  -1.992 &    0.307 &   0.762  &  0.111  &             &                  &    \\       
NGC~6604            &  274.5127 &   -12.2449 &   0.049 &    88 &  -0.453 &   0.208  &  -2.294 &    0.314 &   0.454  &  0.134  &             &                  &    \\       
NGC~6885            &  302.9831 &    26.4935 &   0.137 &   726 &  -3.127 &   0.356  &  -5.471 &    0.413 &   0.439  &  0.245  &       2.378 &           0.333  &   4\\       
Ruprecht~118        &  246.1454 &   -51.9544 &   0.051 &    79 &  -3.152 &   0.188  &  -4.345 &    0.174 &   0.285  &  0.107  &             &                  &    \\       
Ruprecht~123        &  260.7813 &   -37.8977 &   0.055 &    20 &   1.044 &   0.172  &   0.922 &    0.108 &   0.604  &  0.084  &             &                  &    \\       
Ruprecht~55         &  123.1133 &   -32.5815 &   0.064 &   414 &  -2.316 &   0.394  &   2.921 &    0.436 &   0.187  &  0.174  &      64.253 &           1.769  &   2 \\       
SAI~43              &   77.0723 &    49.8645 &   0.035 &   135 &   0.611 &   0.390  &  -0.555 &    0.389 &   0.109  &  0.280  &             &                  &    \\       
Sigma~Orionis       &   84.6860 &    -2.5959 &   0.054 &    45 &   1.336 &   0.388  &  -0.633 &    0.372 &   2.479  &  0.157  &             &                  &    \\       
Stock~3             &   18.0592 &    62.3190 &   0.060 &   114 &  -1.895 &   0.326  &  -0.357 &    0.296 &   0.265  &  0.132  &             &                  &    \\       
Teutsch~132         &   77.5140 &    38.8163 &   0.057 &   112 &   0.326 &   0.506  &  -1.536 &    0.329 &   0.223  &  0.234  &             &                  &    \\       
Trapezium-FG           &   83.8350 &    -5.4095 &   0.352 &   269 &   1.262 &   0.449  &   0.274 &    0.498 &   2.557  &  0.149  &      23.841 &           5.161  &  15\\       
vdBergh~130         &  304.4624 &    39.3404 &   0.049 &    62 &  -3.609 &   0.308  &  -5.075 &    0.292 &   0.521  &  0.154  &             &                  &    \\
\hline
\end{tabular}
\end{center}
\end{table*}

\begin{table*}
\caption[]{Fundamental parameters obtained from the isochrone fits. The last two columns give the distances estimated from parallaxes with a maximum likelihood estimation assuming a normal distribution and taking into account individual parallax uncertainties. The standard errors provided in the distance from parallaxes were estimated considering the calculated 5th and 95th percentile confidence intervals assuming a symmetric distribution so that $\sigma = r_{95} - r_{5}/(2\times1.645)$, which is equivalent to 1$\sigma$ Gaussian uncertainty. The 0.029 mas correction  \citep{Lindegren2018} to the mean parallaxes was previously added. 
 }
 
\label{tab:photometric}
\begin{center}
\begin{tabular}{lccccrccccc}
\hline
   Name                      &    $dist$           &            $\sigma_{dist}$              &             $age$          &              $\sigma_{age}$          &                \multicolumn{1}{c}{$[Fe/H]$}        &               $\sigma_{[Fe/H]}$          &                  $A_{V}$         &                $\sigma_{A_{V}}$     &           $dist_{\pi}$ &    $\sigma_{dist_{\pi}}$ \\
   &    $(pc)$   &   $(pc)$  &   $(dex)$  & $(dex)$  &  \multicolumn{1}{c}{$(dex)$}  &   $(dex)$ &  $(mag)$ &   $(mag)$ &  $(pc)$ &  $(pc)$ \\
\hline     
BH~88                        &   2011        &     321          &       8.766        &     0.435      &        -0.141      &  0.228         &        1.612      &       0.262   &           1936 &    1115  \\     
Berkeley~64                  &   4547        &     378          &       8.926        &     0.046      &        -0.203      &  0.171         &        2.951      &       0.043   &           4889 &     812  \\     
Bica~2                       &   1550        &      85          &       6.746        &     0.041      &         0.100      &  0.214         &        4.126      &       0.186   &           1665 &      40  \\     
Bochum~10                    &   2365        &      19          &       7.167        &     0.061      &         0.179      &  0.172         &        1.175      &       0.066   &           2390 &      18  \\     
Collinder~104                &   1599        &      35          &       7.197        &     0.081      &        -0.104      &  0.171         &        2.104      &       0.195   &           1609 &      25  \\     
Czernik~43                   &   2350        &     113          &       8.088        &     0.245      &        -0.023      &  0.160         &        1.931      &       0.103   &           2616 &      74  \\     
DC~3                         &   7934        &     607          &       9.474        &     0.076      &        -0.146      &  0.158         &        1.042      &       0.082   &           8744 &    3184  \\     
Dolidze~1                    &   2860        &      63          &       7.090        &     0.032      &         0.054      &  0.177         &        2.049      &       0.056   &           2949 &      95  \\     
Dolidze~35                   &   2334        &     98          &       7.952        &     0.466      &         0.204      &  0.180         &        3.987      &       0.050   &           2603 &     122  \\     
ESO~123~26                   &    914        &      55          &       8.616        &     0.089      &         0.065      &  0.191         &        0.451      &       0.084   &            948 &       9  \\     
ESO~332~08                   &   1693        &      22          &       6.911        &     0.085      &         0.230      &  0.187         &        1.234      &       0.053   &           1723 &      15  \\     
ESO~332~13                   &   1487        &      84          &       6.840        &     0.139      &         0.168      &  0.183         &        1.380      &       0.027   &           1673 &      51  \\     
ESO~392~13                   &   1032        &      85          &       8.656        &     0.428      &         0.074      &  0.185         &        1.906      &       0.201   &           1057 &      22  \\     
ESO~429~02                   &   2875        &     322          &       7.113        &     0.291      &        -0.120      &  0.195         &        1.233      &       0.108   &           3141 &     170  \\     
FSR~0224                     &   1706        &     127          &       6.739        &     0.106      &         0.242      &  0.266         &        3.061      &       0.091   &           1659 &      29  \\     
FSR~0236                     &   1610        &      91          &       6.877        &     0.076      &         0.091      &  0.198         &        3.564      &       0.062   &           1678 &      28  \\     
FSR~0377                     &   3563        &     297          &       7.085        &     0.317      &        -0.065      &  0.175         &        2.159      &       0.098   &           4124 &     162  \\     
FSR~0441                     &   3473        &     196          &       7.079        &     0.285      &        -0.153      &  0.177         &        2.419      &       0.039   &           3579 &     183  \\     
FSR~0591                     &   2930        &      52          &       7.014        &     0.182      &        -0.187      &  0.205         &        2.270      &       0.052   &           3014 &      39  \\     
FSR~0674                     &   2944        &     656          &       8.782        &     0.134      &        -0.140      &  0.156         &        3.107      &       0.150   &           3106 &     193  \\     
FSR~0761                     &   2485        &     312          &       8.770        &     0.258      &        -0.112      &  0.205         &        1.568      &       0.368   &           3107 &     193  \\     
FSR~1443                     &   3303        &     157          &       8.703        &     0.416      &        -0.019      &  0.159         &        1.819      &       0.159   &           3444 &      60  \\     
FSR~1698                     &   3122        &     118          &       7.136        &     0.068      &         0.228      &  0.163         &        2.907      &       0.035   &           3341 &      95  \\     
Hogg~16                      &   1943        &     131          &       7.494        &     0.262      &         0.110      &  0.206         &        1.422      &       0.107   &           2190 &      68  \\     
Hogg~22                      &   2354        &     171          &       7.076        &     0.060      &         0.120      &  0.170         &        2.097      &       0.040   &           2749 &     123  \\     
IC~1442                      &   2710        &     112          &       7.665        &     0.151      &        -0.100      &  0.160         &        1.271      &       0.277   &           3378 &    9075  \\     
Majaess~65                   &    944        &      10          &       8.207        &     0.167      &         0.006      &  0.160         &        0.768      &       0.105   &            945 &       4  \\     
NGC~133                      &   3308        &     311          &       8.201        &     0.427      &        -0.133      &  0.163         &        2.310      &       0.222   &           3615 &     142  \\     
NGC~1977                     &    381        &       9          &       6.721        &     0.064      &        -0.184      &  0.170         &        0.344      &       0.148   &            388 &      27  \\     
NGC~1980                     &    316        &      19          &       6.970        &     0.049      &        -0.242      &  0.175         &        0.129      &       0.060   &            384 &      18  \\     
NGC~2384                     &   2494        &     179          &       7.318        &     0.228      &        -0.147      &  0.257         &        0.976      &       0.098   &           2775 &      99  \\     
NGC~6200                     &   2352        &     205          &       7.138        &     0.060      &         0.166      &  0.193         &        1.858      &       0.038   &           2821 &     152  \\     
NGC~6444                     &   1492        &      88          &       8.632        &     0.262      &         0.177      &  0.191         &        1.298      &       0.130   &           1823 &      54  \\     
NGC~6530                     &   1206        &      39          &       6.728        &     0.045      &         0.373      &  0.203         &        1.163      &       0.037   &           1265 &      18  \\     
NGC~6604                     &   1885        &      75          &       6.807        &     0.118      &         0.104      &  0.222         &        2.804      &       0.057   &           2007 &      59  \\     
NGC~6885                     &   1453        &      95          &       8.092        &     0.124      &         0.055      &  0.192         &        1.927      &       0.123   &           1671 &    1466  \\     
Ruprecht~118                 &   2224        &     125          &       8.425        &     0.503      &         0.386      &  0.196         &        1.144      &       0.041   &           3004 &      88  \\     
Ruprecht~123                 &   1511        &      74          &       8.682        &     0.147      &         0.188      &  0.224         &        1.909      &       0.169   &           1622 &      48  \\     
Ruprecht~55                  &   4238        &     286          &       7.328        &     0.148      &        -0.226      &  0.154         &        1.639      &       0.056   &           4430 &   43070  \\     
SAI~43                       &   4451        &     131          &       8.410        &     0.124      &        -0.198      &  0.172         &        1.538      &       0.075   &           5009 &     480  \\     
Sigma~Orionis                &    303        &      26          &       6.997        &     0.114      &        -0.092      &  0.158         &        0.166      &       0.040   &            402 &      25  \\     
Stock~3                      &   2747        &     281          &       7.226        &     0.531      &        -0.100      &  0.168         &        2.355      &       0.073   &           3051 &      84  \\     
Teutsch~132                  &   3474        &      81          &       6.992        &     0.266      &        -0.160      &  0.206         &        2.217      &       0.069   &           3567 &     267  \\     
Trapezium-FG                    &    381        &      12          &       6.778        &     0.069      &        -0.146      &  0.160         &        0.246      &       0.089   &            386 &       1  \\     
vdBergh~130                  &   1456        &     240          &       6.974        &     0.091      &        -0.029      &  0.222         &        2.356      &       0.042   &           1714 &     563  \\     
\hline
\end{tabular}
\end{center}
\end{table*}

\section{Comparison with the literature}

The comparison presented here has two goals: to provide an extra sanity check of our results and to assess the improvement they bring.
To this end we base the analysis on the widely used DAML and \citet{Kharchenko2013} (hereafter MWSC) catalogues. It is important to note that these are different types of catalogues. On the one hand, the MWSC is the output of a program applied to the PPMXL \citep{Roeser2010} and 2MASS \citep{Skrutskie2006} data. On the other hand, DAML is a compilation, curated by humans, of the best results (judged by the curators) available in the literature. 
The MWSC aims to overcome the non-uniformity in compilations from the literature, which are based on results obtained by different authors using different techniques, models and calibrations. However, as pointed out in \citet{2010IAUS..266..106M} homogeneous methods do not necessarily produce the best results. As an example, for close objects, parallaxes provide the best distances, but at larger distances isochrone fits are better. Assuming the algorithm employed in the MWSC is flawless, the relatively shallow data used in the MWSC limits its usefulness to bright and/or close clusters. 

The DAML catalogue is a compilation of results from the literature. While it is non-homogeneous in nature, it is curated. The curators choose the best results, when more than one is available, and keep public logs of what has changed, of the references to the catalogued parameters and a list of objects that studies have shown not to be real clusters as well as the references to those studies. As a compilation, it also includes results from the MWSC. Thus, the comparison with DAML is also a comparison with individual studies from the literature. 

The cross-identification of our sample with DAML and with the MWSC results in 45 and 40 objects in common, respectively. Since DAML also contains values from the MWSC, the later were not included in the DAML plots to avoid comparing the same points twice. This leaves the DAML comparison sample with 28 clusters. The comparisons of distance, age and $A_{V}$ are shown in Fig.~\ref{fig:comparison}. 

The $A_{V}$ and distances from both catalogs follow the same trend as those obtained with our isochrone fits, although with some considerable scatter (clearly higher in the case of the distances from MWSC) and with a tendency for smaller catalogued distances for the closer ($<$ 3 kpc) sub-sample. The same trend in the distances can be observed when comparing the distances estimated from the parallaxes of clusters published in \citet{Cantat-Gaudin2020A&A...633A..99C} with respect to the DAML and MWSC catalogues.

The age comparison shows a much high dispersion. In the case of the MWSC, ages appear to be almost uncorrelated to the ones here determined, except for a small group of clusters younger than $\sim$ 10 Myr (in the MWSC age scale). 
In the DAML age comparison we find 4 especially discrepant objects. They are ESO~332-08, ESO~429-02, NGC~133 and NGC~6885. 

\begin{figure*}
 \centering
 \includegraphics[scale = 0.44]{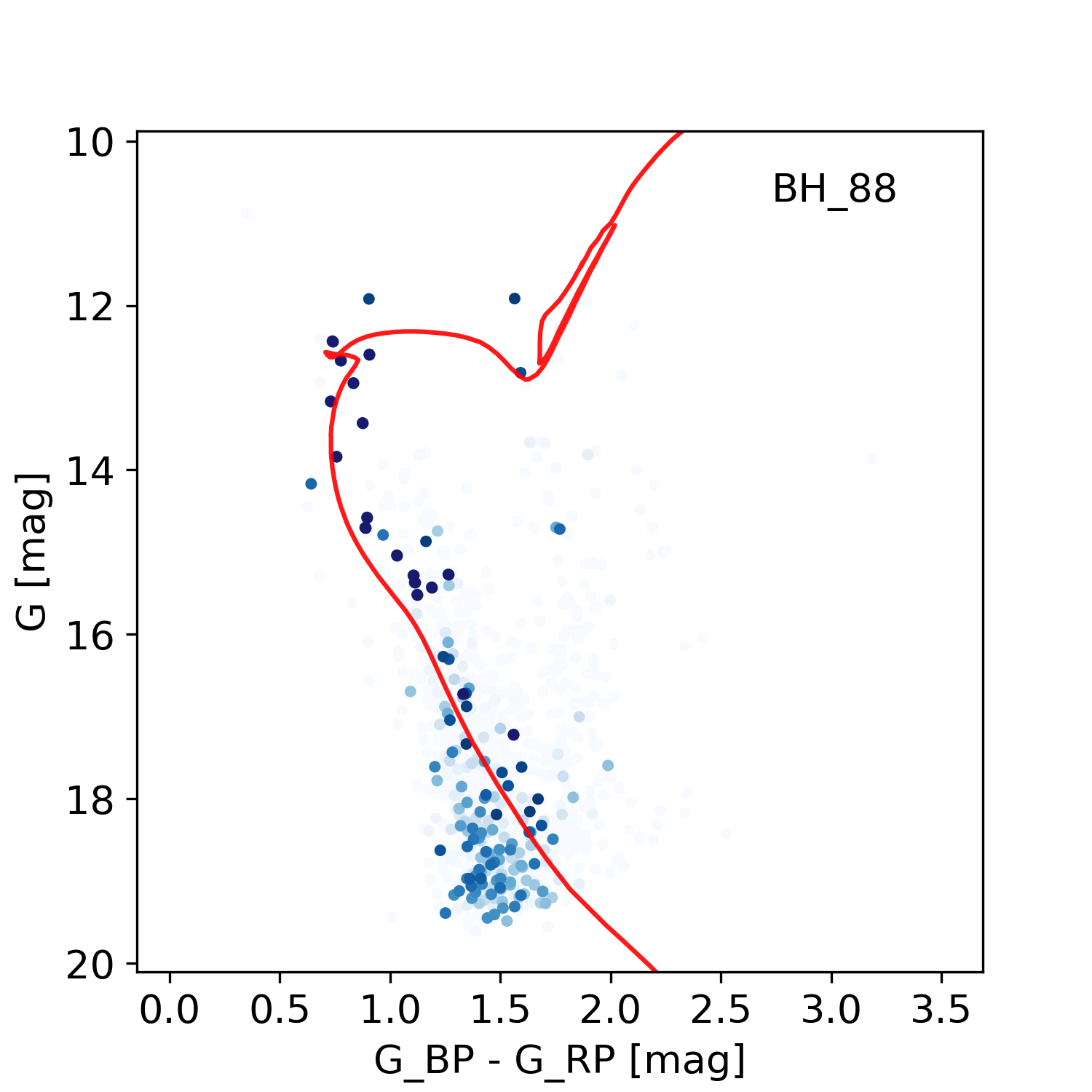}
 \includegraphics[scale = 0.44]{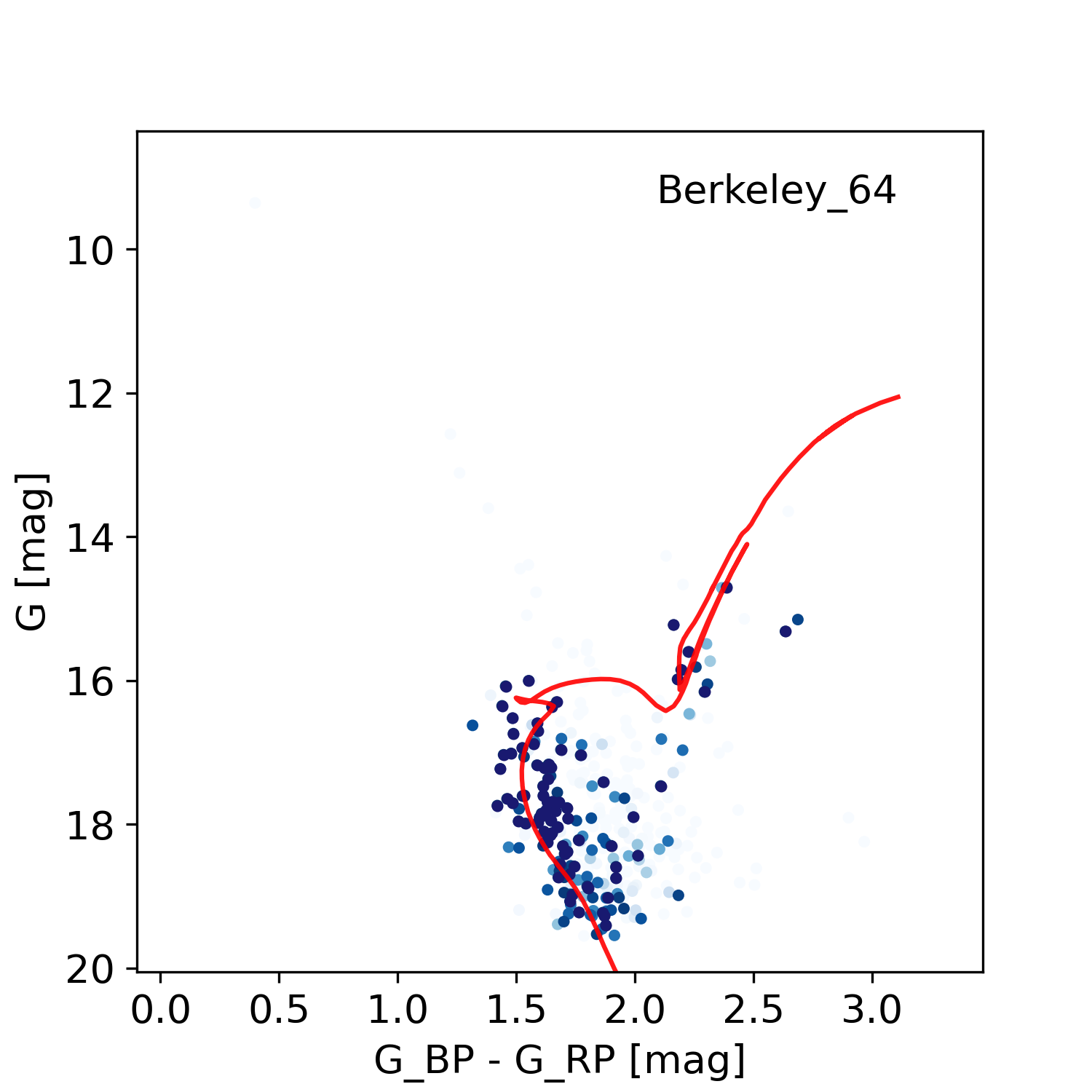}
 \includegraphics[scale = 0.44]{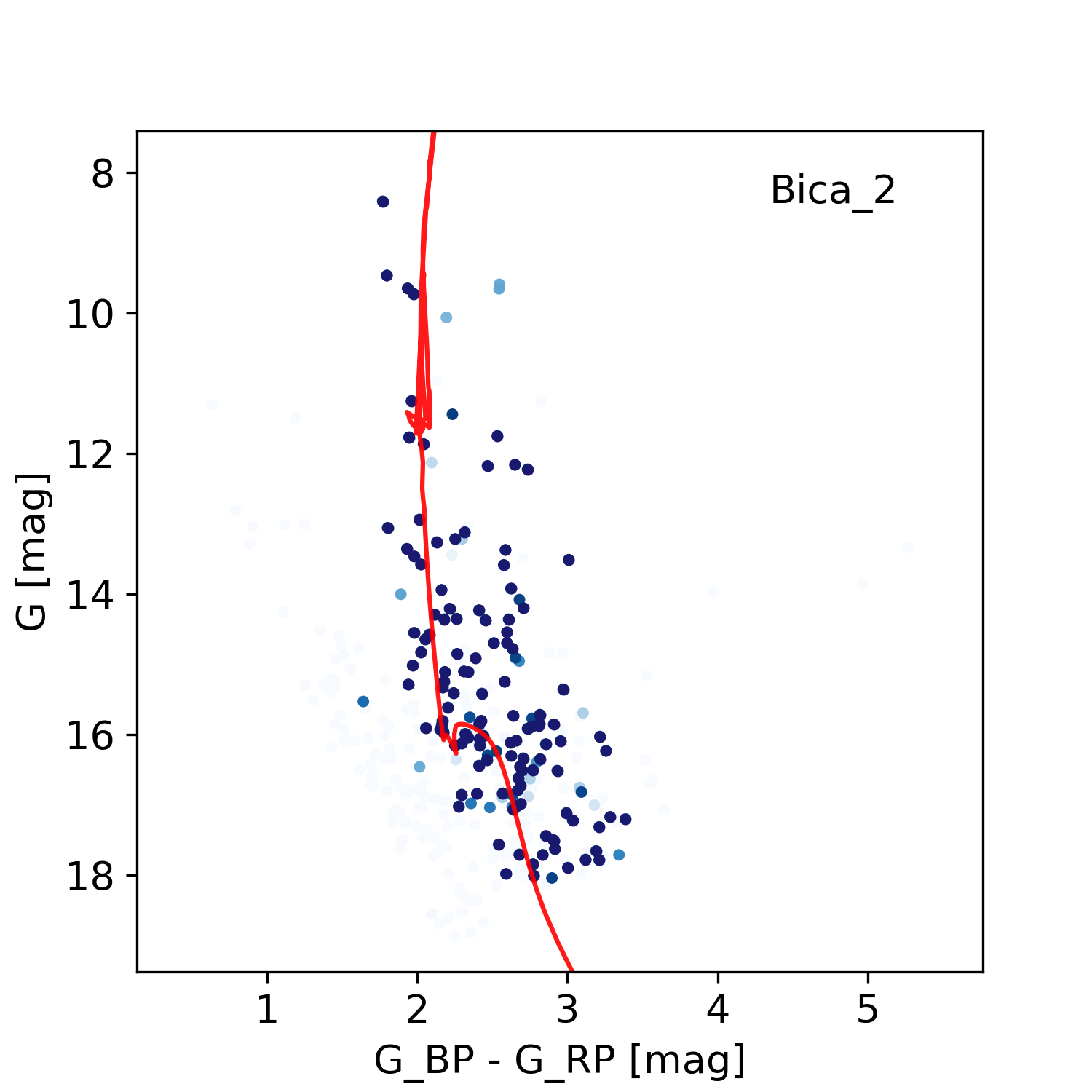} \\
 \includegraphics[scale = 0.44]{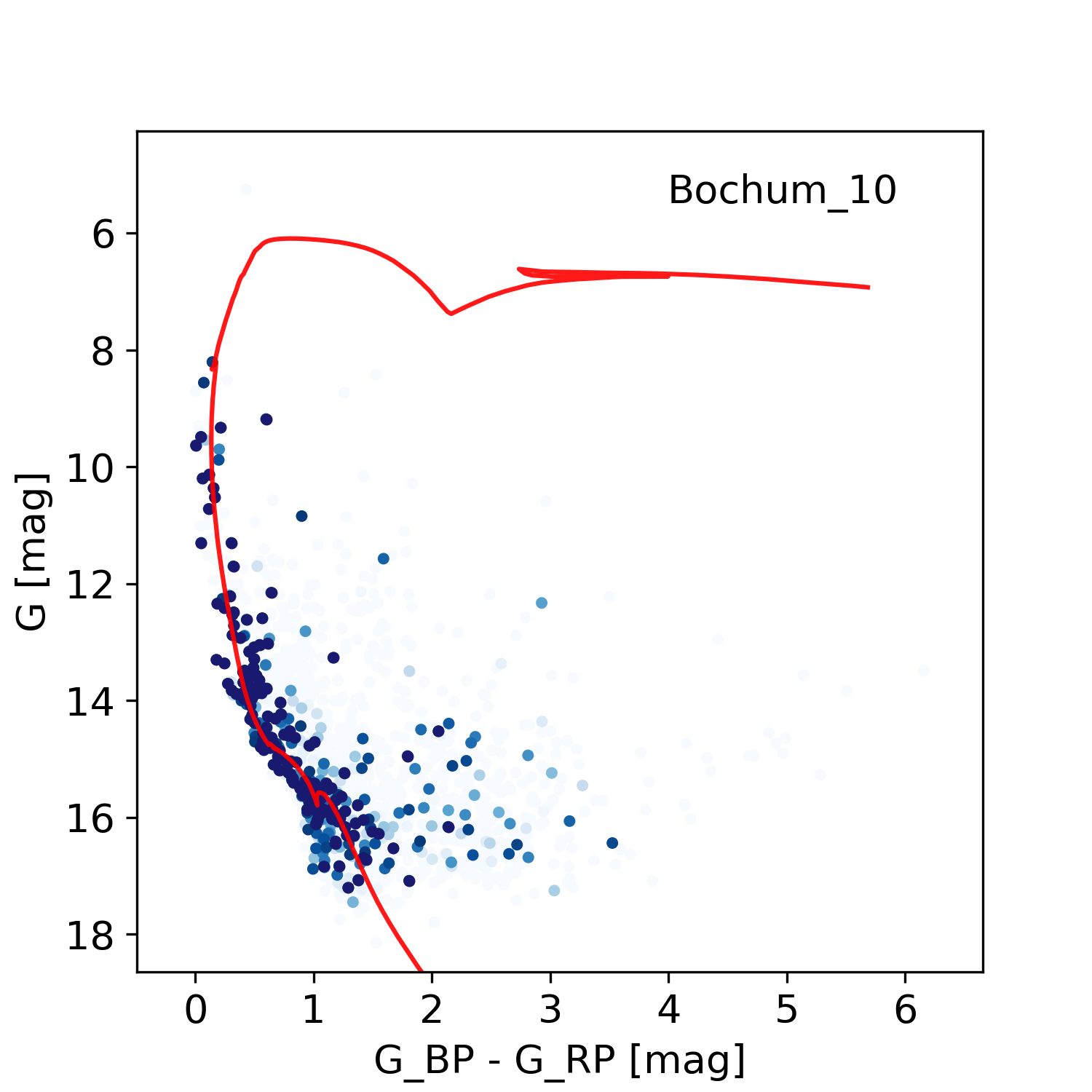}
 \includegraphics[scale = 0.44]{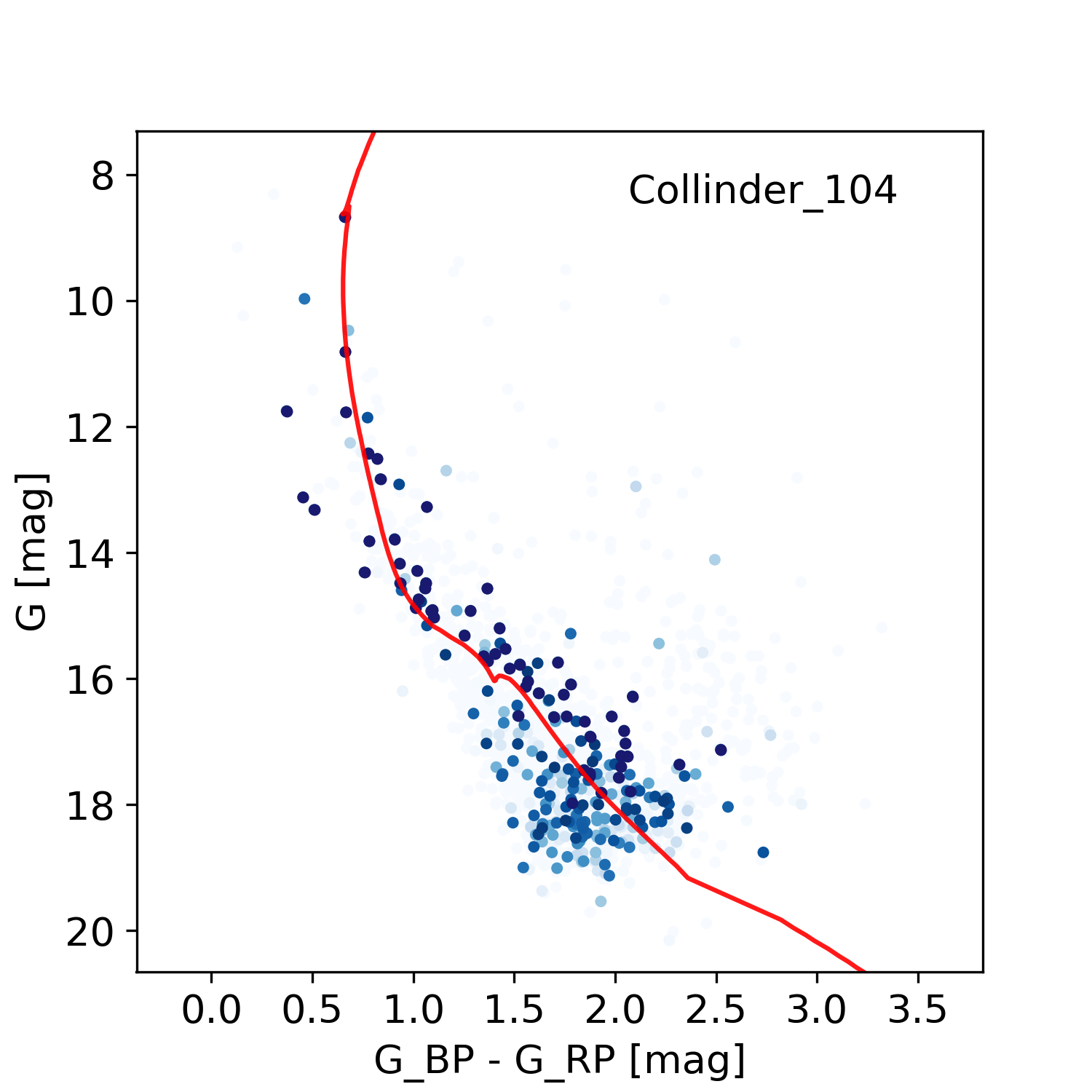}
 \includegraphics[scale = 0.44]{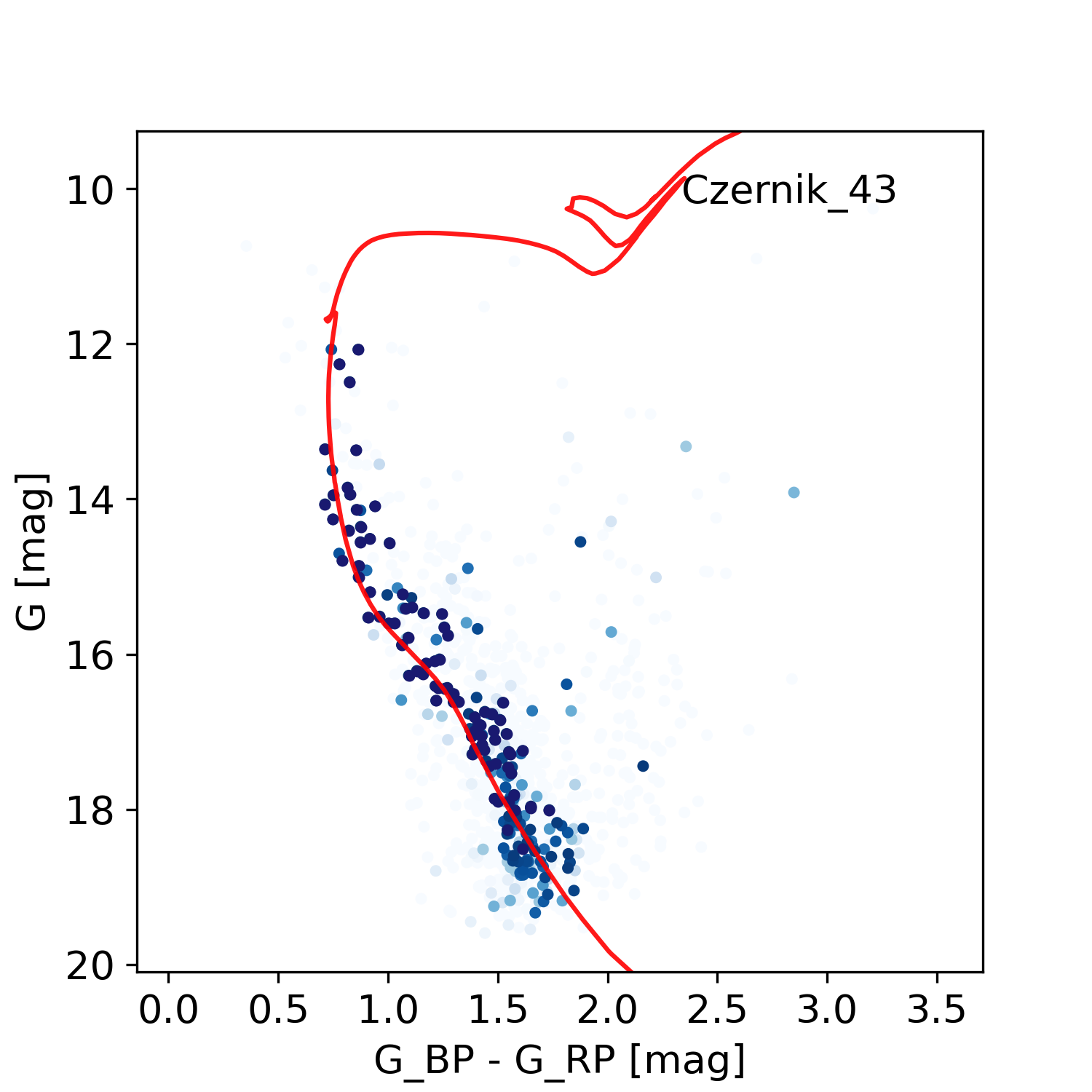} \\
 \includegraphics[scale = 0.44]{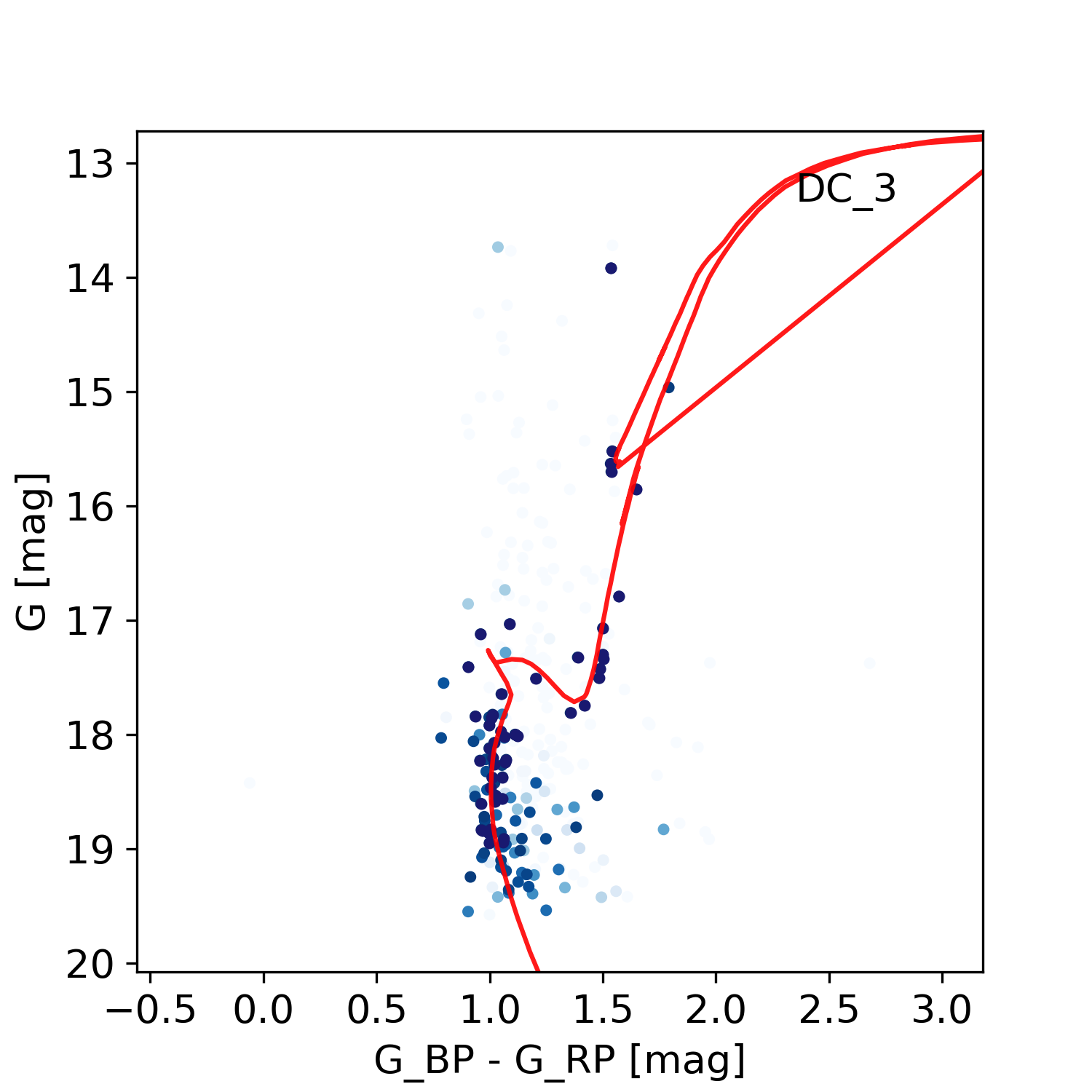}
 \includegraphics[scale = 0.44]{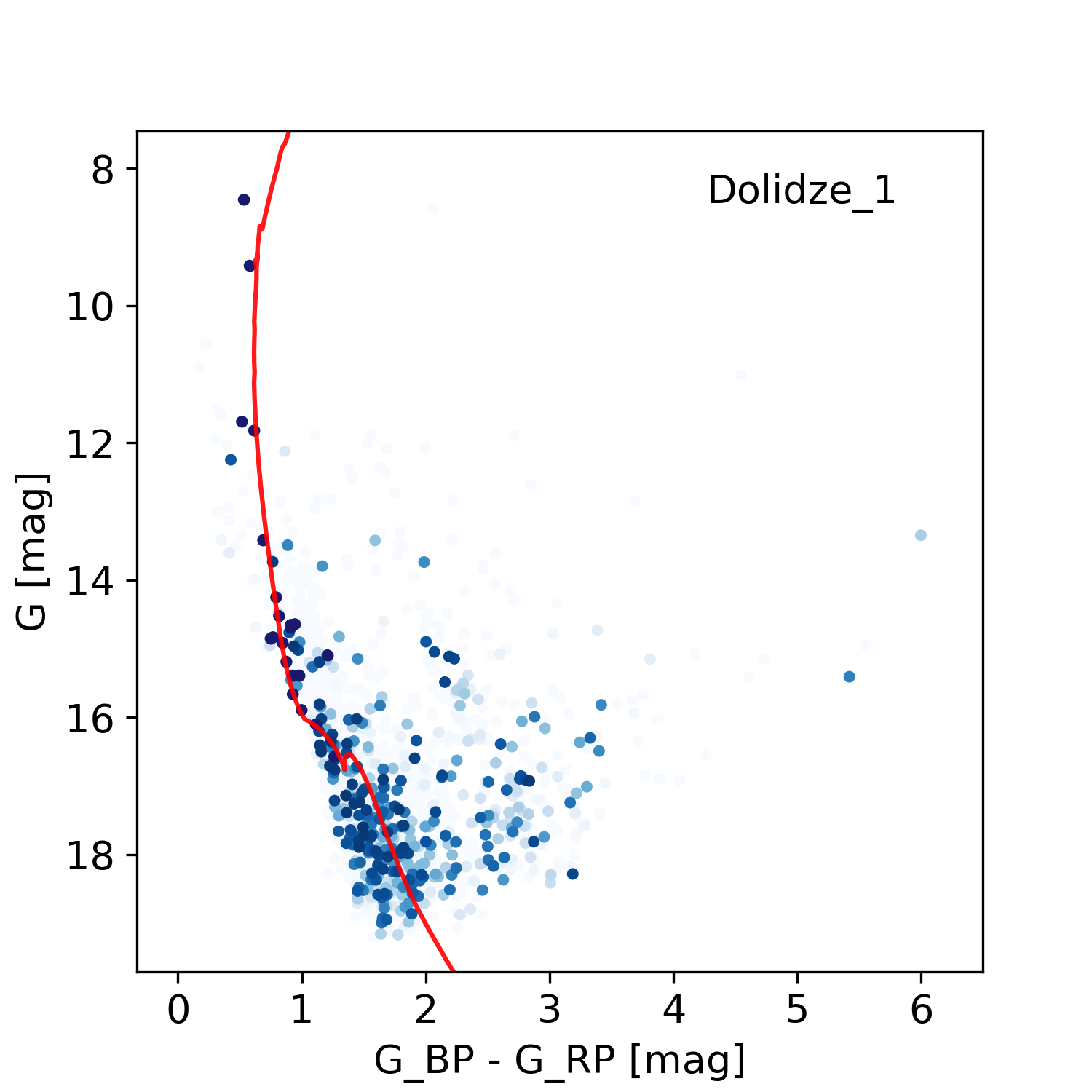}
 \includegraphics[scale = 0.44]{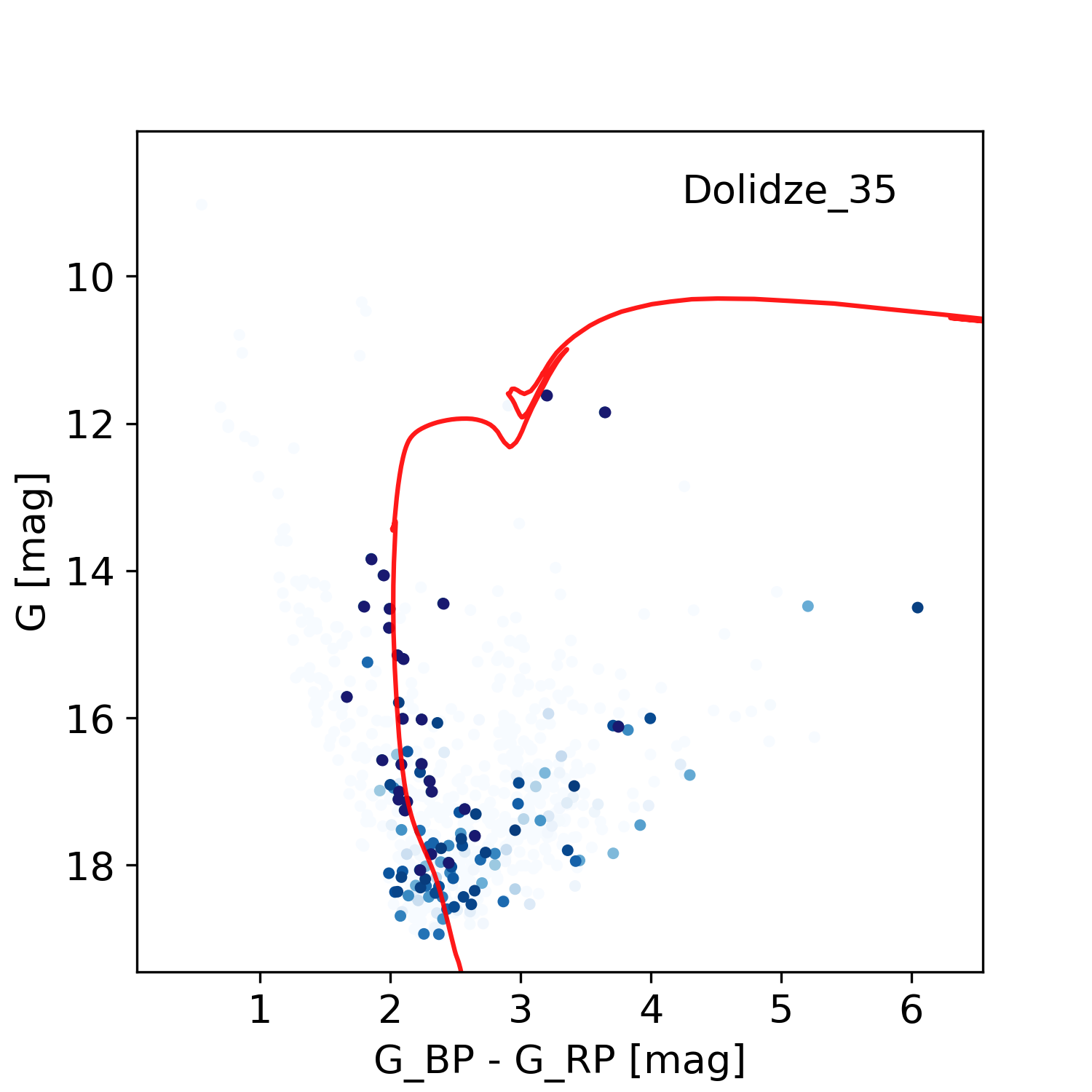} \\
 \includegraphics[scale = 0.44]{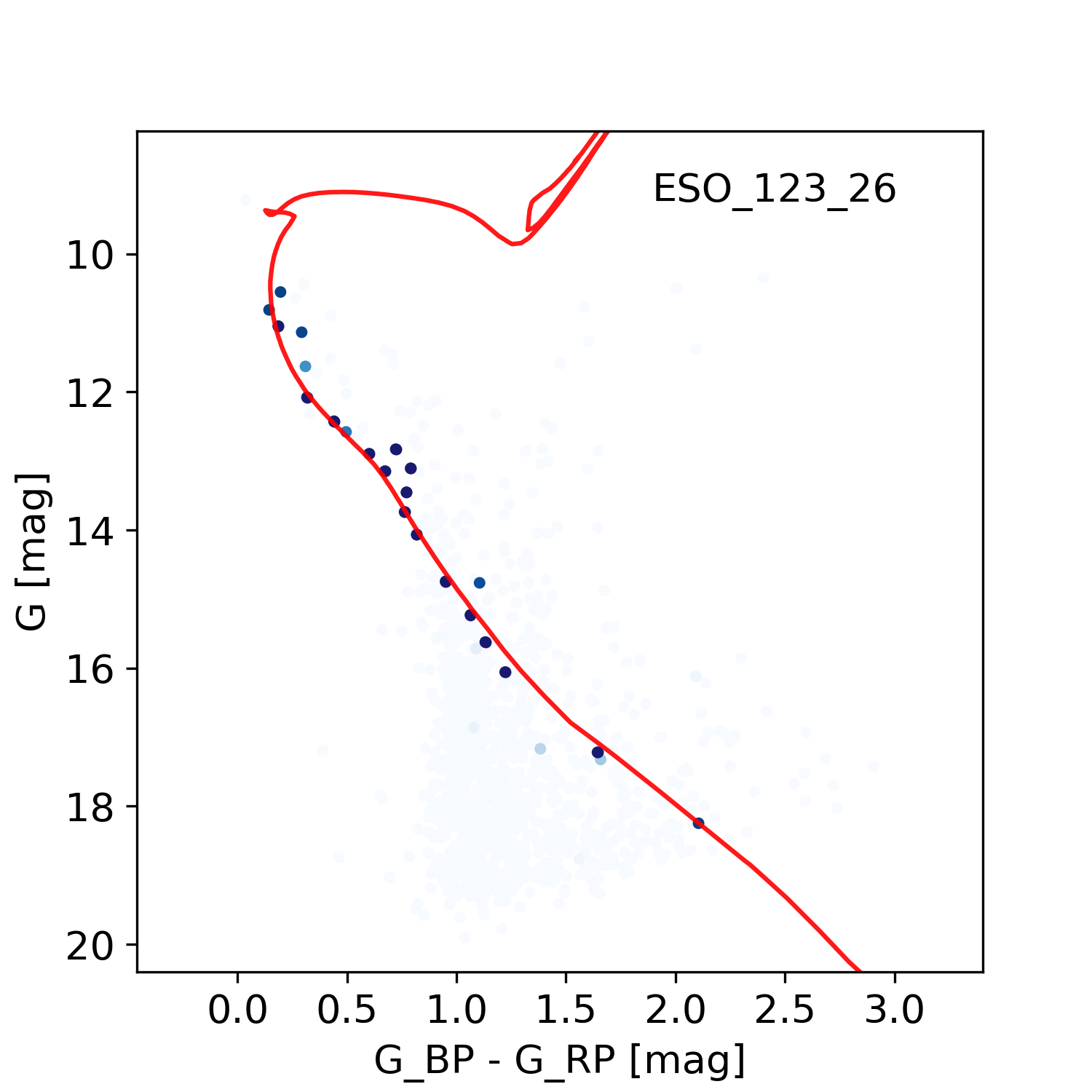}
 \includegraphics[scale = 0.44]{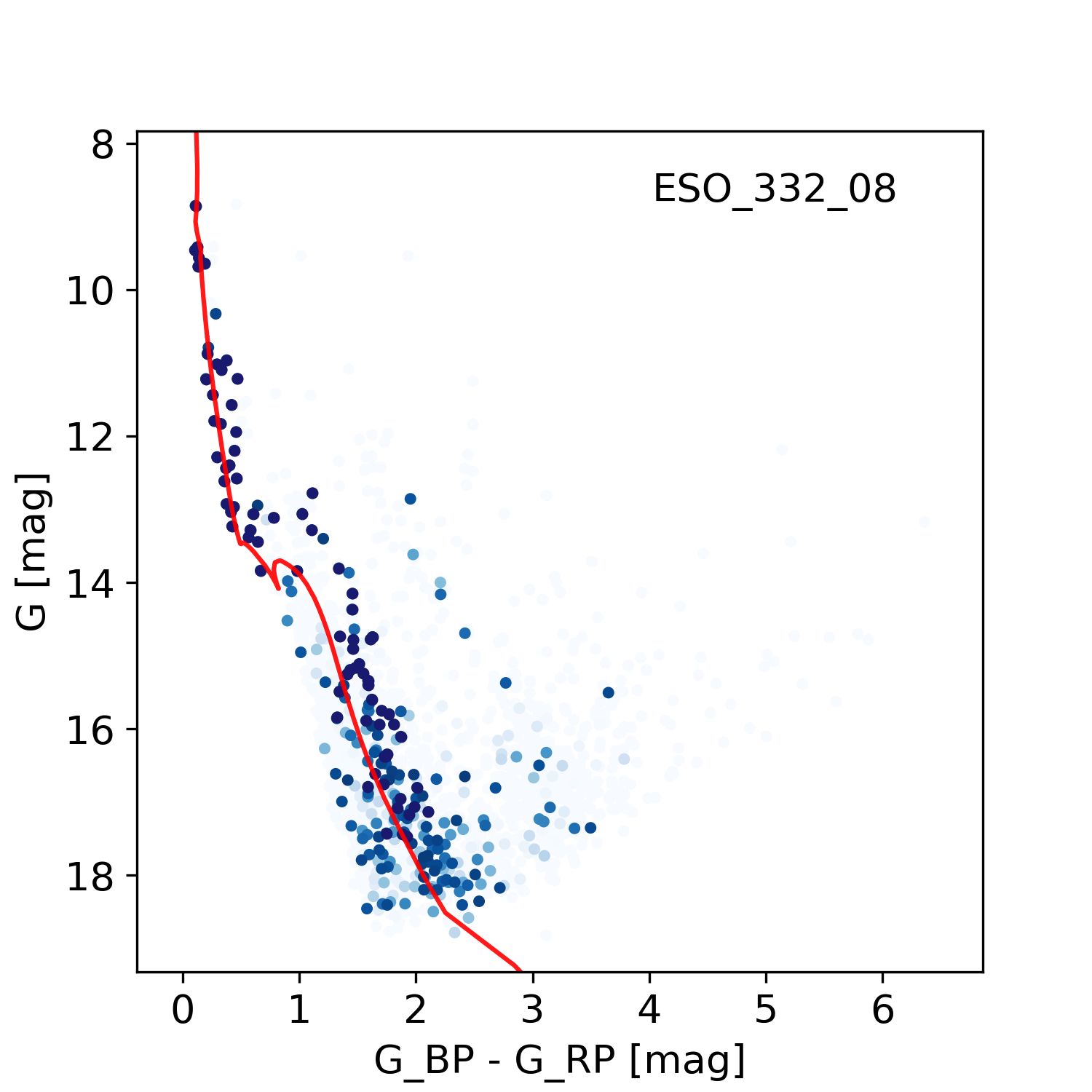}
 \includegraphics[scale = 0.44]{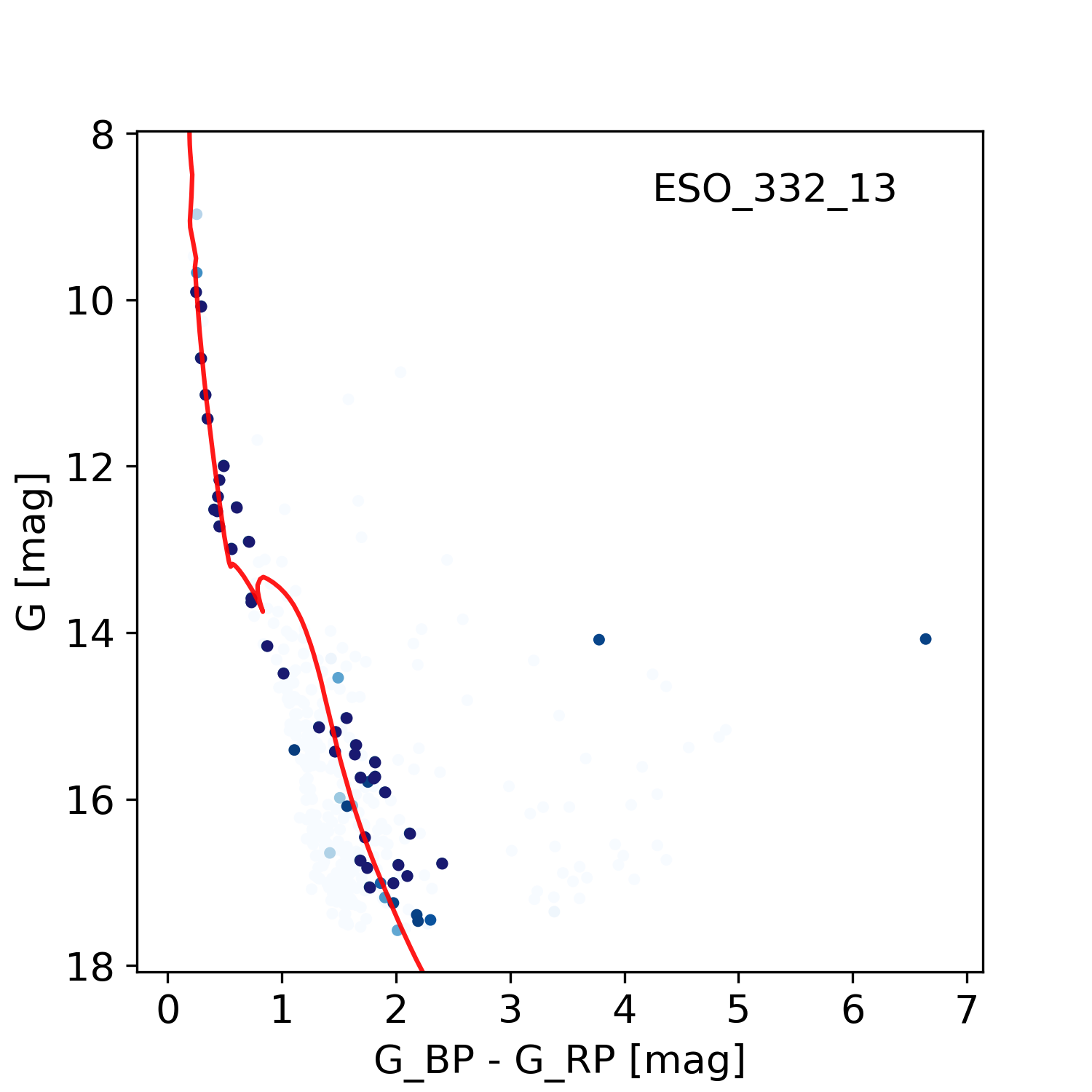} 
 \caption{CMDs and isochrone fits to the Gaia DR2 data for the clusters investigated in this study. Probable member stars are shown in blue dots, with more intense tones indicating higher membership probability. The light-gray dots mark non-member stars in the field.}
 \label{fig:CMD}
\end{figure*}

The cluster sequences for ESO~332-08 and NGC~6885 presented in  Fig.~\ref{fig:CMD}  are well defined and the isochrone fits are clearly adequate.  The parameters in DAML for NGC~6885 are from \citet{Lynga87}. For ESO~332-08 the parameters were taken from \citet{2005A&A...438.1163K}. We note that the same authors later published the MWSC with revised parameters for ESO~332-08, although the ages coincide in both catalogues.  DAML kept the previous version, which listed a larger distance presenting a better fit to CMDs. The isochrone fit in Fig.~\ref{fig:CMD} confirms that the distance in \citet{2005A&A...438.1163K} is  more accurate than the one in the MWSC. 

NGC~133 is the most discrepant cluster in the sample. It is visually identified in DSS images from a small group of bright stars. The CMD in Fig.~\ref{fig:CMD} displays a bifurcation around $G\sim$ 16 mag, leading to a redder evolved branch that our isochrone fit follows, but does not include the bright stars. The blue branch does include the brighter stars, which is what \citet{2002A&A...387..479C} identifies as NGC~133 and results in the parameters listed in DAML. A possibility would be that we are looking at different objects along the same line of sight. However, both branches display the same proper motions and have probable members, which together with the sparseness of the blue branch indicates that the apparently younger sequence is composed of blue stragglers in NGC~133. We conclude that the cluster now revealed by {\it Gaia} DR2 is in fact older than previously estimated.

ESO 429-02 is an interesting case. The cluster sequence revealed by its members is sparse, but still clearly young  with a pronounced pre-main-sequence (PMS) well fitted by the $logt =$ 7.1 isochrone in Fig.~\ref{fig:CMD} once taking into account the relatively high (variable) extinction and variability in the PMS phase. The parameters in DAML come from the analysis of 2MASS photometry and UCAC2 astrometry done by  \citet{2007A&A...468..139P}. Despite the above mentioned limitations of these data-sets, their work reveals a CMD that although poor, can  be  plausibly reproduced by an older $logt =$ 8.6 isochrone.  An inspection of the {\it Gaia} DR2 proper motion vector point diagram reveals two over-densities, in which the stronger peak corresponds to the sequence identified in our work. It is a possible case of two different objects along the same line of sight. 

Of the 11 mean radial velocities of open clusters determined here, 6 are in common with DAML (Bochum 10, NGC 6885, Trapezium-FG, NGC 1980, NGC 1977 and Ruprecht 55) published by \citet{Dias2014}. The comparison of this small sample shows discrepancies ranging from -29 $km s^{-1}$ to 4 $km s^{-1}$. Considering that the memberships presented in this work are superior to those published in \citet{Dias2014}, we believe the radial velocity estimates in this work are more reliable.

In the previous sections we validated our cluster parameter determination procedure by comparing distances with those from from {\it Gaia} DR2 parallaxes and metallicities with those from high resolution spectroscopy. In this section we confirm that in general, while following the same trend as those from pre-{\it Gaia} studies, our determinations, represent a substantial improvement over the previous values. 
It is also interesting how the comparisons clearly show that in this case, a non-homogeneous compilation of parameters (DAML) can provide a more accurate data-set than an homogeneously derived catalogue (MWSC). We note, however, that this is seen because we removed the the MWSC values from the DAML sample.

\section{Conclusions}

We have investigated 45 open clusters with {\it Gaia} DR2. From the astrometric data (proper motions and parallaxes) we determined their stellar membership probabilities, taking into account the full co-variance matrix of the data. 

For all clusters we estimated mean proper motion and mean parallax considering the member stars (membership $\ge$ 0.51). Mean radial velocities were determined for 12 clusters, 7 of them for the first time, although based on small numbers of members with available Gaia DR2 radial velocity measurements. The fundamental parameters age, distance and $A_V$ were estimated with a new version of the global optimization code presented in \citet{Monteiro2017} applied to $G_{BP}$ and $G_{RP}$ photometry using a revised extinction polynomial law for {\it Gaia} DR2 and the Galactic abundance gradient as a prior for metallicity. The new procedure was validated using a sample of clusters in the literature for which high resolution spectroscopy was available. Our isochrone fitting results for a high resolution spectroscopy sample are also presented. 
We verify that the PMS portions of the PARSEC isochrones fit well the cluster sequences, consistently with the main sequence fit, indicating that they are suitable for analyses of young clusters (down to 4 Myr) at least in the Gaia photometric bands.

This study provides the first determination of distance and age for the cluster Majaess~65 and of age for Ruprecht~123. The cluster DC~3 is found to be one of the oldest (5.6 Gyr) and most distant ($\sim$7900 pc) known open clusters.

We assessed the quality of our results by comparing with distances from parallaxes, metallicities from high resolution spectroscopy and a critic inspection of the literature.   
In the process, we identified several clusters reported as new discoveries in recent papers based on {\it Gaia} DR2 that were already known clusters listed in DAML. We find that our cluster parameter determinations, represent a substantial improvement over the previous values.

This work is part of an ongoing project that will bring DAML to the {\it Gaia} era.

\addtocounter{figure}{-1}
 \begin{figure*}
 \centering
 \includegraphics[scale = 0.45]{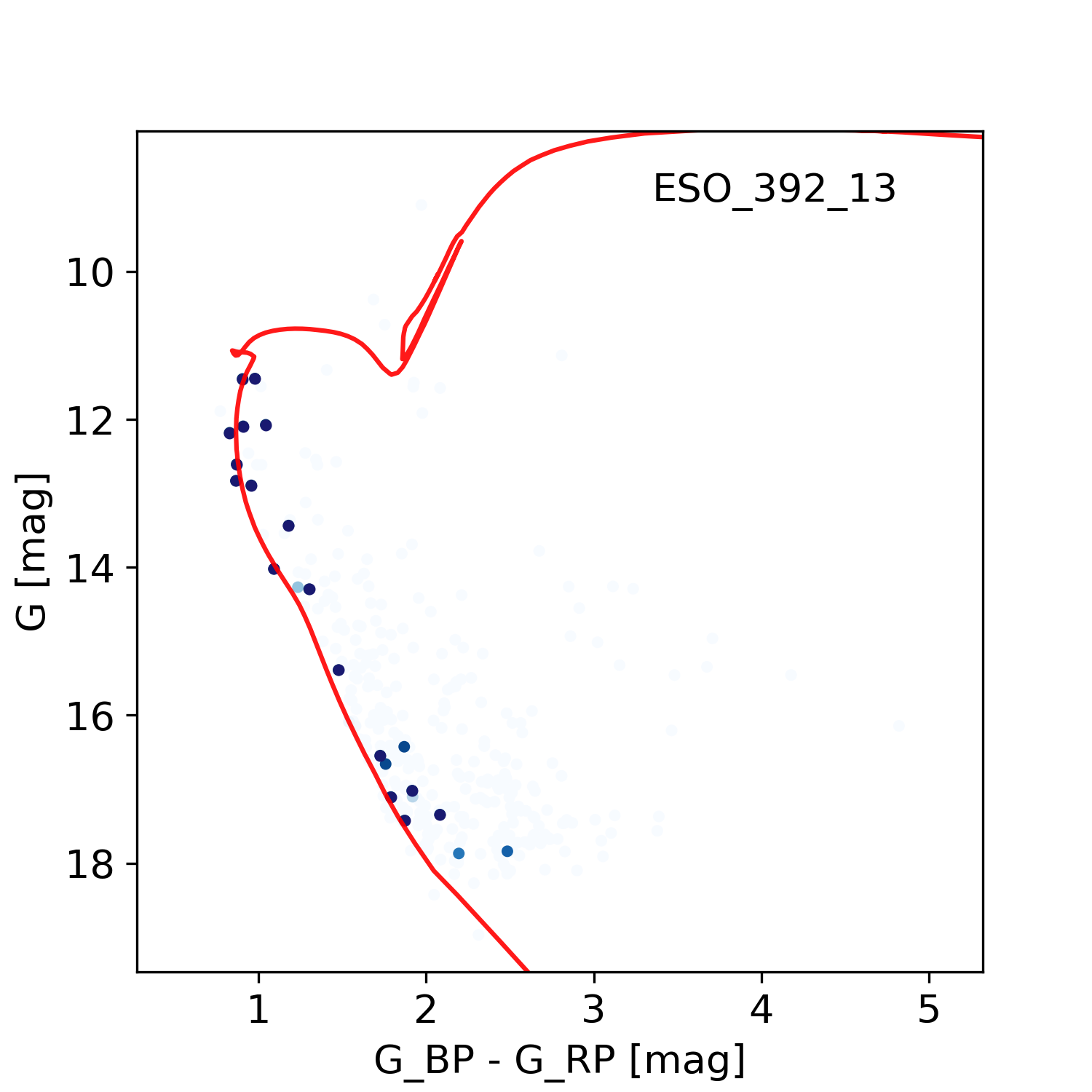}
 \includegraphics[scale = 0.45]{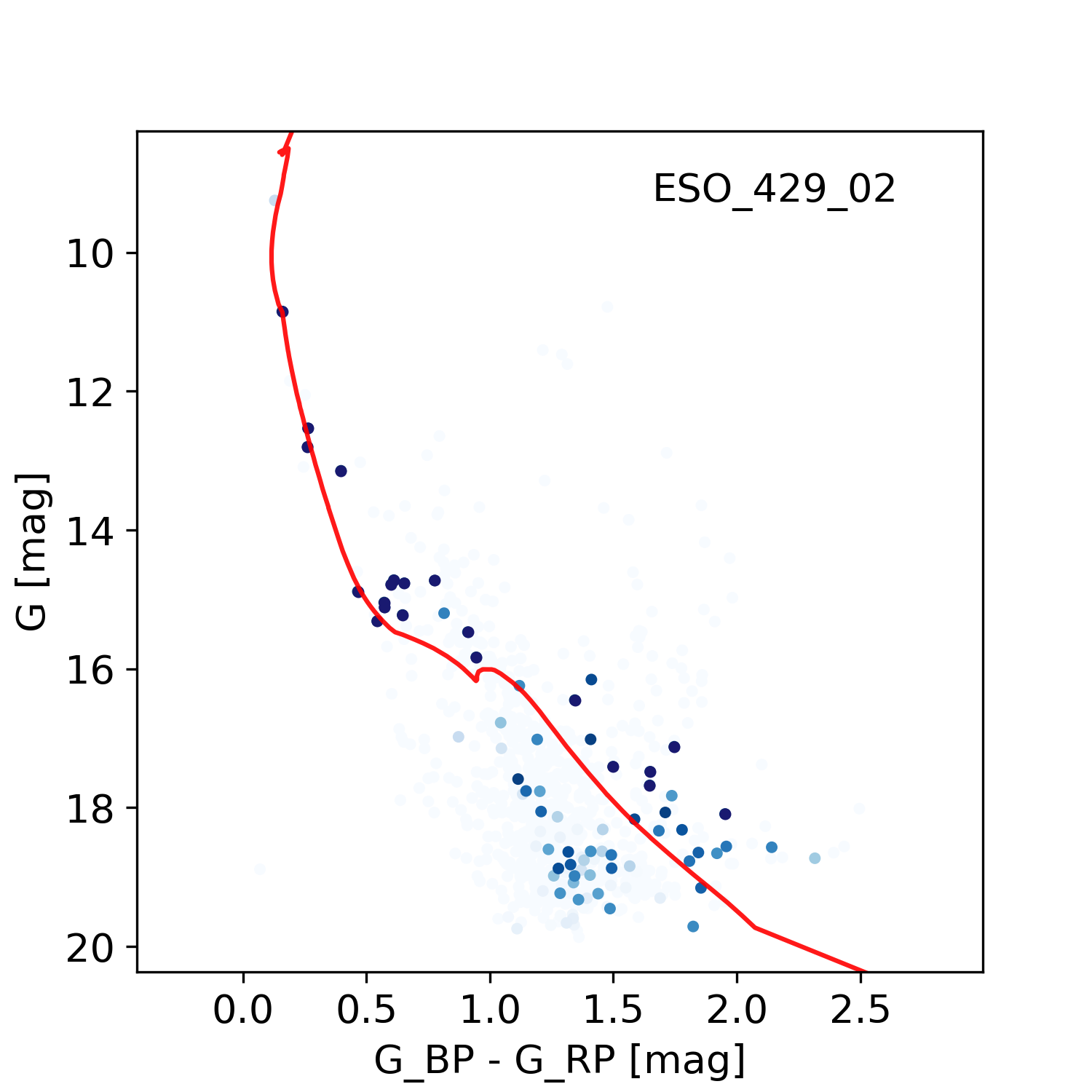}
 \includegraphics[scale = 0.45]{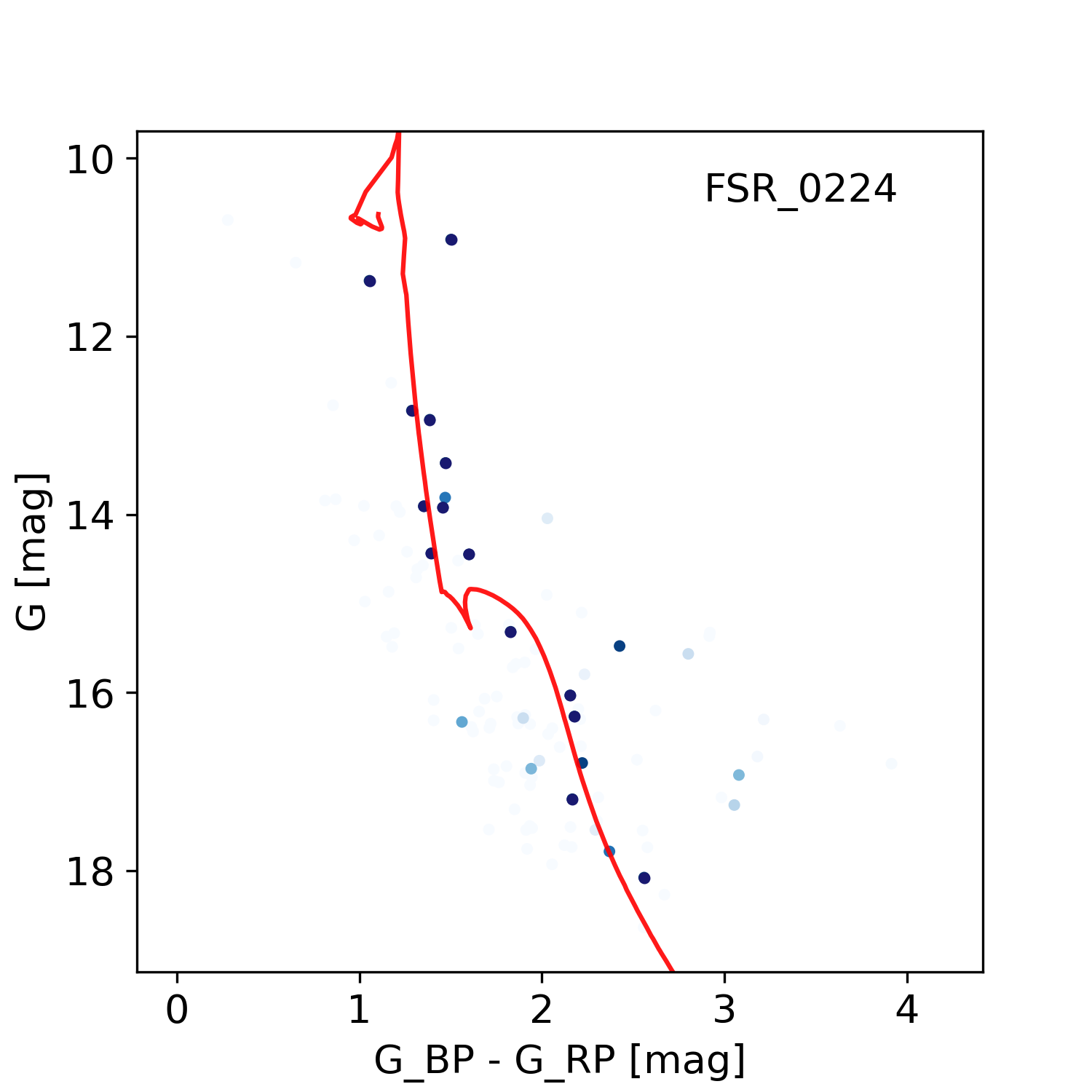} \\
 \includegraphics[scale = 0.45]{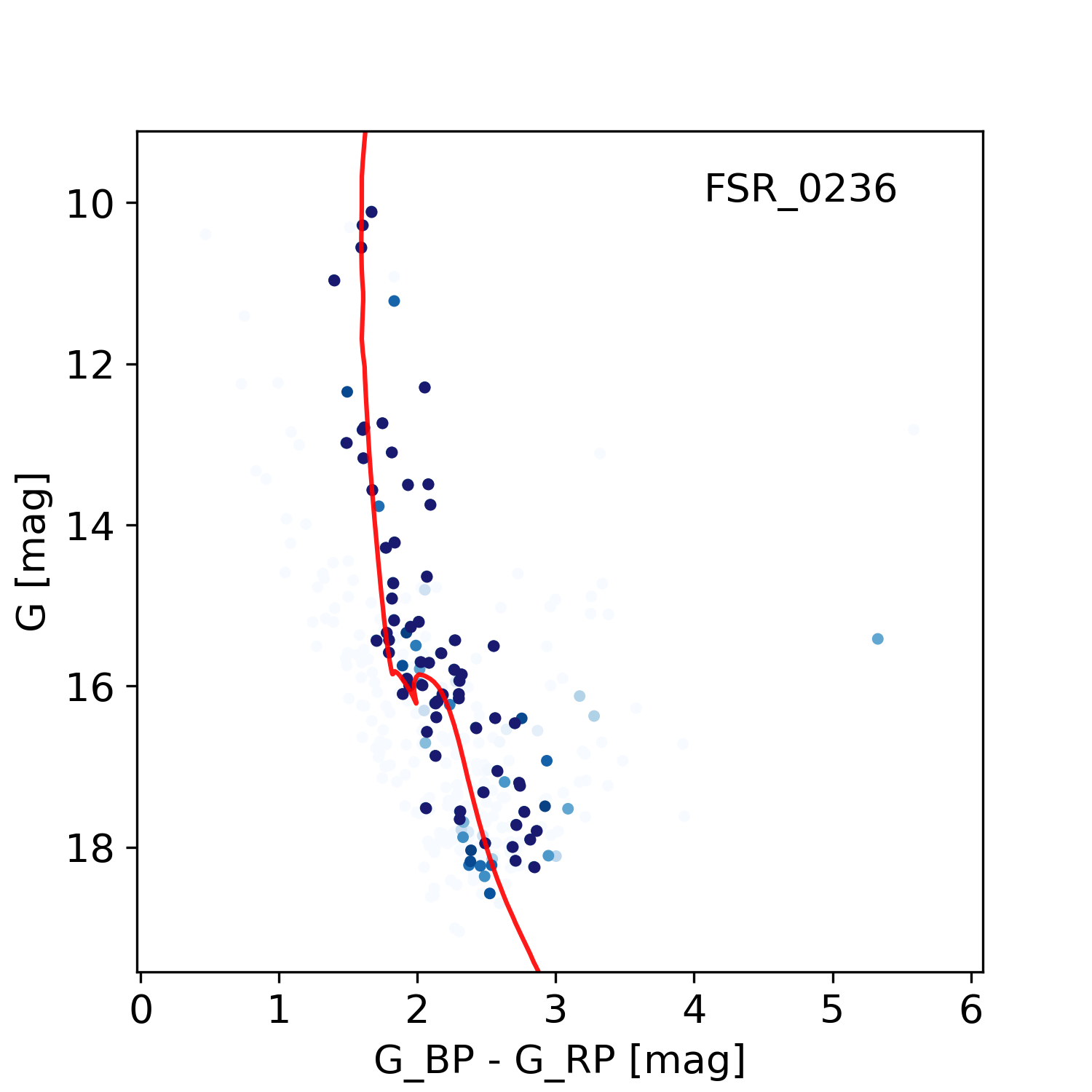}
 \includegraphics[scale = 0.45]{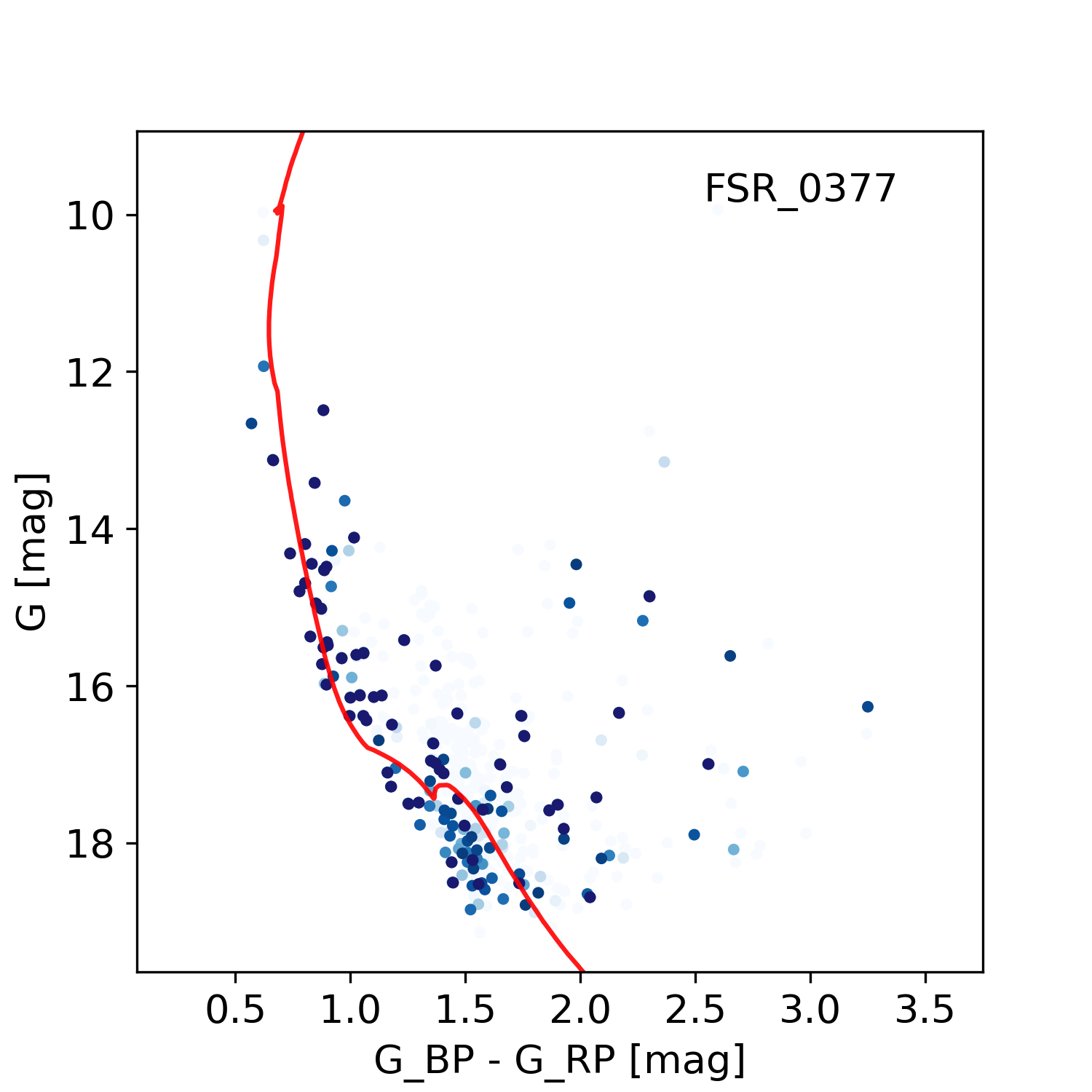}
 \includegraphics[scale = 0.45]{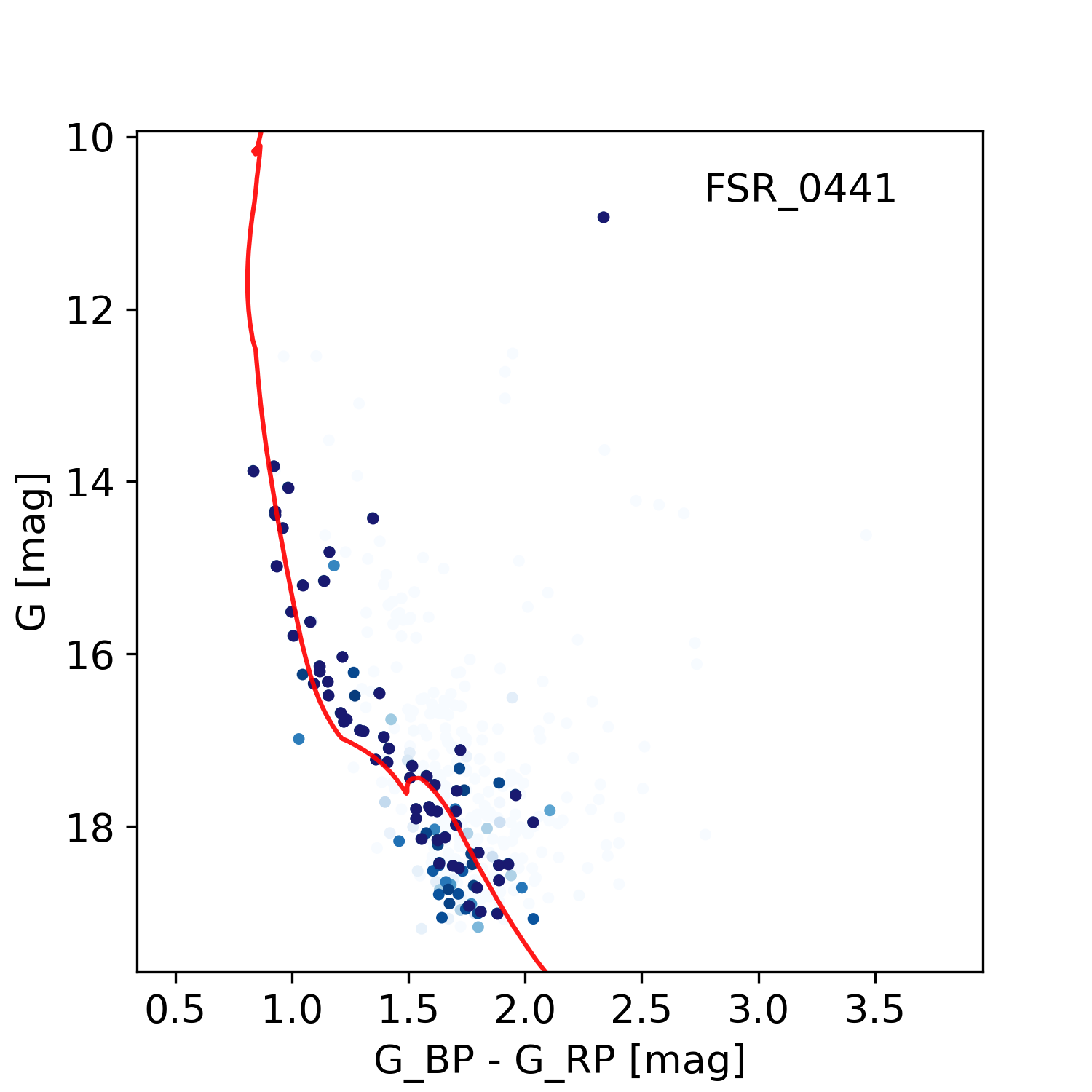} \\
 \includegraphics[scale = 0.45]{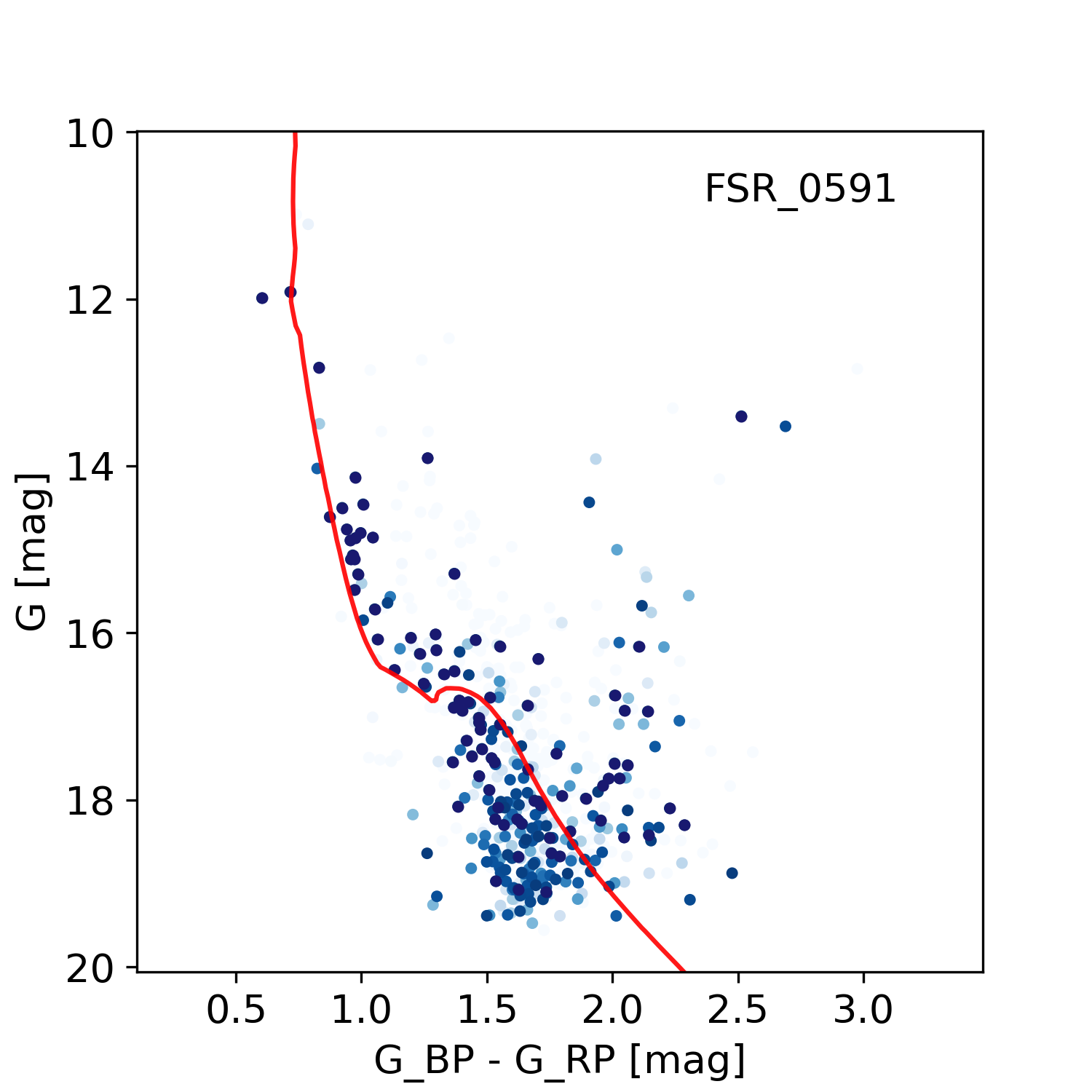}
 \includegraphics[scale = 0.45]{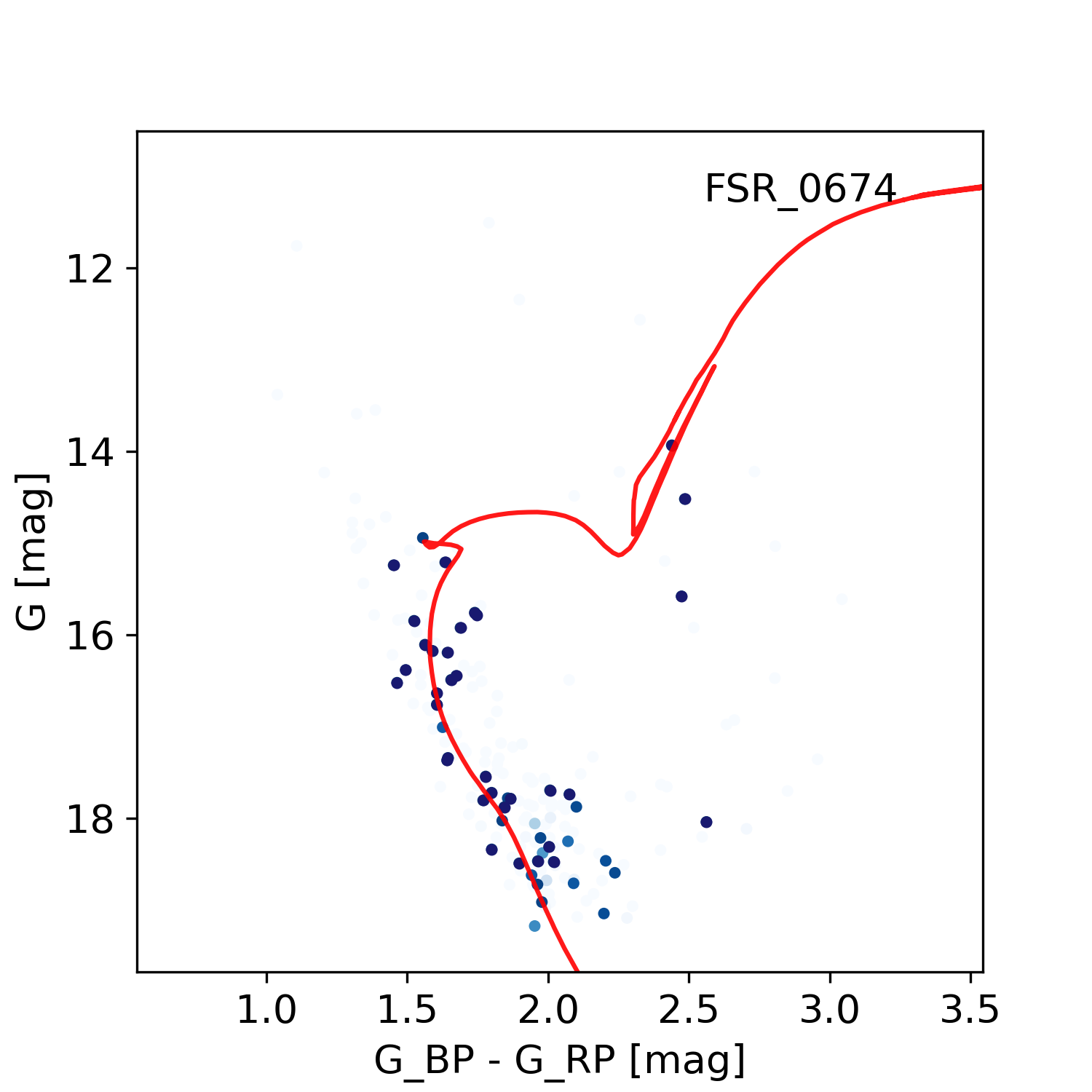}
 \includegraphics[scale = 0.45]{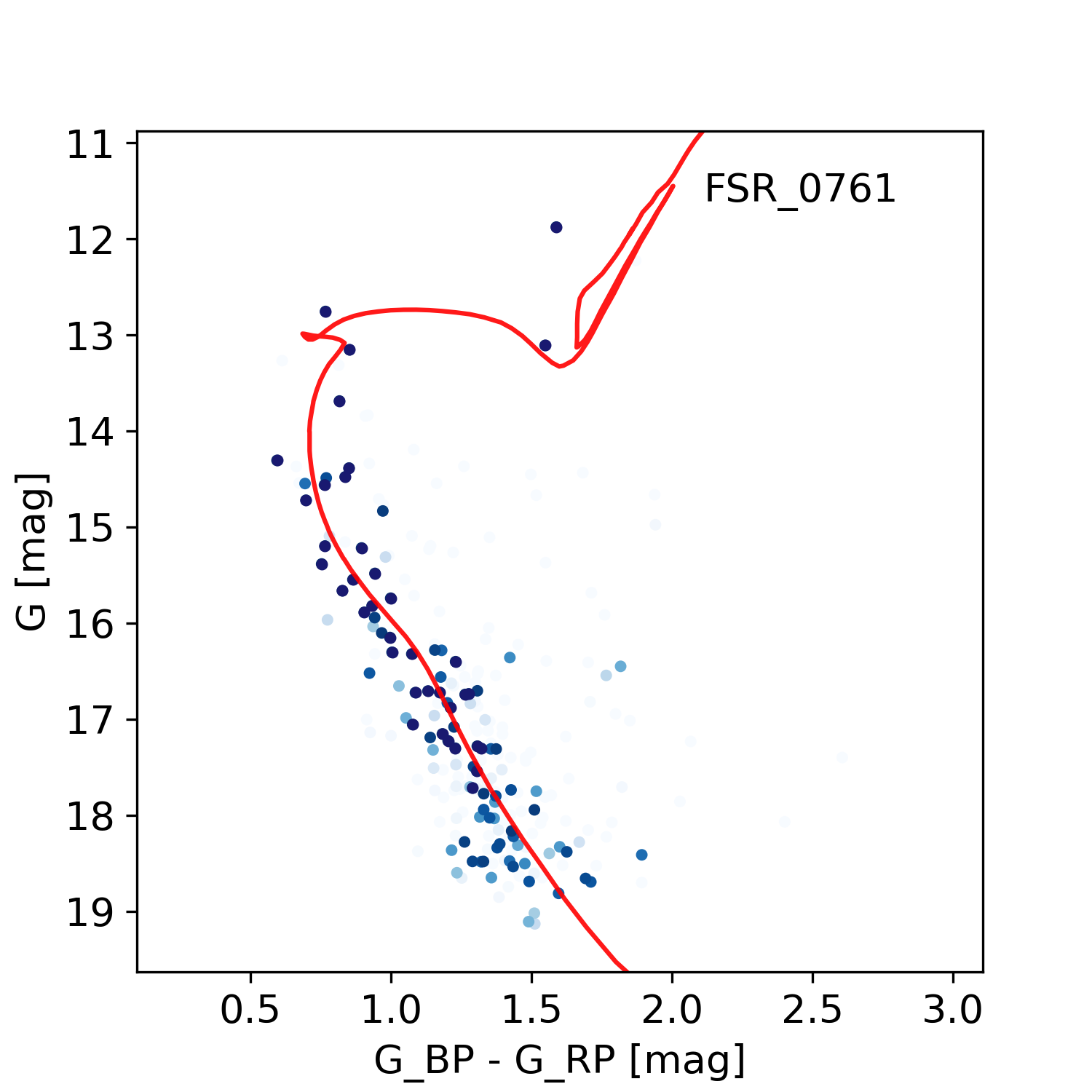} \\
 \includegraphics[scale = 0.45]{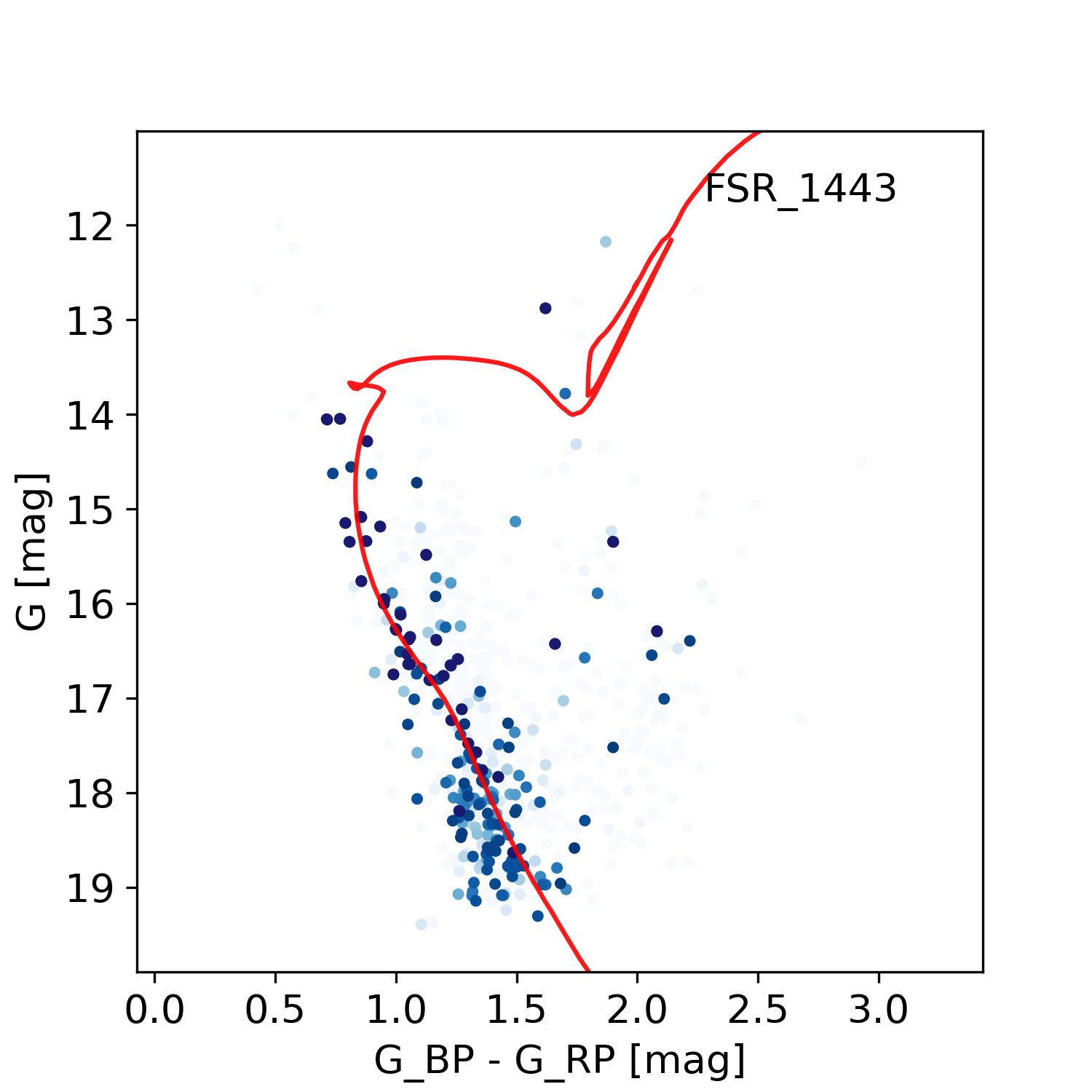}
 \includegraphics[scale = 0.45]{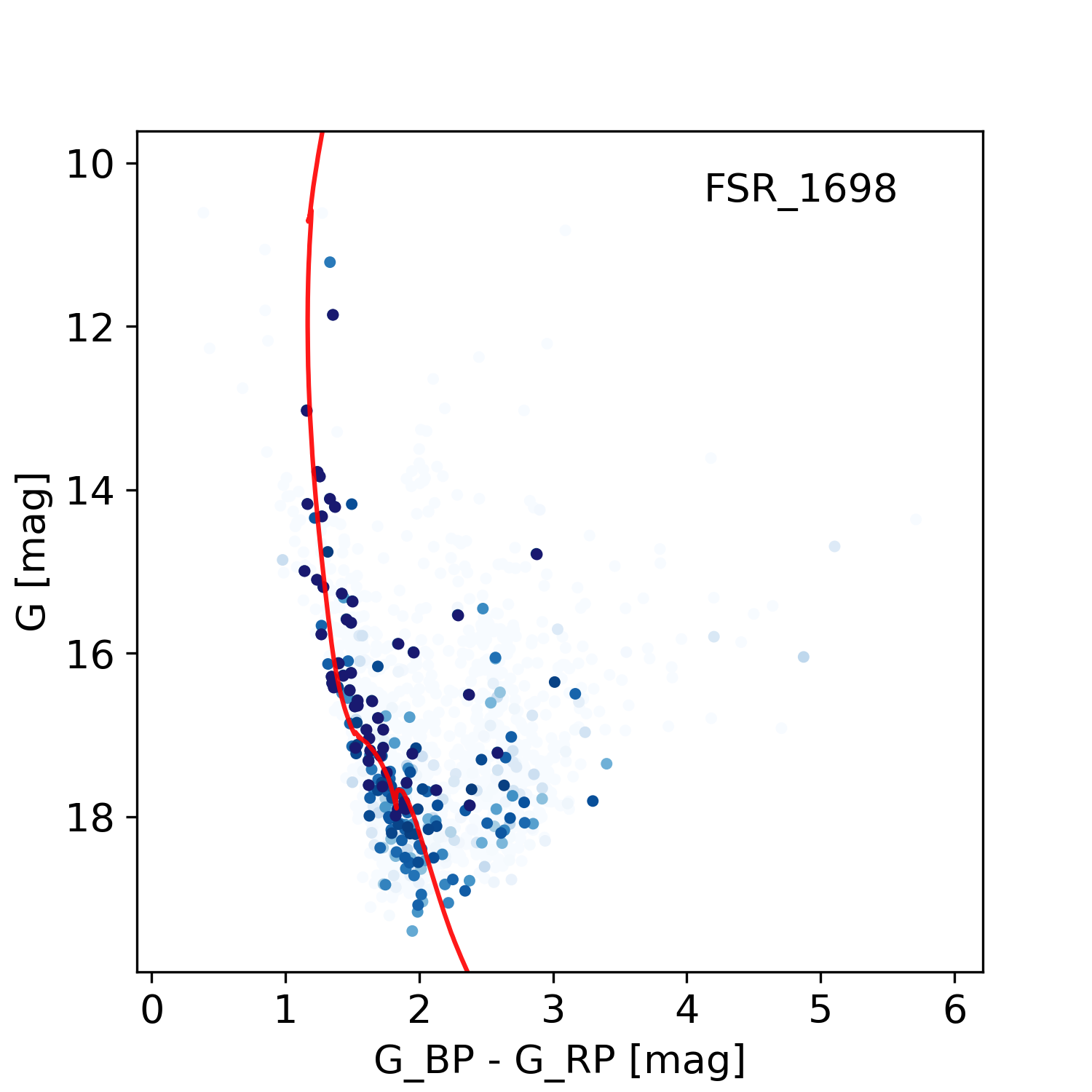}
 \includegraphics[scale = 0.45]{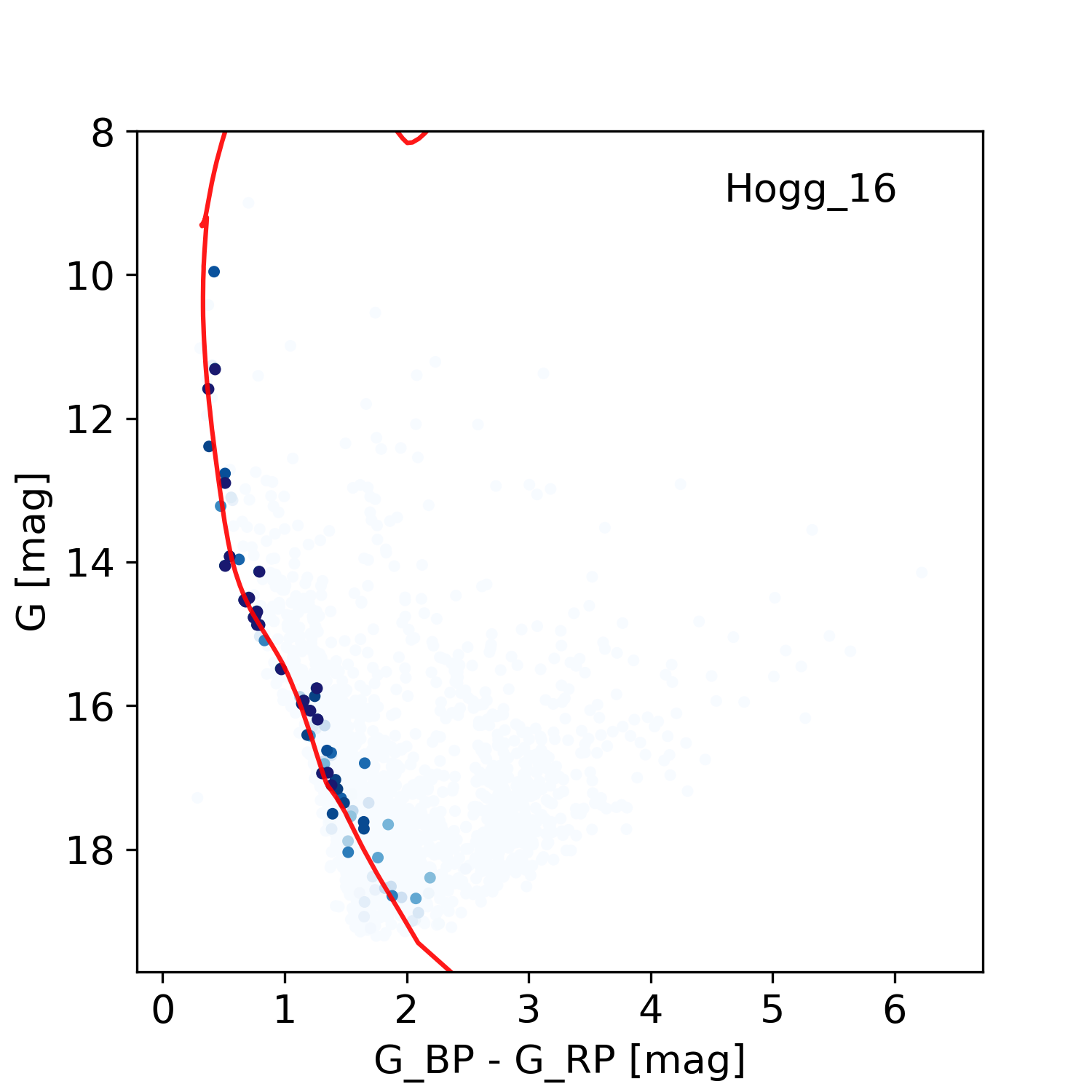}
 \caption{CMDs and isochrone fits (continued)}
 \end{figure*}

\addtocounter{figure}{-1}
 \begin{figure*}
 \centering
 \includegraphics[scale = 0.45]{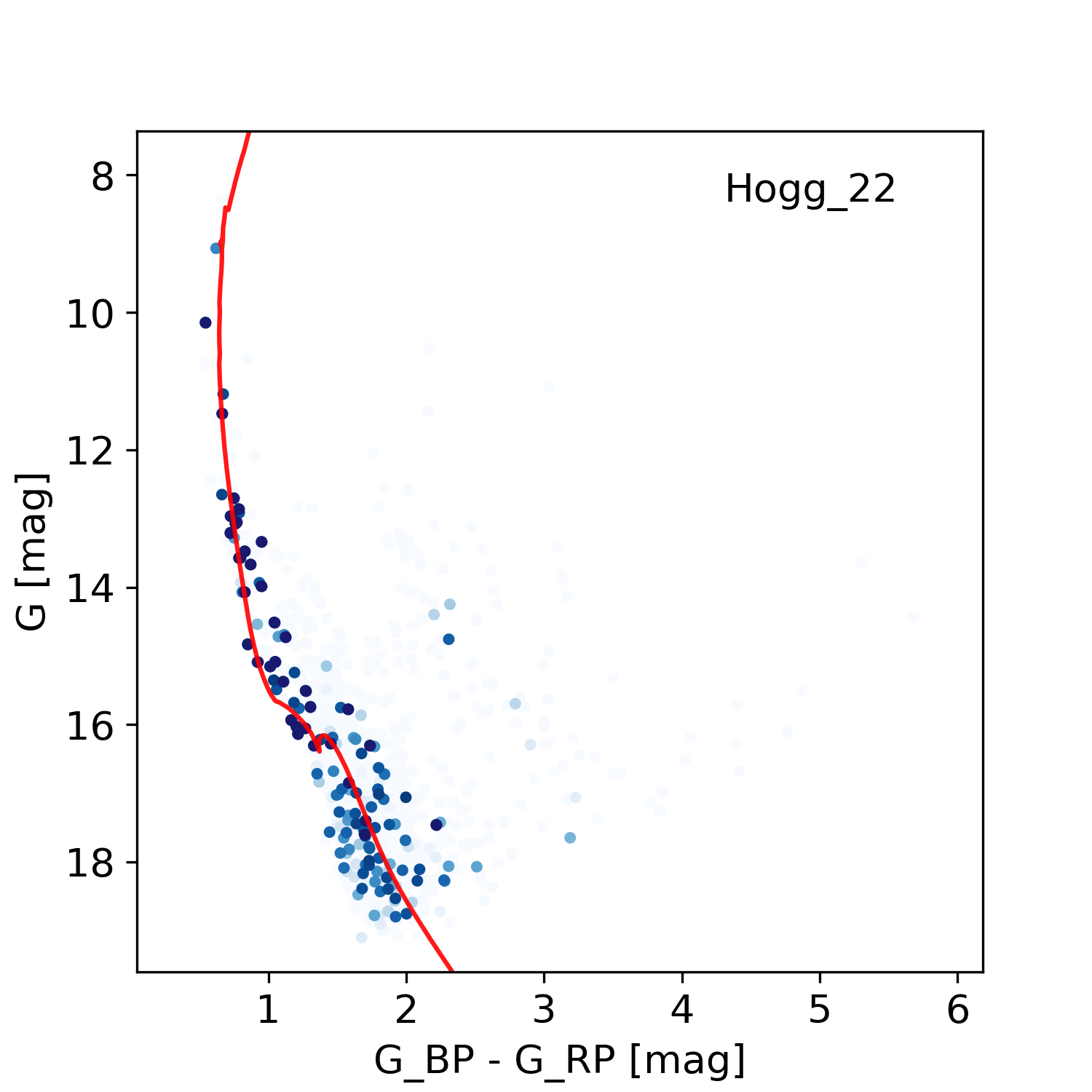}
 \includegraphics[scale = 0.45]{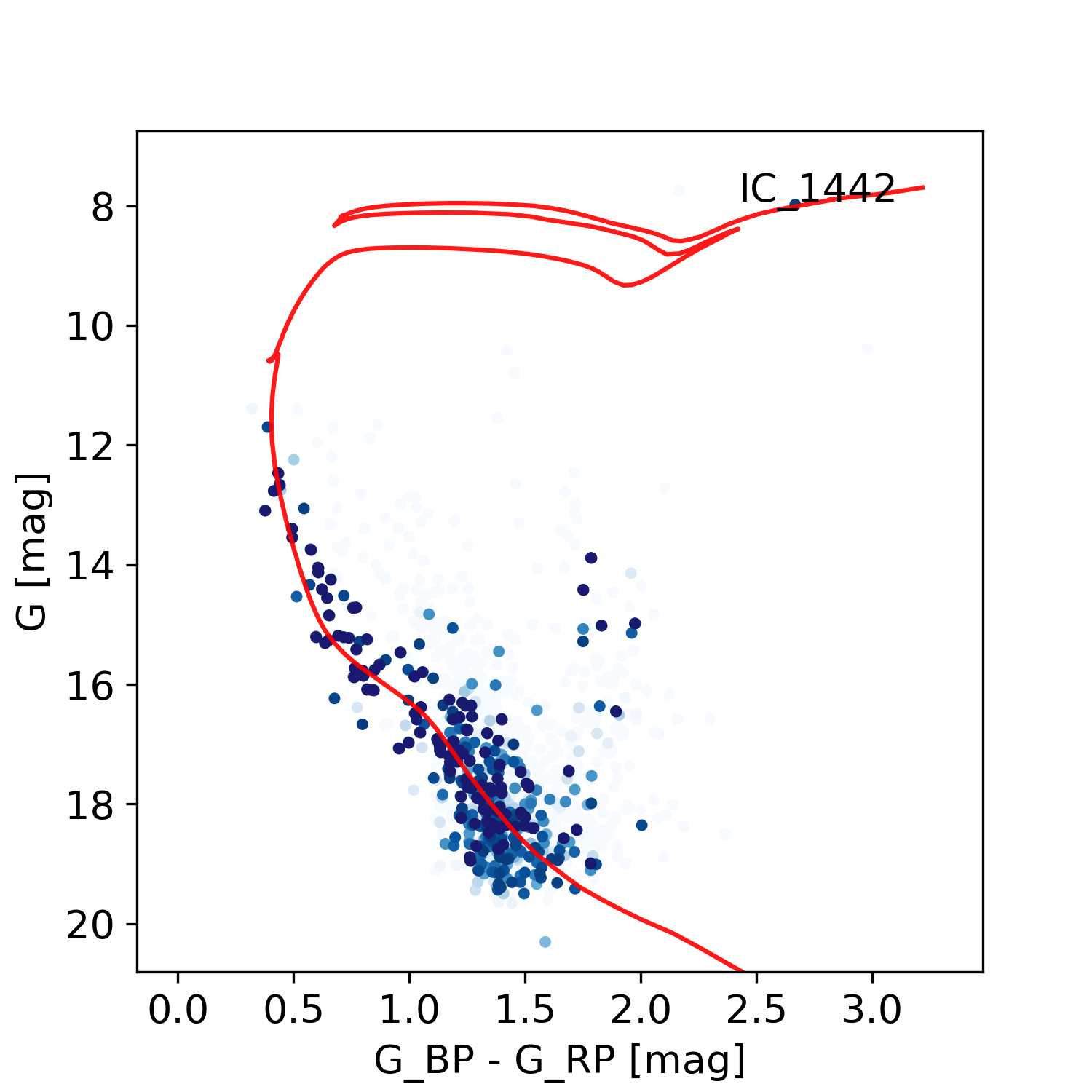}
 \includegraphics[scale = 0.45]{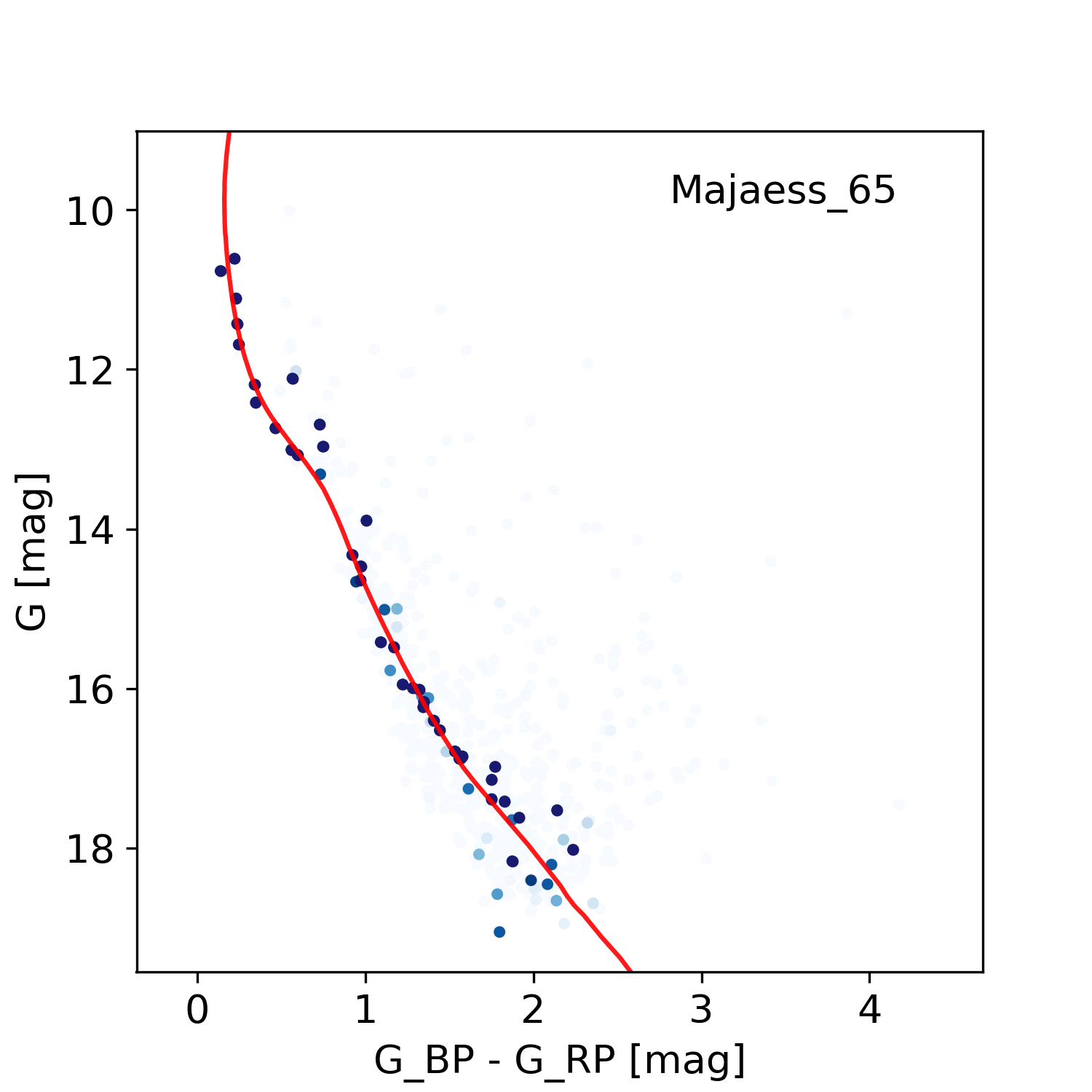} \\
 \includegraphics[scale = 0.45]{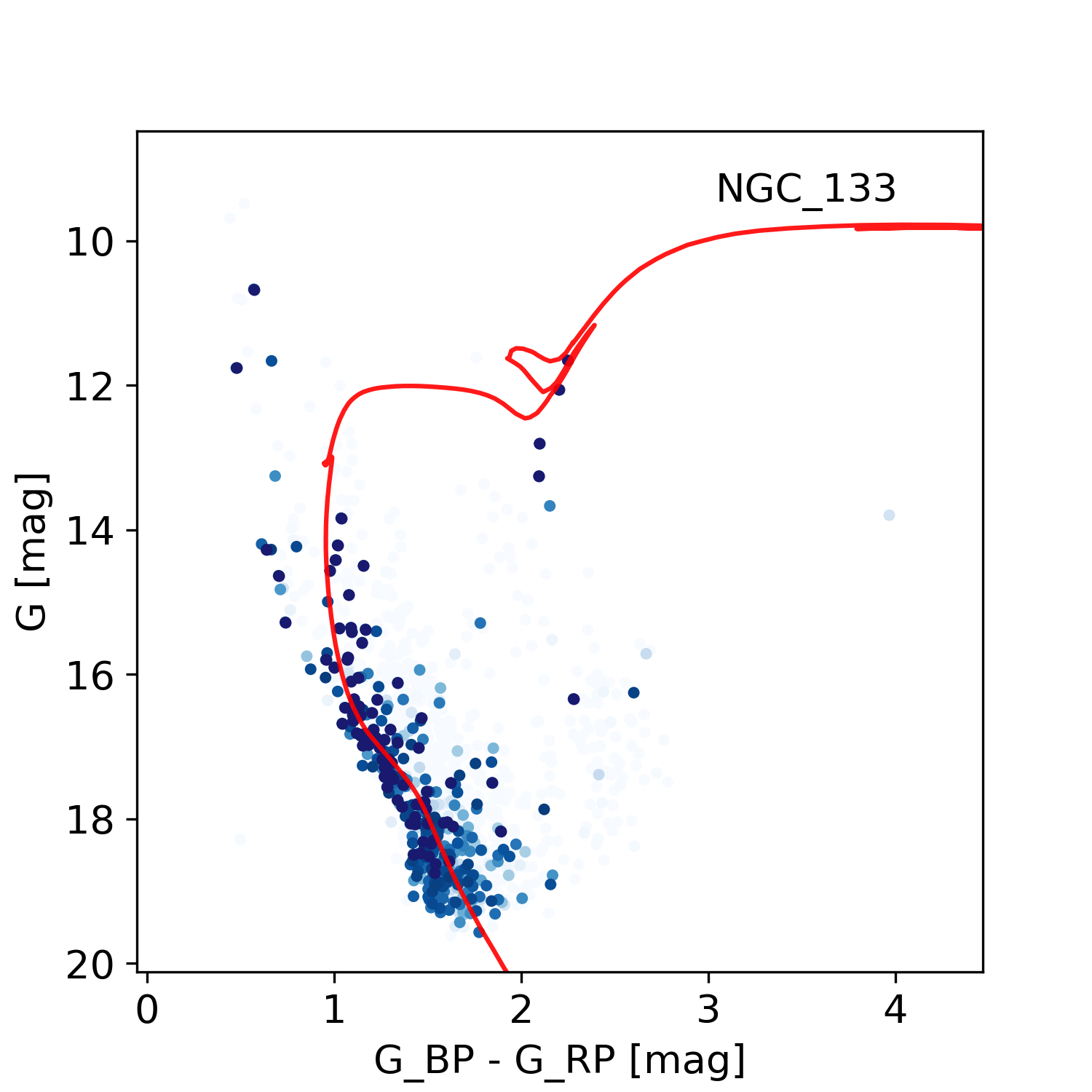}
 \includegraphics[scale = 0.45]{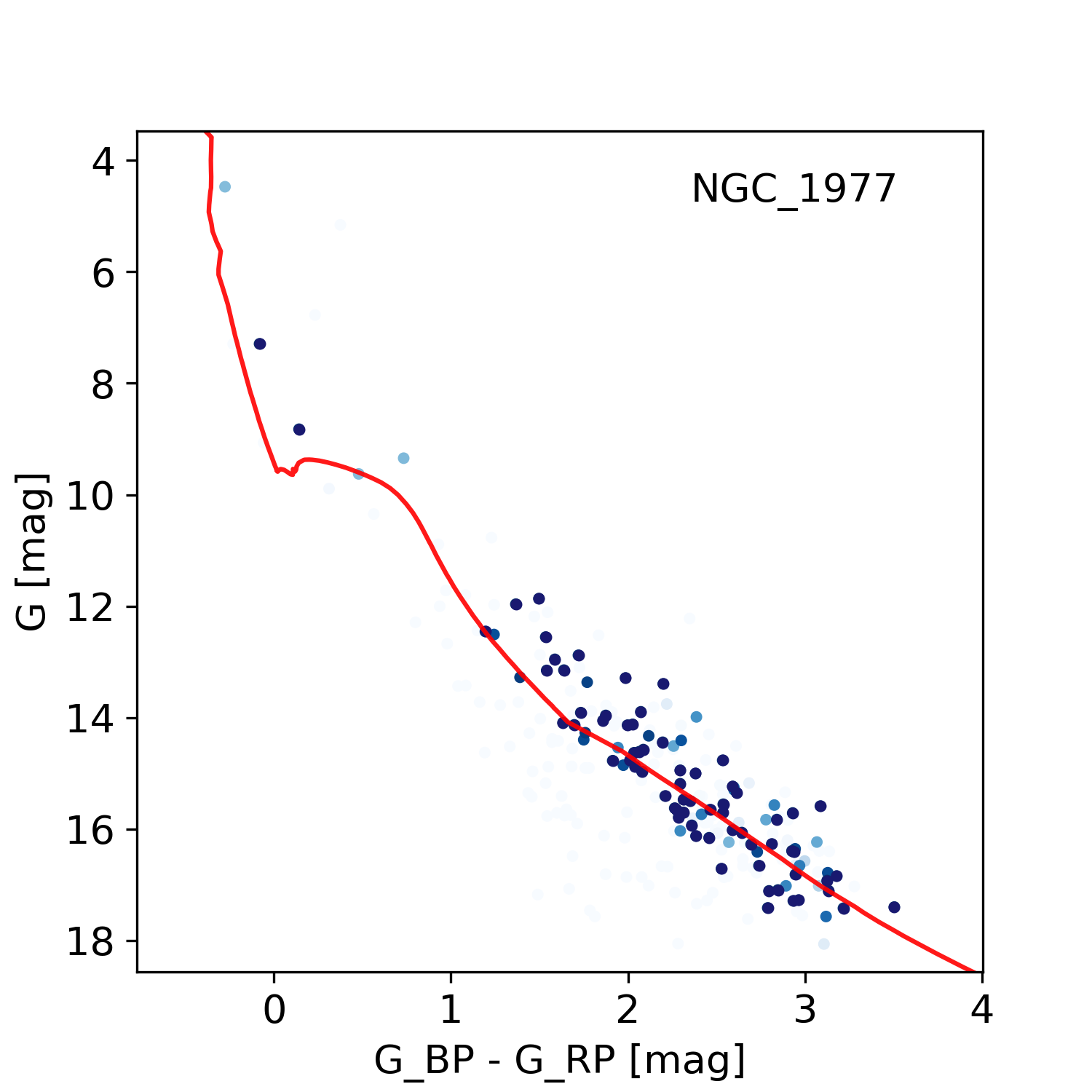}
 \includegraphics[scale = 0.45]{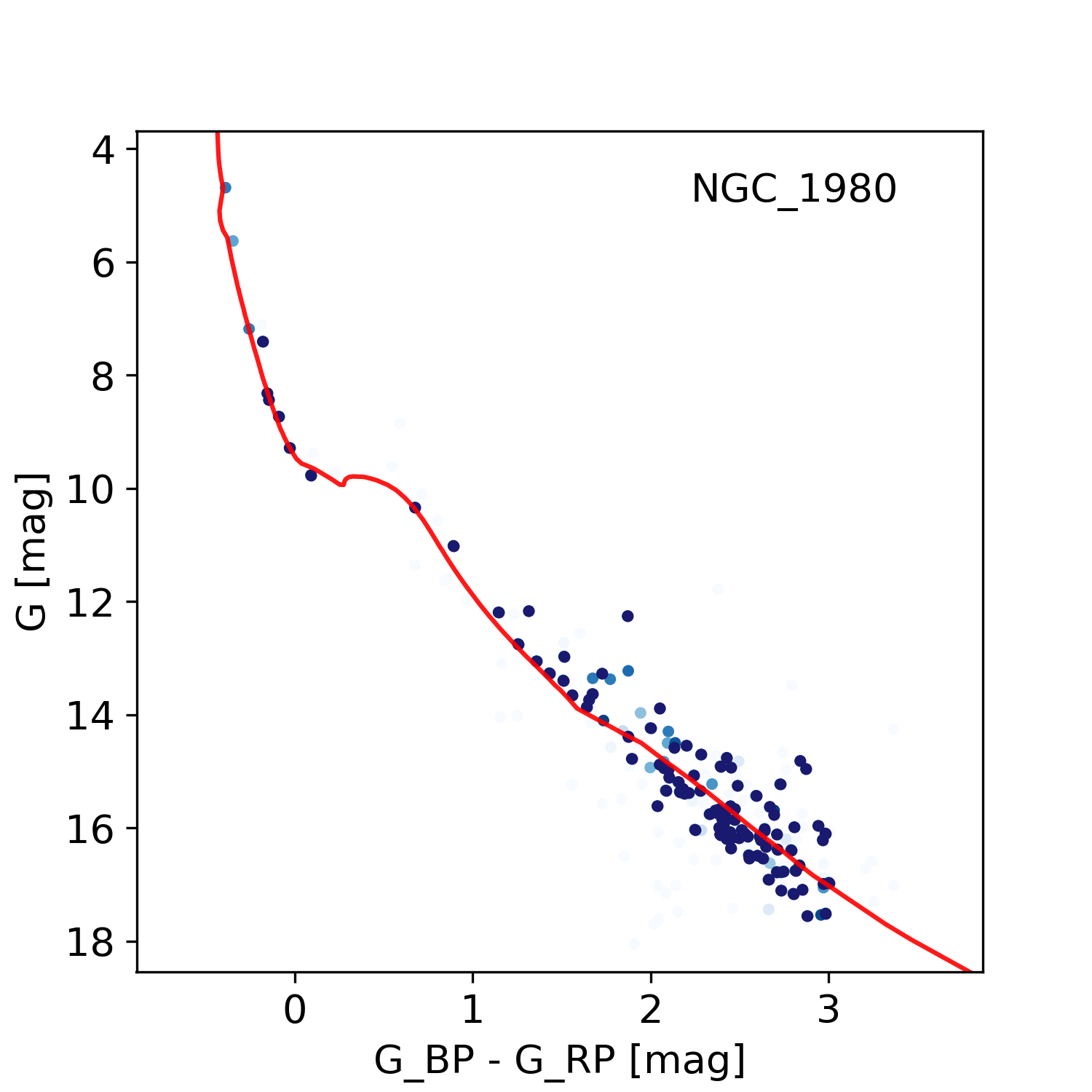} \\
 \includegraphics[scale = 0.45]{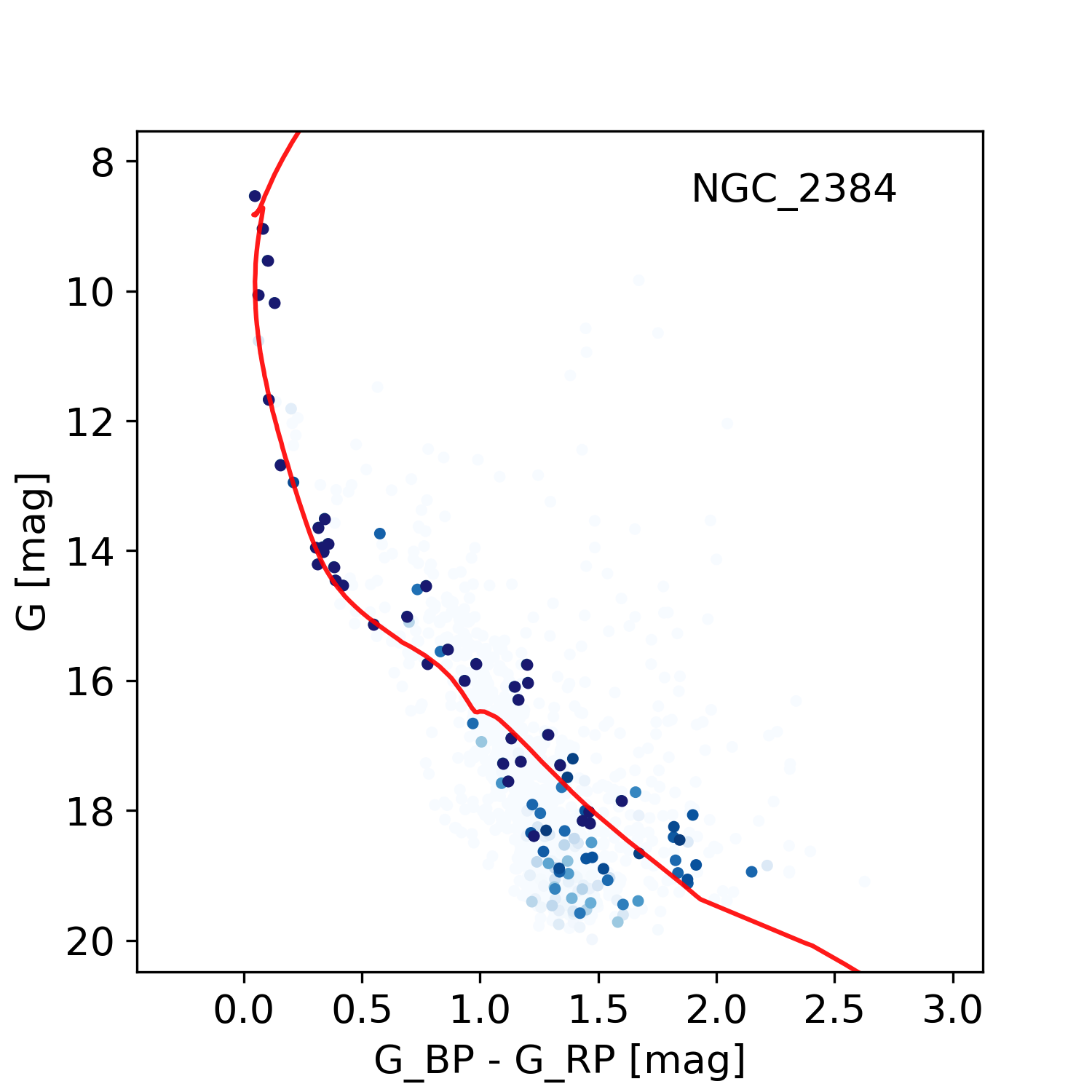}
 \includegraphics[scale = 0.45]{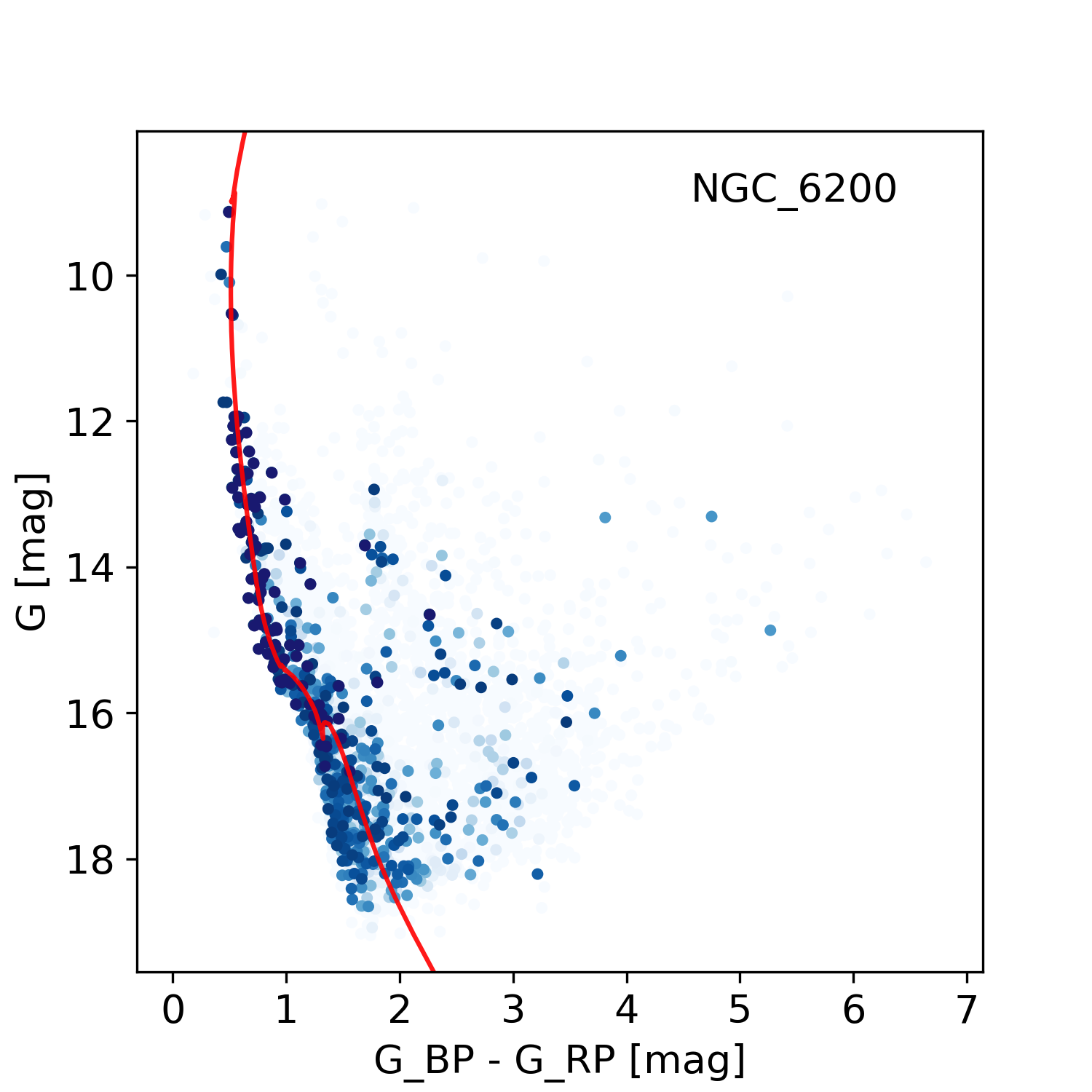}
 \includegraphics[scale = 0.45]{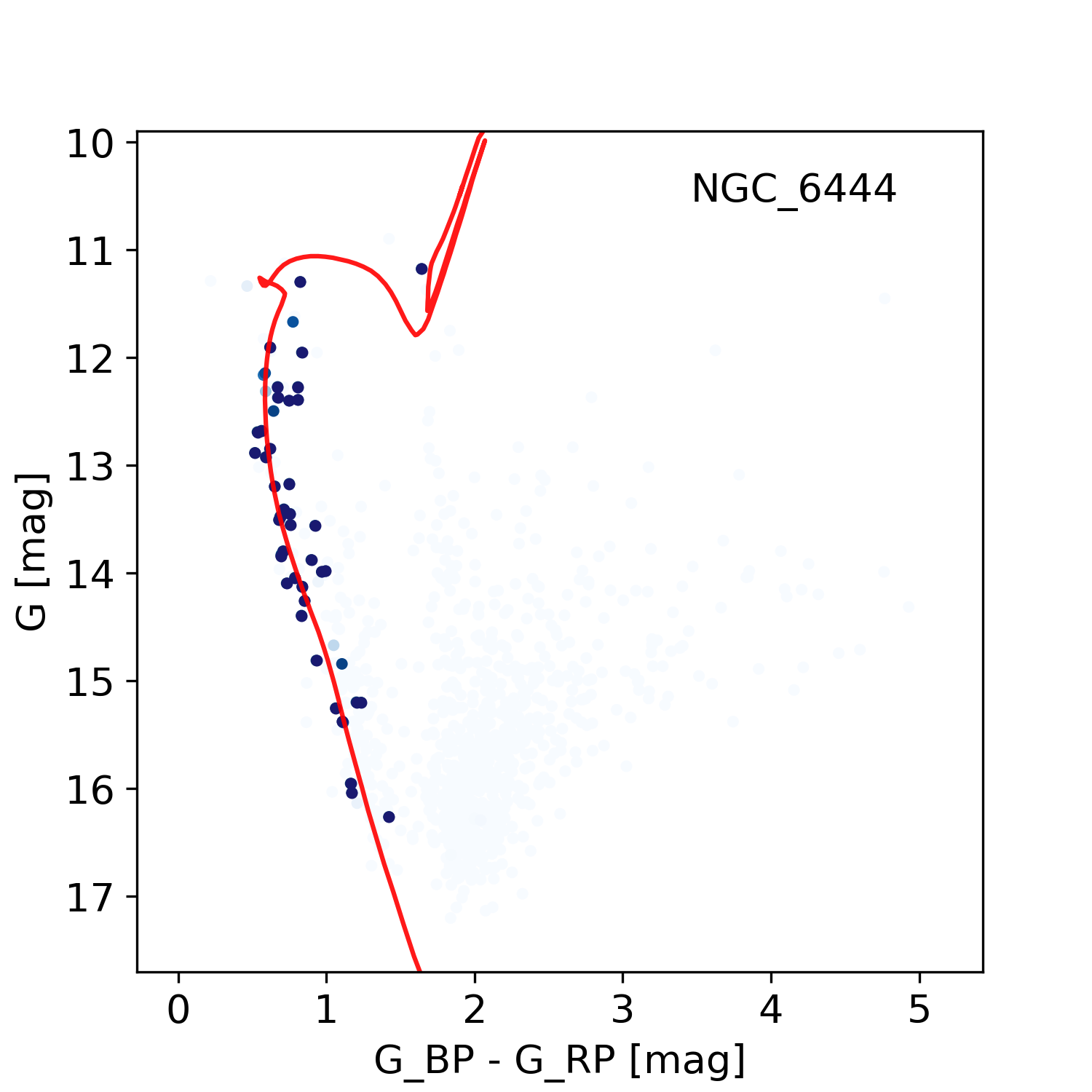}\\
 \includegraphics[scale = 0.45]{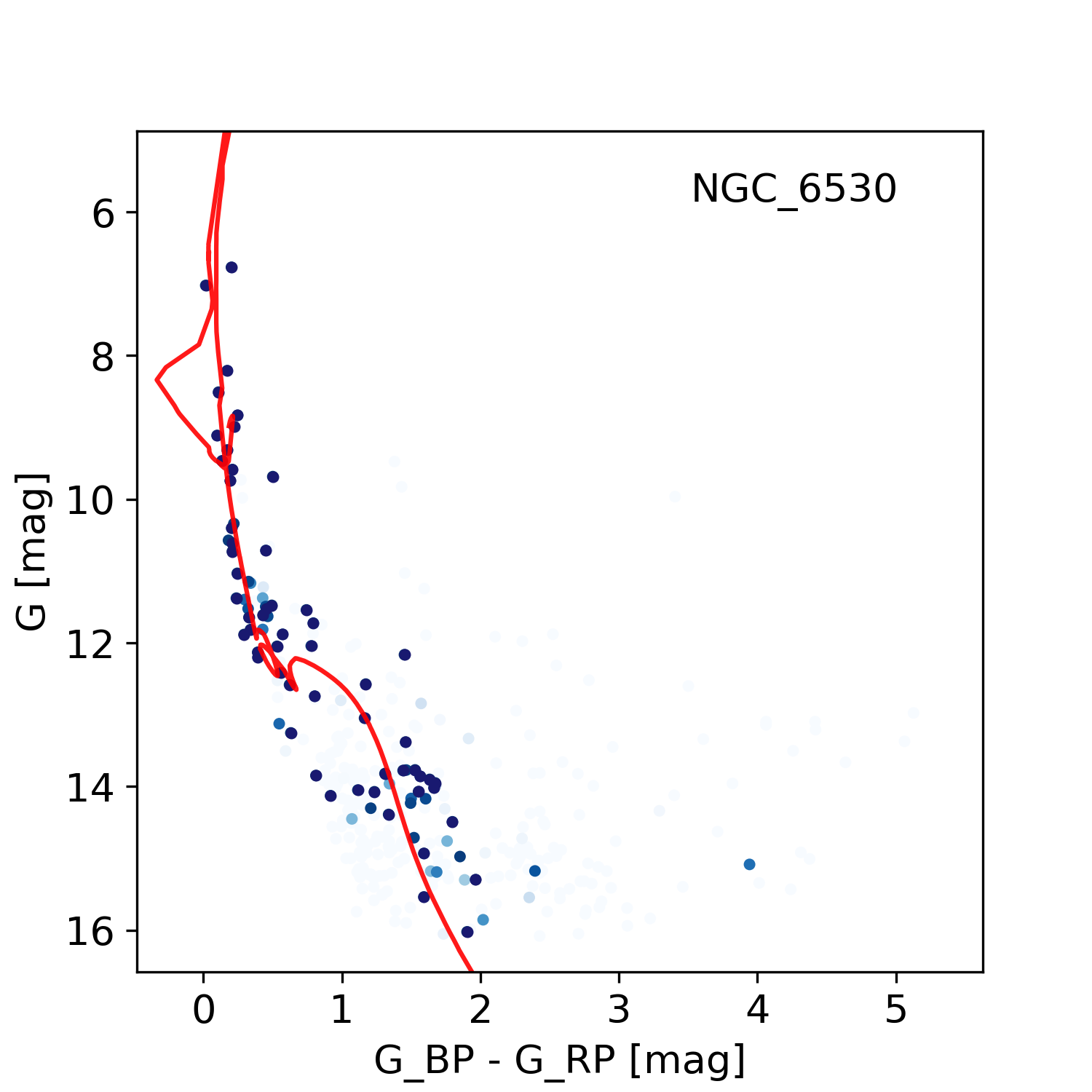}
 \includegraphics[scale = 0.45]{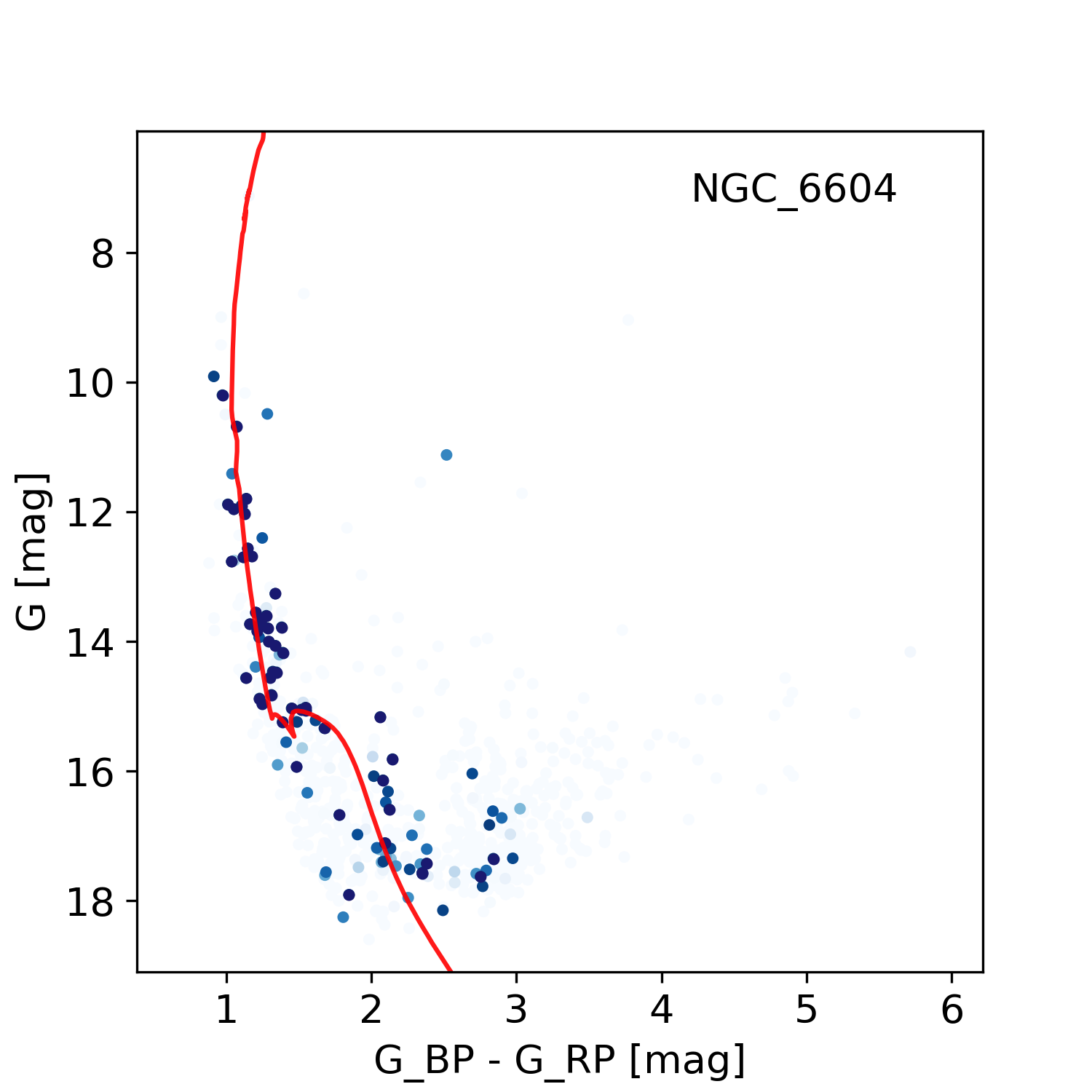}
 \includegraphics[scale = 0.45]{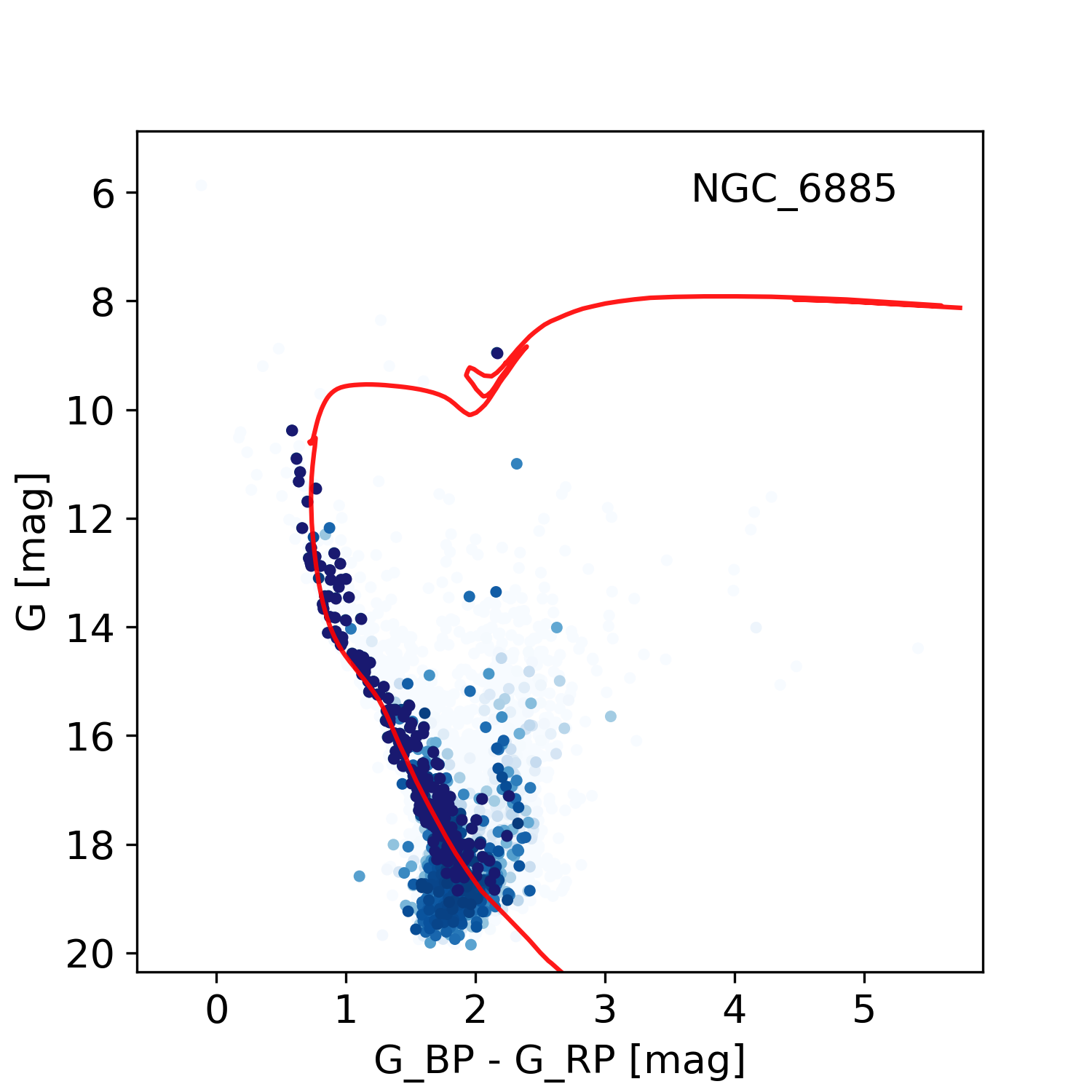} 
 \caption{CMDs and isochrone fits (continued)}
 \end{figure*}

\addtocounter{figure}{-1}
 \begin{figure*}
 \raggedright
 \includegraphics[scale = 0.45]{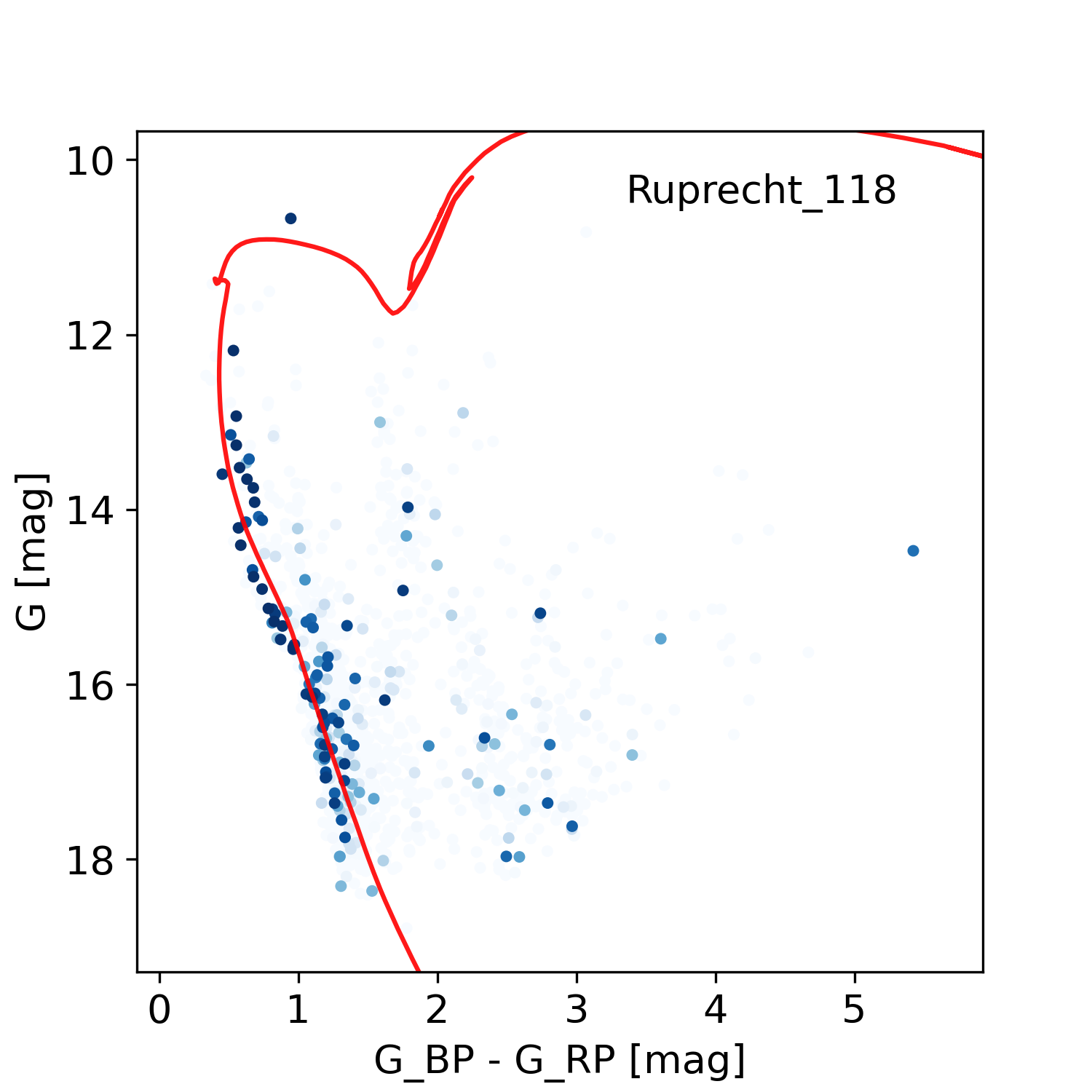}
 \includegraphics[scale = 0.45]{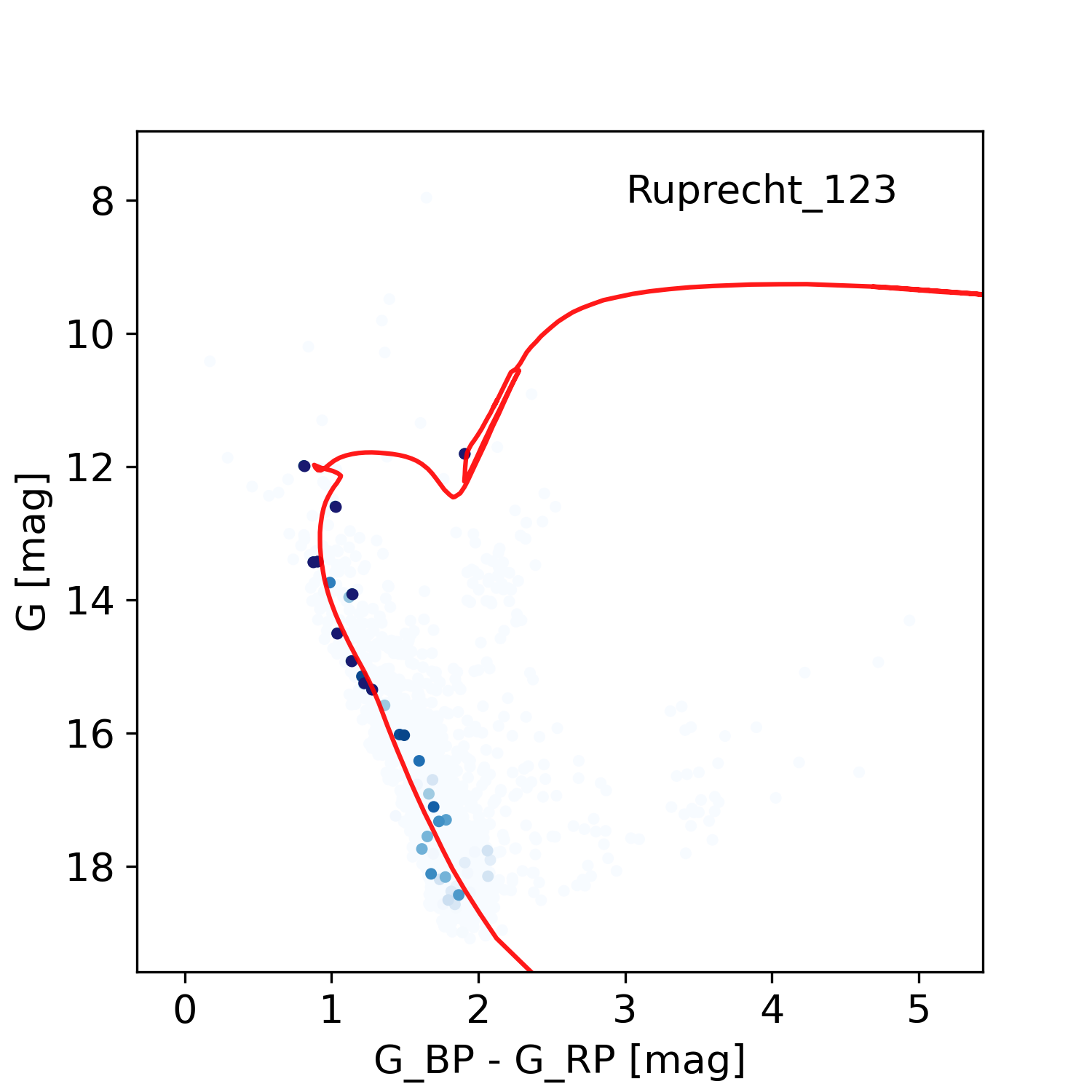}
 \includegraphics[scale = 0.45]{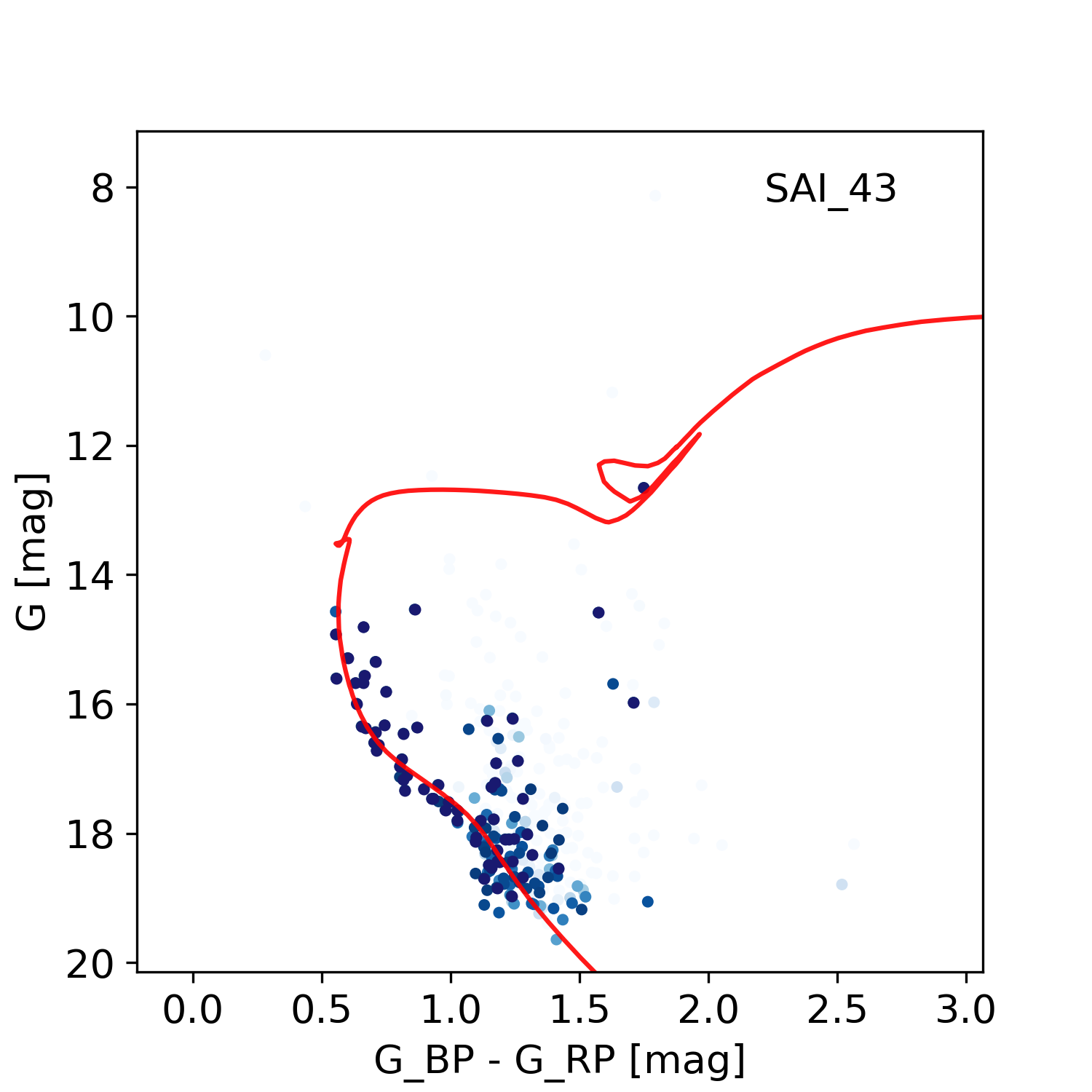}\\
 
 \includegraphics[scale = 0.45]{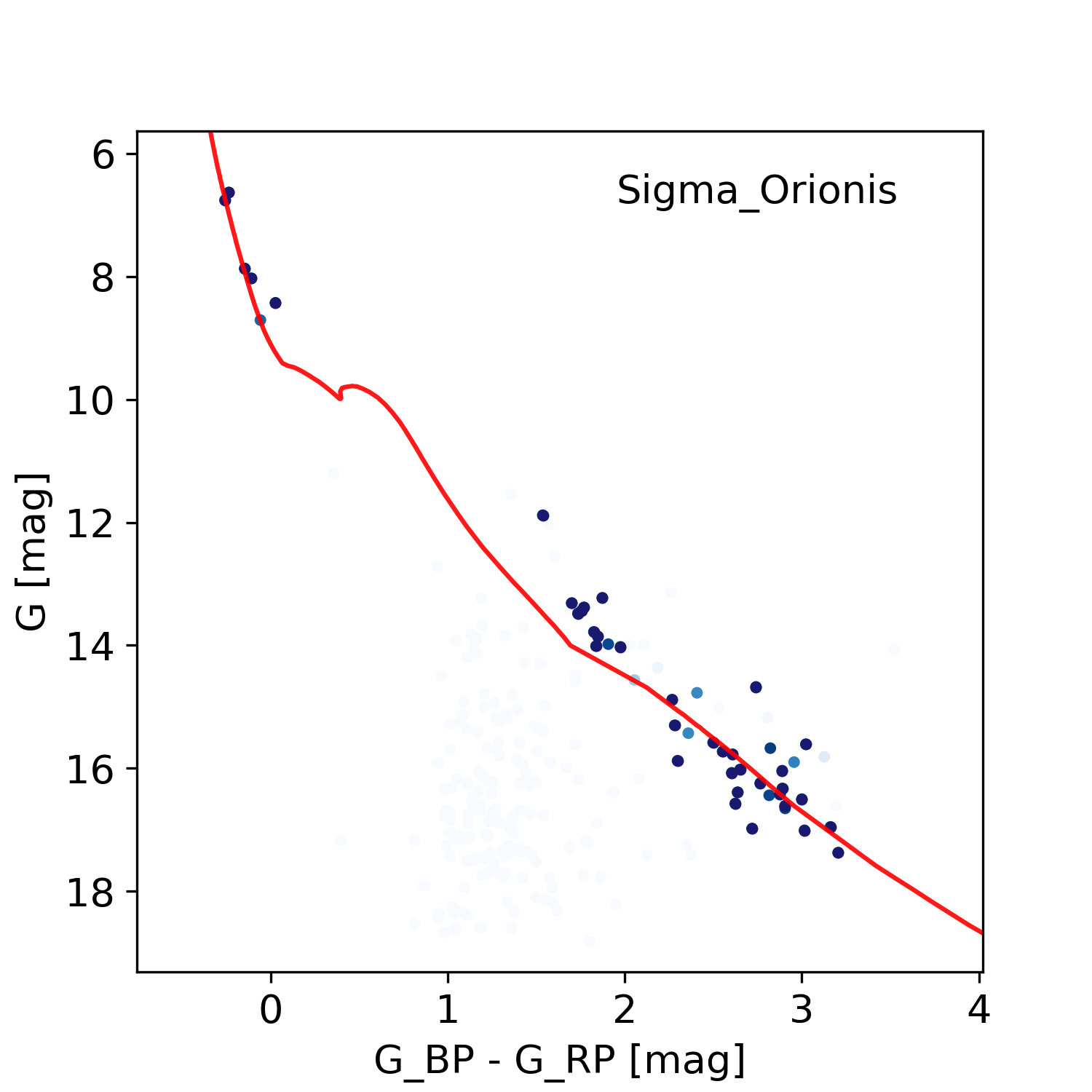}
 \includegraphics[scale = 0.45]{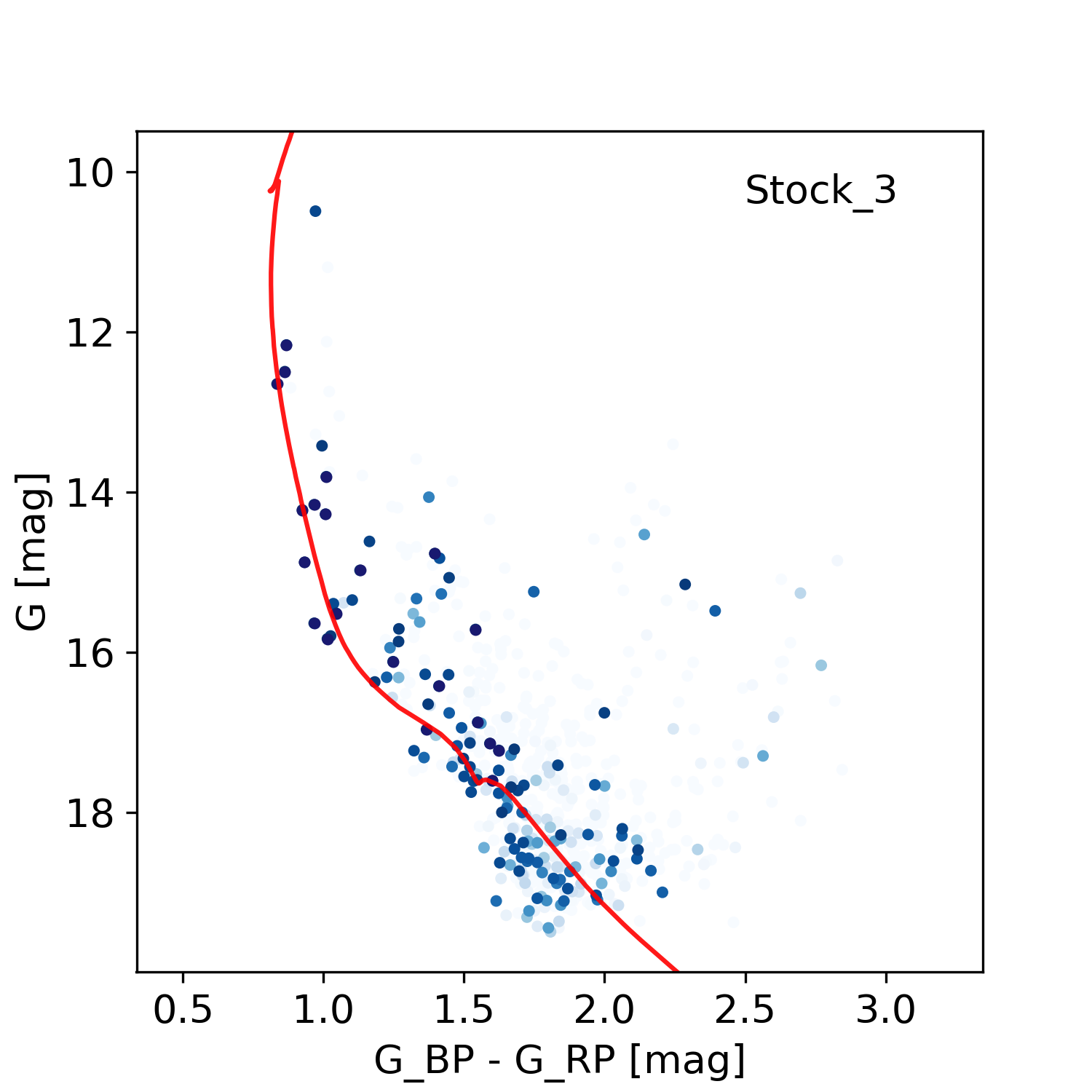} 
 \includegraphics[scale = 0.45]{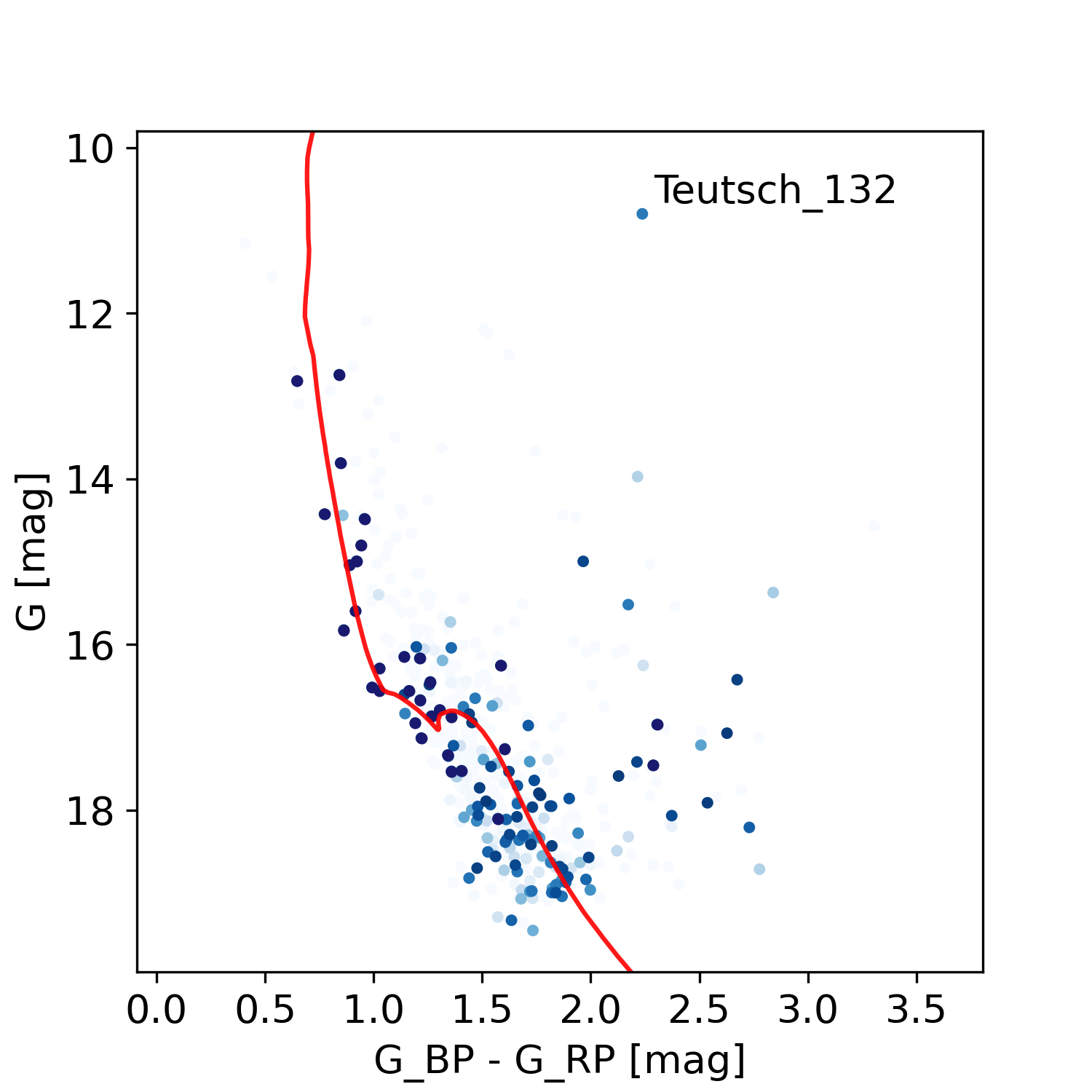}\\
 
 \includegraphics[scale = 0.45]{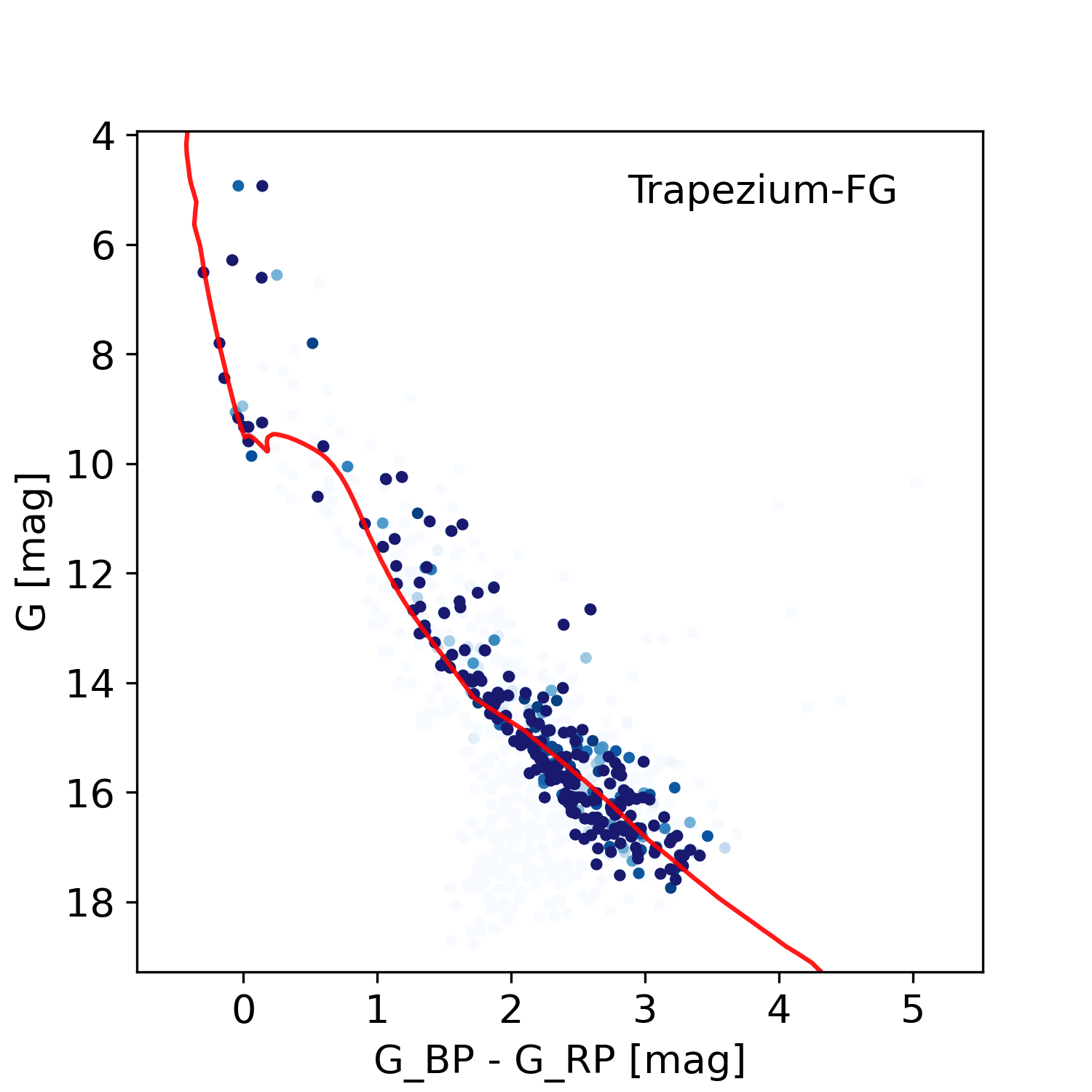}
 \includegraphics[scale = 0.45]{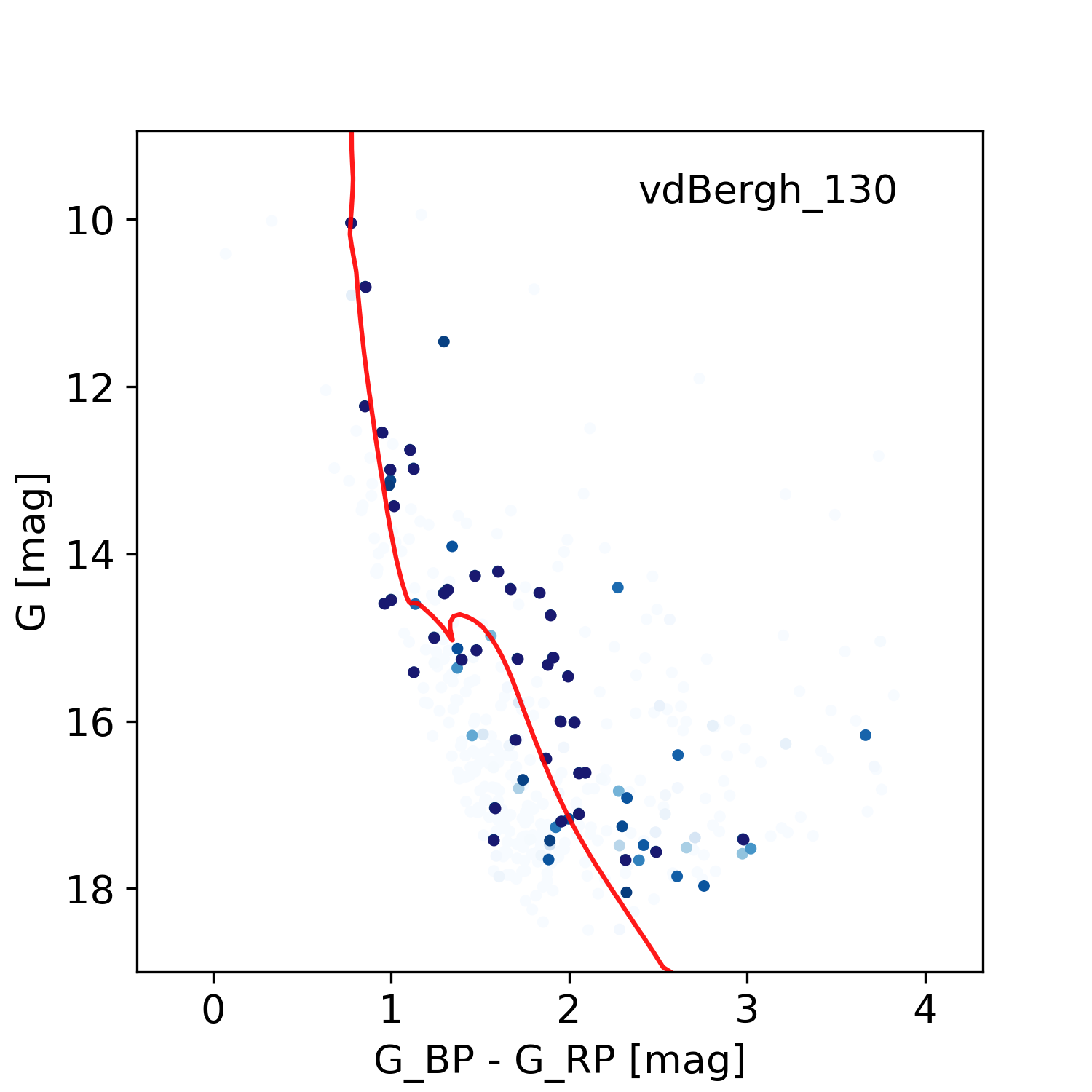}
 \caption{CMDs and isochrone fits (continued)}
 \end{figure*}

\section*{Data Availability Statement}
The data underlying this article, that support the plots
and other findings, are available in the article and in its online supplementary material.  

This work has made use of data from the European Space Agency(ESA) Gaia (http://www.cosmos.esa.int/gaia) mission, processedby the Gaia Data Processing and Analysis Consortium (DPAC,http://www.cosmos.esa.int/web/gaia/dpac/consortium).

We also employed catalogs from CDS/Simbad (Strasbourg)
and Digitized Sky Survey images from the Space Telescope Science
Institute (US Government grant NAG W-2166)

\section*{Acknowledgements}
W. S. Dias acknowledges the S\~ao Paulo State Agency
FAPESP (fellowship 2013/01115-6). H. Monteiro would like to thank
FAPEMIG grants APQ-02030-10 and CEX-PPM-00235-12. AM acknowledges the support from the Portuguese FCT Strategic Programme UID/FIS/00099/2019 for CENTRA. 
This research was performed using the facilities of the Laborat\'orio de Astrof\'isica Computacional da Universidade Federal de Itajub\'a (LAC-UNIFEI).




\bibliographystyle{mnras}
\bibliography{refs} 




\appendix

\onecolumn
\section{Results for the high resolution spectroscopy validation sample}

\begin{landscape}
\setlength\LTcapwidth{\linewidth}
\begin{longtable}{@{\extracolsep{4pt}}lccccccrcrcrcrc@{}}
\caption{Results of isochrone fits done using the $[Fe/H]$ prior based on the metallicity gradient from \citet{OCCAMgradient20} for the control sample described in section \ref{sec:validation}. The available $[Fe/H]$ values and uncertainties from the literature used in the comparisons are also presented. The \emph{h} and \emph{l} suffixes in values from \citet{Netopil16} denote their high and low quality samples, respectively.}\label{tab:fitsfeh}
\endfirsthead
\multicolumn{15}{c}%
{Table {\thetable}: continued from previous page} \\
\hline
& & & & & & & & & 
\multicolumn{2}{c}{\citet{apogee_sample}}      & \multicolumn{4}{c}{\citet{Netopil16}}      \\
\cline{10-11} \cline{12-15}
Name           & $dist$& $\sigma_{dist}$ & $age$   & $\sigma_{age}$ & $A_V$    & $\sigma_{A_V}$ & \multicolumn{1}{c}{$[Fe/H]$}   & $\sigma_{[Fe/H]}$   & \multicolumn{1}{c}{$[Fe/H]$}   & $\sigma_{[Fe/H]}$   & \multicolumn{1}{c}{$[Fe/H]_h$}   & $\sigma_{[Fe/H]_h}$   & \multicolumn{1}{c}{$[Fe/H]_l$}   & $\sigma_{[Fe/H]_l}$  \\
\hline
\endhead

\hline
\endfoot
\hline
& & & & & & & & & 
\multicolumn{2}{c}{\citet{apogee_sample}}      & \multicolumn{4}{c}{\citet{Netopil16}}      \\
\cline{10-11} \cline{12-15}
Name           & $dist$& $\sigma_{dist}$ & $age$   & $\sigma_{age}$ & $A_V$    & $\sigma_{A_V}$ & \multicolumn{1}{c}{$[Fe/H]$}   & $\sigma_{[Fe/H]}$   & \multicolumn{1}{c}{$[Fe/H]$}   & $\sigma_{[Fe/H]}$   & \multicolumn{1}{c}{$[Fe/H]_h$}   & $\sigma_{[Fe/H]_h}$   & \multicolumn{1}{c}{$[Fe/H]_l$}   & $\sigma_{[Fe/H]_l}$  \\
\hline

ASCC 21       & 344  & 2    & 7.062 & 0.039 & 0.201 & 0.029 & -0.026 & 0.165 & 0.01   & 0.09 &    -  &   -  &    -  &   -  \\
Basel 11b     & 1663 & 71   & 8.451 & 0.068 & 1.856 & 0.094 & -0.090  & 0.184 & 0.01   & 0.05 &    -   &   -  &    -  &   -  \\
Berkeley 17   & 3278 & 105  & 9.791 & 0.110  & 1.923 & 0.090  & -0.173 & 0.157 & -0.10   & 0.04 & -0.06   &   -  &    -  &   -  \\
Berkeley 19   & 6393 & 790  & 9.271 & 0.042 & 1.493 & 0.134 & -0.431 & 0.187 & -0.22  &   -  &     -   &   -  &    -  &   -  \\
Berkeley 31   & 7019 & 329  & 9.502 & 0.027 & 0.540  & 0.032 & -0.302 & 0.157 & -0.31  & 0.04 &     -   &   -  &    -  &   -  \\
Berkeley 33   & 4467 & 277  & 8.520  & 0.051 & 2.085 & 0.043 & -0.268 & 0.181 & -0.23  & 0.11 &     -   &   -  & -0.26 & 0.05 \\
Berkeley 43   & 1994 & 93   & 7.660  & 0.338 & 4.774 & 0.047 & 0.171  & 0.156 & 0.00      &   -  &     -   &   -  &    -  &   -  \\
Berkeley 53   & 4525 & 263  & 8.885 & 0.021 & 4.391 & 0.044 & -0.090  & 0.171 & -0.02  & 0.03 &     -   &   -  &    -  &   -  \\
Berkeley 66   & 8738 & 1318 & 8.637 & 0.088 & 3.886 & 0.072 & -0.110  & 0.180  & -0.12  & 0.01 &     -   &   -  &    -  &   -  \\
Berkeley 71   & 3203 & 138  & 8.827 & 0.030  & 3.045 & 0.052 & -0.100   & 0.157 & -0.20   & 0.03 &     -   &   -  &    -  &   -  \\
Berkeley 9    & 1720 & 135  & 9.187 & 0.060  & 2.976 & 0.037 & -0.100   & 0.198 & -0.17  & 0.18 &     -   &   -  &    -  &   -  \\
Berkeley 98   & 3391 & 78   & 9.504 & 0.032 & 0.753 & 0.022 & -0.090  & 0.154 & 0.03   & 0.02 &     -   &   -  &    -  &   -  \\
Collinder 69  & 398  & 1    & 6.880  & 0.043 & 0.405 & 0.045 & -0.100   & 0.160  & -0.01  & 0.06 &     -   &   -  &    -  &   -  \\
Collinder 95  & 661  & 8    & 6.791 & 0.152 & 0.886 & 0.643 & -0.065 & 0.176 & -0.03  & 0.02 &     -   &   -  &    -  &   -  \\
Czernik 21    & 3900 & 510  & 8.915 & 0.123 & 3.211 & 0.160  & -0.268 & 0.180  & -0.24  & 0.01 &     -   &   -  &    -  &   -  \\
Czernik 23    & 3070 & 172  & 8.474 & 0.544 & 1.761 & 0.104 & -0.100   & 0.177 & -0.25  &   -  &     -   &   -  &    -  &   -  \\
Czernik 30    & 5729 & 365  & 9.466 & 0.023 & 0.976 & 0.098 & -0.289 & 0.213 & -0.28  & 0.02 &     -   &   -  &    -  &   -  \\
FSR 0496      & 1506 & 70   & 8.814 & 0.023 & 3.112 & 0.054 & -0.130  & 0.166 & -0.07  &   -  &     -   &   -  &    -  &   -  \\
FSR 0542      & 5506 & 709  & 8.889 & 0.089 & 3.514 & 0.107 & -0.177 & 0.290  & -0.19  &   -  &     -   &   -  &    -  &   -  \\
FSR 0667      & 1100 & 28   & 8.655 & 0.129 & 1.502 & 0.167 & 0.154  & 0.227 & 0.03   & 0.01 &     -   &   -  &    -  &   -  \\
FSR 0716      & 3388 & 169  & 9.043 & 0.051 & 1.242 & 0.122 & -0.243 & 0.183 & -0.30   &   -  &     -   &   -  &    -  &   -  \\
FSR 0941      & 4029 & 37   & 8.826 & 0.085 & 2.445 & 0.057 & -0.100   & 0.225 & -0.23  &   -  &     -   &   -  &    -  &   -  \\
FSR 0942      & 3151 & 510  & 8.840  & 0.139 & 2.414 & 0.178 & -0.122 & 0.172 & -0.28  &   -  &     -   &   -  &    -  &   -  \\
Gulliver 6    & 415  & 2    & 7.137 & 0.079 & 0.304 & 0.077 & 0.031  & 0.183 & -0.10   &   -  &     -   &   -  &    -  &   -  \\
Haffner 4     & 3758 & 260  & 8.950  & 0.118 & 1.430  & 0.107 & -0.326 & 0.191 & -0.13  &   -  &     -   &   -  &    -  &   -  \\
IC 1369       & 2683 & 284  & 7.773 & 0.695 & 2.594 & 0.122 & -0.018 & 0.159 & -0.02  & 0.01 &     -   &   -  &    -  &   -  \\
IC 1805       & 1849 & 113  & 6.805 & 0.081 & 2.296 & 0.022 & -0.056 & 0.189 & 0.32   &   -  &     -   &   -  &    -  &   -  \\
King 15       & 2727 & 140  & 8.493 & 0.452 & 1.926 & 0.083 & -0.100   & 0.243 & -0.05  &   -  &     -   &   -  &    -  &   -  \\
Kronberger 57 & 2211 & 663  & 6.738 & 0.799 & 4.336 & 0.148 & 0.383  & 0.308 & 0.02   &   -  &     -   &   -  &    -  &   -  \\
Melotte 20    & 174  & 3    & 7.858 & 0.025 & 0.386 & 0.040  & 0.036  & 0.162 & 0.08   & 0.07 & 0.14    & 0.11 &    -  &   -  \\
Melotte 22    & 136  & 1    & 8.090  & 0.097 & 0.154 & 0.051 & 0.127  & 0.167 & 0.06   & 0.08 & -0.01   & 0.05 &    -  &   -  \\
Melotte 71    & 1966 & 47   & 9.097 & 0.031 & 0.512 & 0.060  & -0.100   & 0.176 & -0.09  & 0.02 & -0.27   &   -  &    -  &   -  \\
NGC 1193      & 5166 & 206  & 9.713 & 0.057 & 0.674 & 0.028 & -0.221 & 0.159 & -0.25  & 0.01 & -0.22   &   -  &    -  &   -  \\
NGC 1245      & 2636 & 42   & 9.096 & 0.016 & 0.871 & 0.026 & -0.100   & 0.153 & -0.06  & 0.03 & 0.02    & 0.03 & -0.05 & 0.06 \\
NGC 136       & 4648 & 282  & 8.376 & 0.681 & 2.124 & 0.133 & -0.124 & 0.172 & -0.22  &   -  &     -   &   -  &    -  &   -  \\
NGC 1664      & 1197 & 23   & 8.790  & 0.022 & 0.918 & 0.068 & -0.127 & 0.156 & -0.01  &   -  &     -   &   -  &    -  &   -  \\
NGC 1798      & 4741 & 243  & 9.139 & 0.014 & 1.725 & 0.086 & -0.294 & 0.191 & -0.18  & 0.02 &     -   &   -  &    -  &   -  \\
NGC 1817      & 1544 & 39   & 9.078 & 0.017 & 0.785 & 0.067 & -0.119 & 0.166 & -0.09  &   -  & -0.11   & 0.03 & -0.16 & 0.03 \\
NGC 1857      & 2506 & 114  & 8.377 & 0.398 & 1.679 & 0.087 & -0.192 & 0.176 & -0.12  &   -  &     -   &   -  &    -  &   -  \\
NGC 188       & 1836 & 5    & 9.789 & 0.018 & 0.353 & 0.072 & -0.062 & 0.161 & 0.13   & 0.05 & 0.11    & 0.04 & -0.02 & 0.09 \\
NGC 1907      & 1539 & 54   & 8.681 & 0.142 & 1.672 & 0.147 & -0.268 & 0.174 & -0.05  & 0.01 &     -   &   -  &    -  &   -  \\
NGC 1912      & 1058 & 22   & 8.479 & 0.123 & 0.937 & 0.068 & 0.048  & 0.164 & -0.07  & 0.02 & -0.10    & 0.14 &    -  &   -  \\
NGC 2158      & 4030 & 306  & 9.381 & 0.065 & 1.495 & 0.067 & -0.268 & 0.156 & -0.15  & 0.03 &     -   &   -  & -0.32 & 0.08 \\
NGC 2168      & 845  & 16   & 8.145 & 0.168 & 0.903 & 0.077 & -0.110  & 0.173 & -0.13  & 0.07 &     -   &   -  & -0.21 & 0.10  \\
NGC 2183      & 786  & 34   & 7.006 & 0.202 & 1.670  & 0.473 & -0.100   & 0.199 & -0.08  & 0.08 &     -   &   -  &    -  &   -  \\
NGC 2243      & 4005 & 106  & 9.542 & 0.044 & 0.168 & 0.024 & -0.358 & 0.154 & -0.42  &   -  &     -   &   -  & -0.50  & 0.08 \\
NGC 2244      & 1287 & 107  & 7.093 & 0.143 & 1.586 & 0.091 & -0.121 & 0.165 & -0.23  & 0.09 &     -   &   -  &    -  &   -  \\
NGC 2304      & 3814 & 143  & 8.977 & 0.034 & 0.308 & 0.103 & -0.275 & 0.171 & -0.09  & 0.09 &     -   &   -  &    -  &   -  \\
NGC 2318      & 1271 & 41   & 8.878 & 0.076 & 0.839 & 0.128 & 0.078  & 0.158 & 0.01   &   -  &     -   &   -  &    -  &   -  \\
NGC 2324      & 3732 & 70   & 8.749 & 0.036 & 0.814 & 0.073 & -0.215 & 0.156 & -0.15  & 0.05 & -0.22   & 0.07 &    -  &   -  \\
NGC 2355      & 1837 & 20   & 9.086 & 0.034 & 0.329 & 0.015 & 0.042  & 0.153 & -0.11  &   -  & -0.05   & 0.08 & -0.08 & 0.08 \\
NGC 2420      & 2471 & 101  & 9.407 & 0.045 & 0.123 & 0.009 & -0.218 & 0.158 & -0.12  & 0.03 & -0.05   & 0.02 & -0.21 & 0.09 \\
NGC 2682      & 855  & 4    & 9.561 & 0.004 & 0.185 & 0.030  & -0.031 & 0.154 & 0.02   & 0.07 & 0.03    & 0.05 & 0.00     & 0.06 \\
NGC 6705      & 1922 & 50   & 8.440  & 0.163 & 1.502 & 0.071 & 0.046  & 0.16  & 0.16   & 0.03 & 0.12    & 0.09 & 0.25  & 0.05 \\
NGC 6791      & 4422 & 74   & 9.946 & 0.053 & 0.391 & 0.052 & 0.221  & 0.165 & 0.40    & 0.07 & 0.42    & 0.05 & 0.35  & 0.07 \\
NGC 6811      & 1097 & 17   & 9.021 & 0.022 & 0.232 & 0.048 & 0.015  & 0.161 & -0.01  & 0.04 & 0.03    & 0.01 &    -  &   -  \\
NGC 6819      & 2310 & 141  & 9.459 & 0.037 & 0.507 & 0.026 & 0.011  & 0.166 & 0.10    & 0.04 & 0.09    & 0.01 & -0.04 & 0.08 \\
NGC 6866      & 1392 & 44   & 8.844 & 0.034 & 0.477 & 0.076 & 0.183  & 0.156 & 0.04   & 0.02 &     -   &   -  &    -  &   -  \\
NGC 7058      & 362  & 2    & 7.860  & 0.584 & 0.291 & 0.183 & -0.100   & 0.175 & 0.12   & 0.04 &     -   &   -  &    -  &   -  \\
NGC 7062      & 2109 & 168  & 8.643 & 0.397 & 1.730  & 0.154 & -0.081 & 0.173 & 0.04   &   -  &     -   &   -  &    -  &   -  \\
NGC 752       & 444  & 4    & 9.179 & 0.019 & 0.166 & 0.061 & -0.037 & 0.159 & 0.01   &   -  & -0.03   & 0.06 & -0.09 & 0.13 \\
Teutsch 12    & 3939 & 331  & 8.948 & 0.036 & 1.969 & 0.118 & -0.118 & 0.185 & -0.14  & 0.02 &     -   &   -  &    -  &   -  \\
Teutsch 51    & 5387 & 443  & 8.817 & 0.068 & 3.311 & 0.083 & -0.285 & 0.181 & -0.28  & 0.04 &     -   &   -  &    -  &   -  \\
Tombaugh 4    & 3127 & 215  & 8.918 & 0.062 & 3.218 & 0.065 & -0.103 & 0.154 & -0.47  &   -  &     -   &   -  &    -  &   -  \\
Trumpler 26   & 1336 & 76   & 8.669 & 0.114 & 1.703 & 0.108 & 0.175  & 0.172 & 0.28   & 0.05 &     -   &   -  &    -  &   -  \\
Trumpler 3    & 663  & 8    & 8.094 & 0.079 & 0.931 & 0.042 & 0.156  & 0.150  & -0.22  &   -  &     -   &   -  &    -  &   -  \\
Trumpler 5    & 3275 & 56   & 9.536 & 0.025 & 1.846 & 0.066 & -0.152 & 0.164 & -0.36  & 0.02 & -0.44   & 0.07 & -0.47 & 0.05 \\
\end{longtable}
\end{landscape}

\clearpage
\section{Removed clusters}
Here we identify the clusters that have been removed from our studied sample, as discussed in section~\ref{sec:results}.

\begin{table*}[!H]
\caption[]{Removed clusters. Central coordinates and radii from DAML.}
\label{tab:removed}
\begin{center}
\begin{tabular}{lrrc}
\hline
   Name             &  
   \multicolumn{1}{c}{$RA_{J2000}$} & 
   \multicolumn{1}{c}{$DE_{J2000}$}  &  
   $radius$  \\
   $ $            &  
   \multicolumn{1}{c}{$(deg)$} & 
   \multicolumn{1}{c}{$(deg)$} & 
   \multicolumn{1}{c}{$(deg)$} \\
\hline
ASCC 94 & 273.9000 & -14.9900 & 0.250 \\ 
BH 4 & 114.4333 & -36.0667 & 0.017 \\ 
Bochum 1 & 96.3542 & 19.7667 & 0.217 \\ 
Collinder 347 & 266.5750 & -29.3333 & 0.083 \\ 
Collinder 92 & 95.7250 & 5.1167 & 0.092 \\ 
Dolidze 13 & 12.4250 & 64.1264 & 0.133 \\ 
Dolidze 24 & 101.1708 & 1.6847 & 0.157 \\ 
Dolidze 35 & 291.3500 & 11.6583 & 0.058 \\ 
Dolidze 41 & 304.7042 & 37.7500 & 0.092 \\ 
Dolidze 49 & 101.7667 & -0.0069 & 0.018 \\ 
ESO 522 05 & 273.2208 & -24.3639 & 0.037 \\ 
FSR 0182 & 297.9417 & 33.5119 & 0.010 \\ 
FSR 0258 & 311.2083 & 43.9150 & 0.013 \\ 
FSR 0354 & 332.8000 & 57.6994 & 0.043 \\ 
FSR 0453 & 356.8542 & 63.2264 & 0.037 \\ 
FSR 0522 & 13.4583 & 65.7933 & 0.006 \\
FSR 0717 & 71.5250 & 42.1342 & 0.018 \\
FSR 0891 & 94.3708 & 22.4272 & 0.012 \\
FSR 0929 & 96.3833 & 17.7200 & 0.007 \\
FSR 1535 & 151.9792 & -59.1969 & 0.018 \\
Hogg 11 & 167.9042 & -60.4000 & 0.017 \\
Kronberger 39 & 163.5583 & -61.7378 & 0.007 \\
Majaess 50 & 71.3625 & 41.9758 & 0.142 \\
Majaess 95 & 124.4708 & -35.8800 & 0.025 \\
NGC 2013 & 86.0042 & 55.7933 & 0.050 \\
Patchick 78 & 8.2917 & 65.1167 & 0.013 \\
Ruprecht 120 & 248.7917 & -48.2833 & 0.025 \\
Ruprecht 136 & 269.8250 & -24.7000 & 0.025 \\
Ruprecht 59 & 124.8375 & -34.4833 & 0.025 \\
Teutsch 64 & 128.1292 & -41.9881 & 0.038 \\

\hline
\end{tabular}
\end{center}
\end{table*}


\bsp	
\label{lastpage}
\end{document}